\documentclass[12pt,preprint]{aastex}
\newcommand{\ve}[1]{\mbox{\boldmath${#1}$}}
\newcommand{\dif}[2]{{\partial #1 \over \partial #2}}

\def\lesssim{\mathrel{\hbox{\rlap{\hbox{\lower4pt\hbox{$\sim$}}}\hbox{$<$}}}}
\def\gtrsim{\mathrel{\hbox{\rlap{\hbox{\lower4pt\hbox{$\sim$}}}\hbox{$>$}}}}

\slugcomment{Submitted to Astrophysical Journal}

\shorttitle{Evolution of Supernova Remnants}
\shortauthors{Hanayama, H. \& Tomisaka, K., 2005}

\begin{document}

\title{Long-Term Evolution of Supernova Remnants in
 Magnetized Interstellar Medium}

\author{Hidekazu Hanayama\altaffilmark{1}}
\affil{Department of Astronomy, Graduate School of Science,
University of Tokyo, 
Tokyo 113-0033, Japan}
\author{Kohji Tomisaka\altaffilmark{2}}
\affil{Theoretical Astrophysics, National Astronomical Observatory, Mitaka, 
Tokyo 181-8588, Japan}
\altaffiltext{1}{hanayama@th.nao.ac.jp}
\altaffiltext{2}{also School of Physical Sciences, The Graduate University for
 Advanced Studies (SOKENDAI); tomisaka@th.nao.ac.jp}

\begin{abstract}
The evolution of supernova remnants (SNRs) is studied, with particular
 attention to the effect of magnetic fields with axisymmetric 
 two-dimensional magnetohydrodynamical simulations.
The evolution of magnetic SNRs is the same as non-magnetic ones
 in the adiabatic Sedov stage.
After a thin shell is formed, the shell is driven by the pressure of the hot interior gas (bubble).
Evolution in the pressure-driven snow-plow phase is much affected by the magnetic field.
The shell sweeping the magnetic field lines thickens owing to the magnetic pressure force.
After $5\times 10^5{\rm yr}$ - $2\times 10^6{\rm yr}$,
 the inner boundary of the thick shell begins to contract. 
This compresses the hot bubble radially and maintains its thermal pressure.
Thus, the bubble forms a prolate spheroidal shape and becomes thinner and thinner,
 since it expands in a direction parallel to the magnetic field for
 $B_0 \gtrsim 3\mu {\rm G}$.
Finally, the bubble contracts. 
The porosity of the hot low-density gas in ISM is reduced,
 taking the effect of the magnetic field into account.
\end{abstract}

\section{Introduction}

Structure of the interstellar medium (ISM)
 is a key issue in the understanding of our Galaxy. 
A large fraction of the interstellar space is occupied
 with warm ($T\sim 10^4{\rm K}$) and hot ($T\gtrsim 3\times 10^5{\rm K}$)
 media.  
The origin of the hot component is the energy released from supernovae 
 and/or stellar winds.  
Shock waves of the supernova remnants (SNRs) hit the interstellar gas to heat
 it up and to sweep it into a dense shell. 
The evolution of SNRs has been studied for many years \citep[for example,][] 
 {cox72,che74,cio88}.
After the shock expansion speed decelerates below
 $\simeq 260 n^{0.08}{\rm km\,s^{-1}}$, 
 the gas passed through the shock front is quickly cooled and a dense shell
 forms.
The expansion of the shell is decelerated as the interstellar gas accumulates.
Finally, the expansion speed of the shock front reaches the random speed of 
 turbulent motion in the ISM and the shock wave decays to a sound wave. 

The interstellar magnetic field is seen at a glance to run in a direction 
parallel to the galactic disk.
As for the strength, combining rotation and dispersion measures,
\citet{ran89} obtained a turbulent field strength of 
 $\simeq 5 \mu {\rm G}$ and its cell size of $\simeq$55 pc.  
The energy density of the magnetic field 
 $B_0^2/8\pi\simeq 10^{-12}{\rm dyn\,cm^{-2}}(B_0/5\mu{\rm G})^2$ attains
 that of the ISM turbulent motion $v_t$ with a density $\rho_0$ as
 $\rho_0 v_t^2/2 \simeq 10^{-12}{\rm dyn\,cm^{-2}}
 (\rho_0/2\times 10^{-24}{\rm g\,cm^{-3}})(v_{t}/10{\rm km\,s^{-1}})^2$ and
 an interstellar thermal energy 
 $p=kT\rho/\mu H\simeq3\times 10^{-12}$ ${\rm dyn\,cm^{-2}}(T/10^4{\rm K})
 (\rho_0/2\times 10^{-24}{\rm g\,cm^{-3}})$.
These three energies are in approximate equipartition.
This seems to indicate the magnetic field plays an important role in the ISM.
At the same time, the late evolutionary phase of SNRs must also be significantly affected by the
 interstellar magnetic field. 

Magnetic SNRs were first studied by \citet{che74}
 using spherically symmetric one-dimensional hydro simulations
 including only the magnetic pressure force. 
\citet{sla92} pointed out the effects of the magnetic field as 
(1) an SNR forms a thick shell supported by the magnetic pressure force;
(2) the hot gas heated by the shock wave in the adiabatic phase
    occupies the interior of the SNR. 
    This hot cavity is effectively re-compressed
    in the later phase by the effect of the 
    magnetic pressure from their spherically symmetric calculations.
Similar one-dimensional simulations have also been done by \cite{smi01}. 
The tension force of the magnetic field has similar effects, broadening 
the shell and compressing the hot bubble, but is not included in these 
 spherically symmetric simulations.
At least two-dimensional magnetohydrodynamical (MHD) simulations are necessary. %\citet{han05} performed a series of two-dimensional magnetohydrodynamic simulations of primordial SNR with the Biermann term and estimated the amplitude and total energy of the produced magnetic fields.
\citet{bal01} have studied the evolution of SNRs expanding 
in the magnetic turbulent ISM using three-dimensional MHD simulations 
and they have explained the evolution of SNRs in the thermal X-ray 
and non-thermal radio continuum.
However, their study is restricted to the adiabatic stage.
In the present paper, we explore the total evolution of SNRs in the magnetized ISM using axisymmetric MHD simulations.

The late phase evolution of SNRs is essential to understanding of the structure of the ISM \citep{koo04}.
For example, hot ionized gas is believed to be confined in the cavities of SNRs, especially in large-volume old SNRs.
However, the volumes of the hot cavities must be strongly affected by the
 interstellar magnetic field strength.
We devote particular attention to the late-phase  evolution of SNRs.

The plan of the paper is as follows: 
in $\S$2, model and numerical method are described.  
Section 3 shows numerical result. 
We pay special attention to the comparison between models with strong
 and weak magnetic field.  
It is shown that the evolution of SNRs is also affected by the ISM density
 and local heating rates.
Expansion laws of the SNR shell and hot cavity are also shown in $\S$3.
Section 4 is devoted to discussion on how the volume fraction of the hot ISM
 is affected by the magnetic field strength.
The reason why the shell density of SNR is determined is also discussed
 in $\S$4.

\section{Model and Numerical Method}

We assume that a uniform interstellar gas with density $n_0{\rm (cm^{-3})}$ and
 magnetic field strength $B_0{\rm (\mu G)}$ is extending.
The half-width at half-maximum of the thick Gaussian component
 and the scale-height of the exponential component of the H{\sc I} disk 
 are equal to 265 pc and 403 pc, respectively \citep{dic90}. 
Since the size of the SNR is smaller than the scale-height of the H{\sc I}, $H_{\rm HI}$, in this paper,
 we ignore the effect of the density stratification in the $z$-direction, 
which becomes important after the size of the bubble
 exceeds several times $H_{\rm HI}$ \citep{tom98}. 
The scale-height of the magnetic field is believed to be much larger than
 that of the gas.
We assume here a uniform density and a uniform magnetic field strength.
      
The basic equations are the ideal MHD equations.
Using the cylindrical coordinates $(z,r,\phi)$,
 assuming that we have no toroidal components of velocity and
 magnetic field $v_\phi=B_\phi=0$ and 
 that variables are independent from the azimuth angle $\phi$ as $\partial/\partial \phi=0$,
the basic equations are as follows: 
\begin{equation}
 \frac{\partial \rho}{\partial t}
+\frac{\partial}{\partial z}\left(\rho v_z\right)
+\frac{1}{r}\frac{\partial}{\partial r}\left(r\rho v_r\right)=0,
 \label{eqn:rho}
\end{equation}
%\begin{eqnarray}
%   \frac{\partial \rho v_z}{\partial t}
%  &+&\frac{\partial}{\partial z}( \rho v_z v_z) +
%  \frac{1}{r}\frac{\partial}{\partial r}(r\rho v_z v_r)= \nonumber \\ 
%  &- & \frac{\partial p}{\partial z}
%     -\frac{1}{4\pi}\left(\frac{\partial B_r}{\partial z}
%                         -\frac{\partial B_z}{\partial r}\right)B_r,
%\label{eqn:eq-motion-z}
%\end{eqnarray}
\begin{eqnarray}
\frac{\partial \rho v_z}{\partial t}
+\frac{\partial}{\partial z}\left( \rho {v_z}^2 + p +\frac{{B_r}^2 - {B_z}^2}{8\pi}\right)
+\frac{1}{r}\frac{\partial}{\partial r}\left\{r\left(\rho v_z v_r - \frac{B_r B_z}{4\pi}\right)\right\}
=0
\label{eqn:eq-motion-z}
\end{eqnarray}
%\begin{eqnarray}
% \frac{\partial \rho v_r}{\partial t}
%  &+&\frac{\partial}{\partial z}( \rho v_r v_z) +
%  \frac{1}{r}\frac{\partial }{\partial r}(r\rho v_r v_r)= \nonumber \\ 
%  &- & \frac{\partial p}{\partial r}
%     +\frac{1}{4\pi}\left(\frac{\partial B_r}{\partial z}
%                         -\frac{\partial B_z}{\partial r}\right)B_z, 
%\label{eqn:eq-motion-r}
%\end{eqnarray}
\begin{eqnarray}
\frac{\partial \rho v_r}{\partial t}
&+&\frac{\partial}{\partial z}\left(\rho v_r v_z-\frac{B_r B_z}{4\pi}\right) \nonumber \\
&+&\frac{1}{r}\frac{\partial}{\partial r}\left[r\left\{\rho {v_r}^2
+ p +\frac{{B_z}^2 - {B_r}^2}{8\pi}\right\}\right] 
= \frac{1}{r}\left(p + \frac{{B_r}^2 + {B_z}^2}{8\pi}\right)
\label{eqn:eq-motion-r}
\end{eqnarray}

%\begin{equation}
%        \frac{\partial \epsilon}{\partial t}
%        +\frac{\partial}{\partial z}(\epsilon v_z)
%        +\frac{1}{r}\frac{\partial}{\partial r}(r\epsilon v_r)=
%                 +\Gamma n-\Lambda (T)n^2 
%                -p \left(\frac{\partial v_z}{\partial z}
%                         +\frac{1}{r}\frac{\partial r v_r}{\partial r}\right),
%\label{eqn:thermal}
%\end{equation}

\begin{eqnarray}
\frac{\partial}{\partial t}(\epsilon+\frac{\ve{B}^2}{8\pi})
&+&\frac{\partial}{\partial z}
\left\{(\epsilon+p)v_z+\frac{B_r(v_zB_r-v_rB_z)}{4\pi}\right\} \nonumber \\
&+&\frac{1}{r}\frac{\partial}{\partial r}
\left[r\left\{(\epsilon+p)v_r-\frac{B_z(v_zB_r-v_rB_z)}{4\pi}\right\}\right]
=\Gamma_0 n_{\rm H}-\Lambda (T){n_{\rm H}}^2 
\label{eqn:thermal}
\end{eqnarray}

\begin{equation}
  \frac{\partial B_z}{\partial t}=\frac{1}{r}\frac{\partial }{\partial r}\{r(v_zB_r-v_rB_z)\},
\label{eqn:induction-eq-z}
\end{equation}

\begin{equation}
  \frac{\partial B_r}{\partial t}=-\frac{\partial }{\partial z}(v_zB_r-v_rB_z),
\label{eqn:induction-eq-r}
\end{equation}
 where the variables have their ordinary meanings as
 $\rho$ (density), $\ve{v}\equiv (v_z,v_r)$ (velocity),
 $p$ (pressure), $T$ (temperature), $\ve{B}\equiv (B_z,B_r)$
 (magnetic flux density), and $\epsilon\equiv p/(\gamma-1)+\rho \ve{v}^2/2$ (internal plus kinetic energy per unit volume).
Equation (\ref{eqn:rho}) is the continuity equation;
equations (\ref{eqn:eq-motion-z}) and (\ref{eqn:eq-motion-r}) are the
 equations of motion;
equation (\ref{eqn:thermal}) represents the energy conservation equation for internal plus kinetic energy.
The induction equations for the poloidal magnetic fields are
 equations (\ref{eqn:induction-eq-z}) and (\ref{eqn:induction-eq-r}).
The terms $\Gamma_0 n_{\rm H}$ and $\Lambda (T) {n_{\rm H}}^2$ in equation (\ref{eqn:thermal}) represent, respectively,
 the heating and radiative cooling rates, where $n_{\rm H}$ means the number density of the Hydrogen atoms. 
We use the radiative cooling rate estimated for the gas 
 in the collisional ionization equilibrium with the solar metallicity
 \citep{ray76}.
   
\subsection{Numerical Method}

We employ here the modified Lax-Wendroff scheme \citep{rub67},
 an explicit finite-difference method, to solve the basic equations 
(\ref{eqn:rho})-(\ref{eqn:induction-eq-r}).
The number of grids and grid spacings are chosen as $(N_z,N_r)=(2500, 2500)$
 and $\Delta z=\Delta r=0.1$ pc.
Thus, our numerical domain covers a region of 250 pc $\times$ 250 pc.
Artificial viscosity is added to smooth the jump at the shock front.
For all the quantities $U\equiv$ ($\rho$, $\rho\ve{v}$, $\epsilon+\ve{B}^2/8\pi$, \ve{B}),
 we added an extra term to express artificial diffusion and solve equations like
\begin{equation}
\dif{U}{t}=\nabla(c \nabla U),
\end{equation}  
where we take 
\begin{equation}
 c=A_v\left[\left(\dif{v_z}{z}\right)^2+\left(\dif{v_z}{r}\right)^2
+\left(\dif{v_z}{z}\right)^2+\left(\dif{v_r}{r}\right)^2\right]^{1/2}\Delta^2, 
\end{equation}
with $\Delta=\Delta z=\Delta r$ and $A_v=4$.
The numerical scheme was tested by comparing known solutions which have been
 analytically and numerically obtained.
We checked the program by comparing 
(1) the adiabatic SNR with the Sedov solution \citep{sed59},
(2) the spherically symmetric SNR with previously calculated \citep{sla92}
and (3) MHD shock tube problems \citep{bri88}.
The analytical solution of (1) is reproduced within a 5\% relative error, for example, pressure and density distributions.
The expansion law of the shock front of hydrodynamical SNRs of (2)
 coincides with previous  work (such as that by \citet{sla92})
 and the difference is not more than 5\% for $t\lesssim 1{\rm Myr}$.  
  
We assume a uniform ISM with number density $n_0$ and
 temperature $T_0$.
A uniform magnetic field is running in the $z$-direction as 
 $\ve{B}_0=(B_z,B_r)=(B_{0},0)$.
As for the radiative cooling rate, we use a spline function fit of cooling curves
 obtained by \citet{ray76} ($T > 10^4{\rm K}$)
 and by \citet{dal72} ($T < 10^4{\rm K}$).
We summarize the parameters used in Table \ref{tab:1}.
The heating coefficient $\Gamma_0$ is chosen as follows: 
in models A-H (see Table \ref{tab:1}), the heating rate $\Gamma_0n_{\rm H}$
 is balanced by the cooling rate $\Lambda(T_0){n_{\rm H}}^2$ in the ambient ISM.
 On the other hand, it is set to be $5\times 10^{-27}{\rm erg\,s^{-1}}$ in models AW-HW (see Table \ref{tab:1}) as minimum heating rate given by cosmic-ray \citep{spi78}.
We begin the simulation by adding thermal energy of $E_0=5 \times 10^{50}{\rm erg}$
 within the sphere of 2pc in radius.  
We assume a 10:1 H to He number ratio,
 which leads to a mean molecular weight for nuclei of $\mu_0=14/11=1.27$ and
 that for fully-ionized ions and electrons of $\mu_t=14/23=0.61$,  
and the number density of the Hydrogen atoms of $n_{\rm H}=(10/11)n$, where $n$ means the number density of the gas. 

\section{Results}

\subsection{Model with $B_0=5\mu{\rm G}$}  
\label{sec:model A}

\begin{figure}
\noindent
{\centering
(a)\hspace*{7.7cm}(b)\\
\plottwo{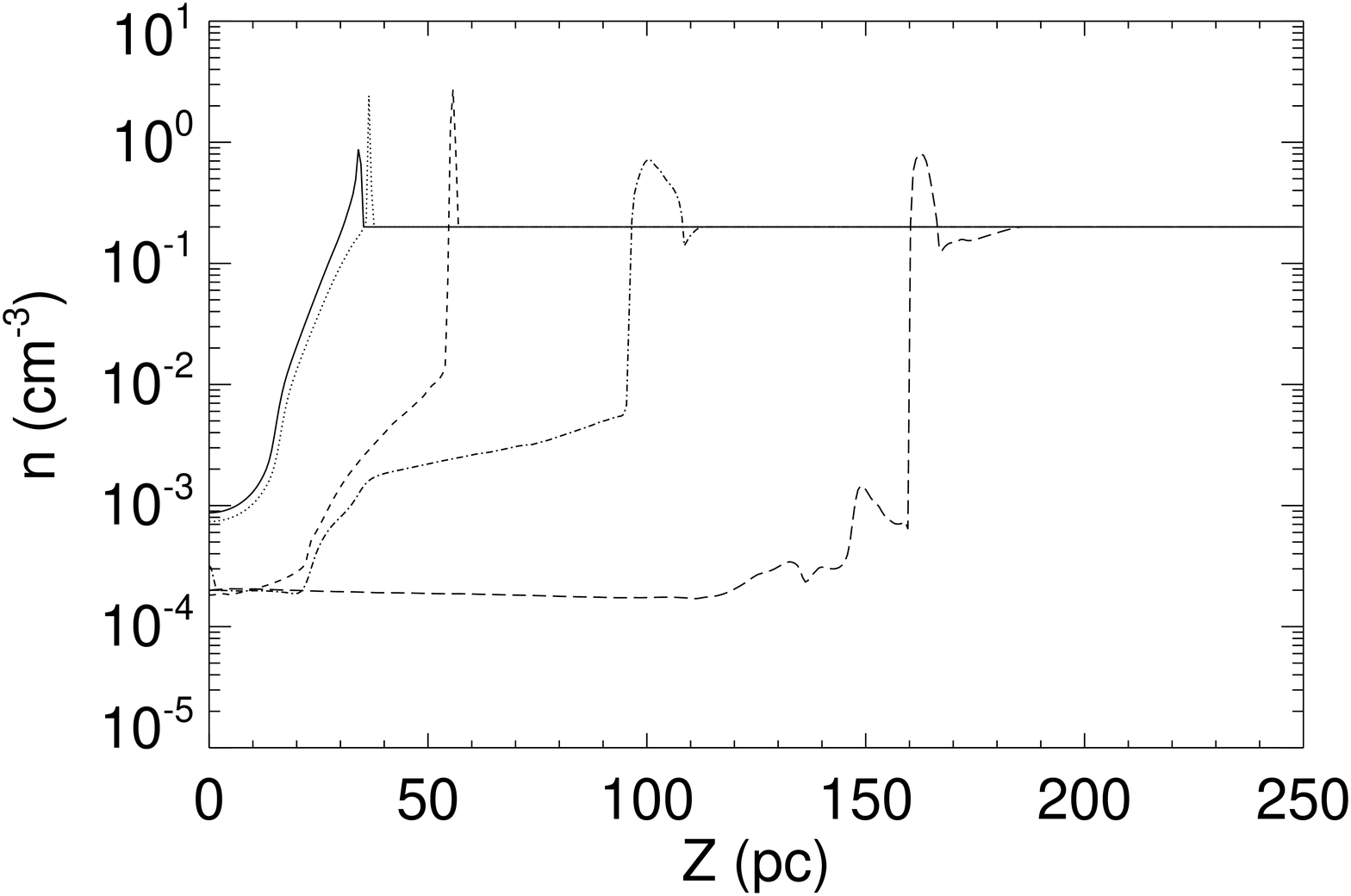}{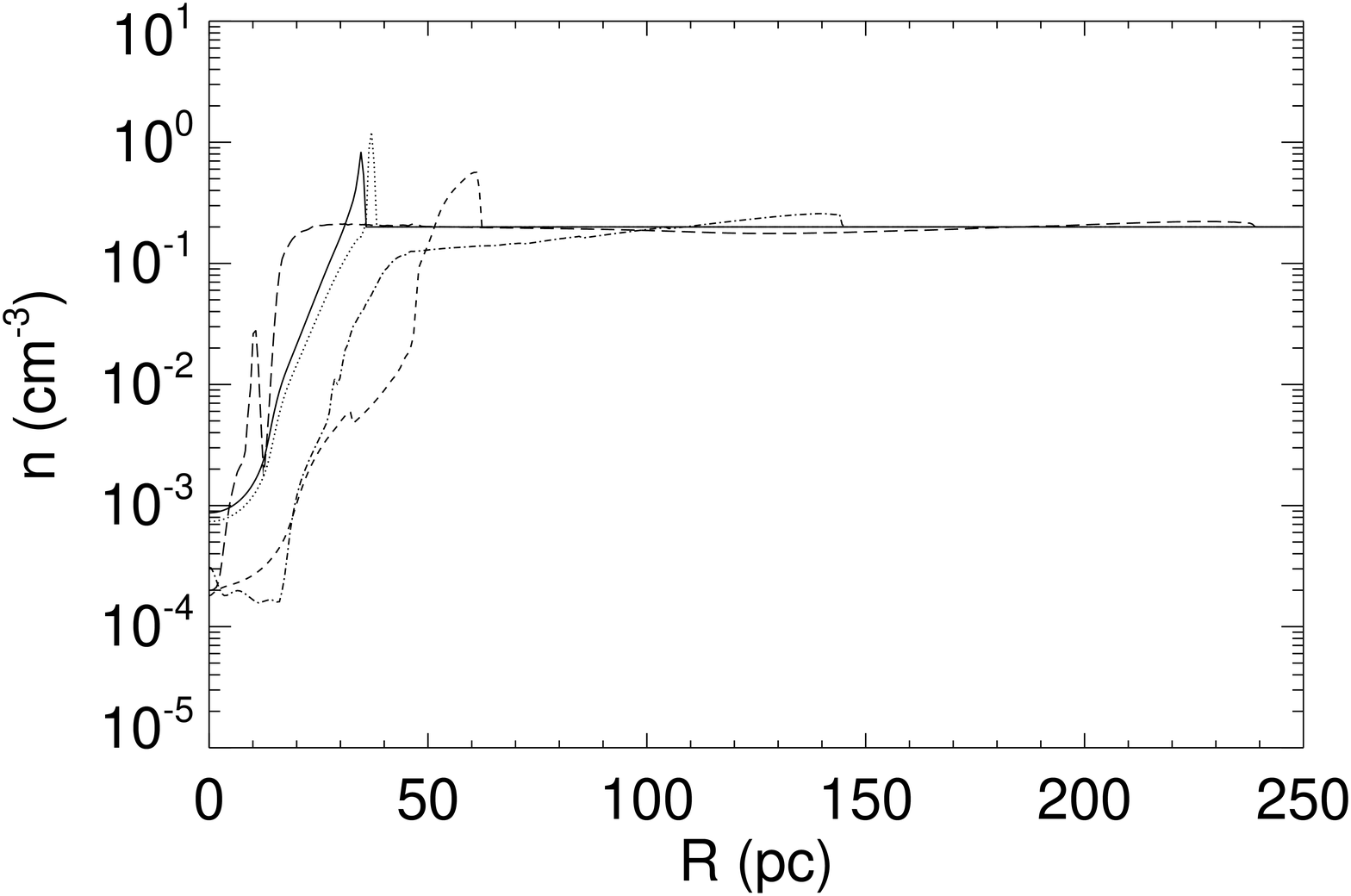}\\
(c)\hspace*{7.7cm}(d)\\
\plottwo{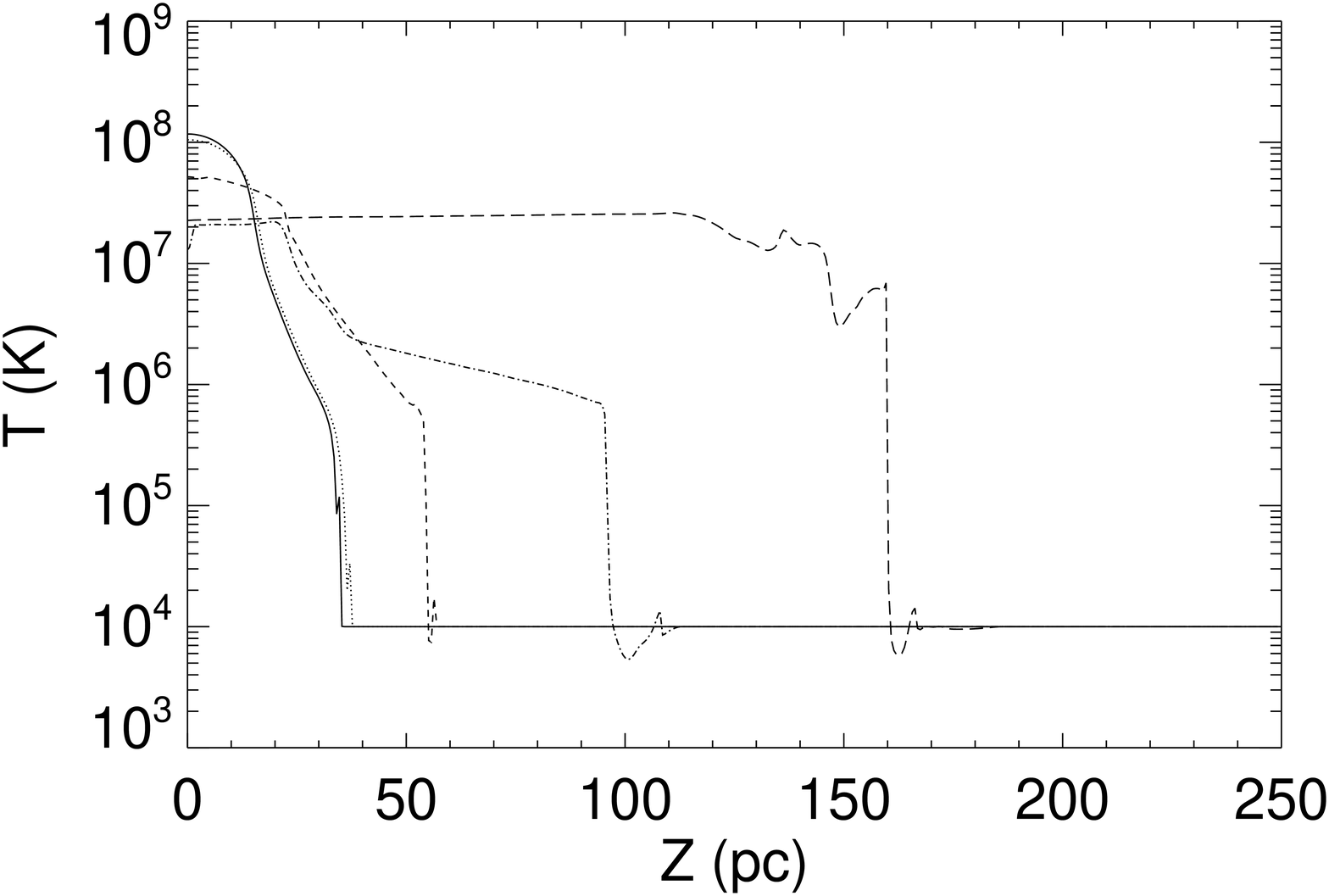}{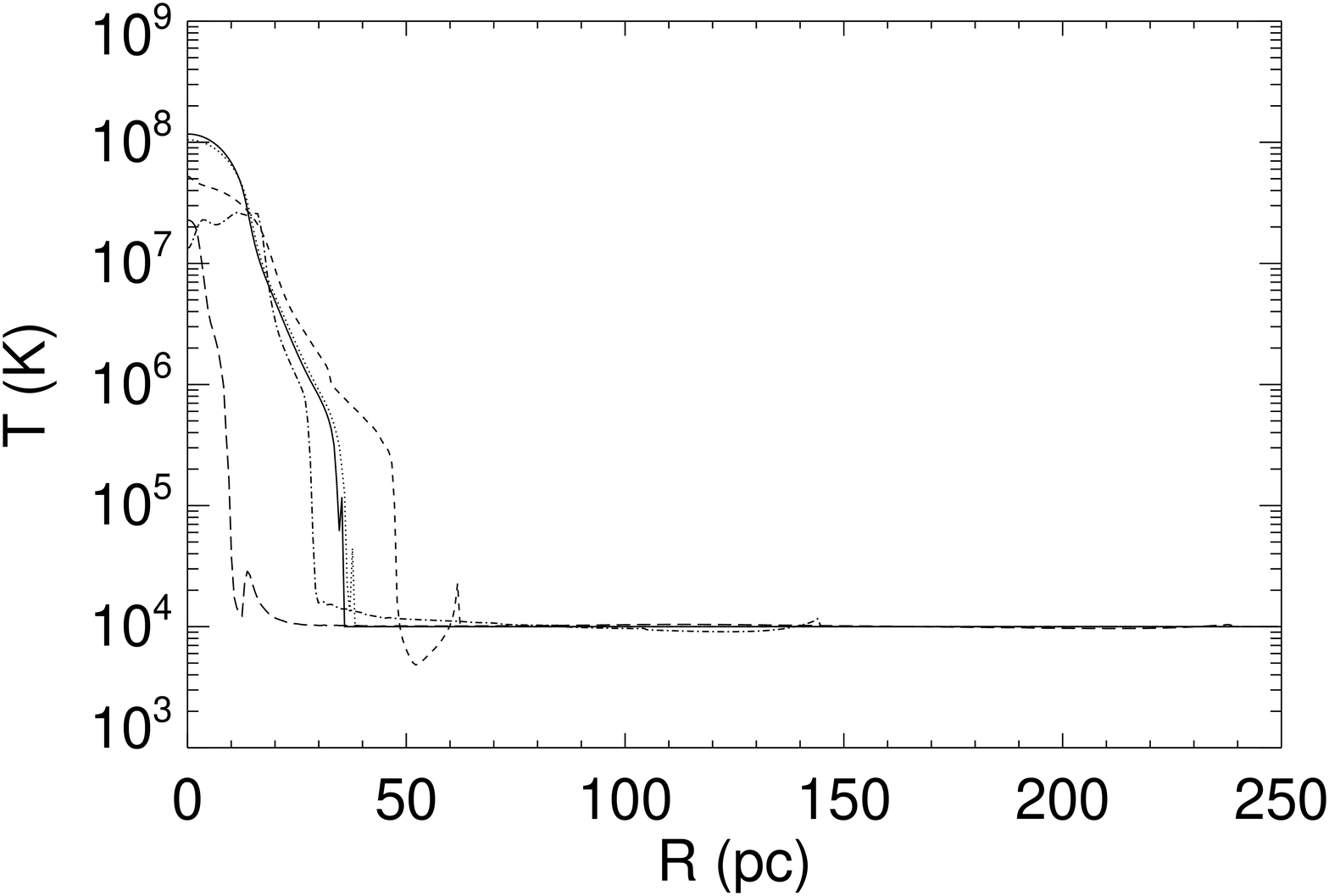}\\
}
\caption{
Density distributions along the $z$- (upper-left:a)
 and $r$-axes (upper-right:b) of SNRs (model A).
Those of temperature along the $z$- and  $r$-axes are also shown
 in {\em c} (lower-left) and  {\em d} (lower-right), respectively.
Five snapshots are taken at the ages of
 $t=7.97\times 10^4{\rm yr}$ (solid line),
 $1\times 10^5{\rm yr}$ (dotted line),
 $3.99\times 10^5{\rm yr}$ (dashed line),
 $2.52\times 10^6{\rm yr}$ (dash-dotted line),
 and $5.64\times 10^6{\rm yr}$ (long dashed line).
}
\label{fig:1}
\end{figure}

In model A, we studied a magnetized ISM with $B_0=5\mu{\rm G}$,
 $n_0=0.2{\rm cm}^{-3}$, and $T_0=10^4{\rm K}$, which mimics the warm ionized ISM. 
With a magnetic pressure of
 $B_0^2/8\pi=10^{-12}{\rm dyn\,cm^{-2}}(B_0/5\mu{\rm G})^2
            =7.2\times 10^3{\rm K\,cm^{-3}}(B_0/5\mu{\rm G})^2 $ and
 a thermal pressure of 
 $p=(\mu_0/\mu_t) n_0 k T_0
   =5.8 \times 10^{-13}$ ${\rm dyn\,cm^{-2}}$ $(n_0/0.2{\rm cm^{-3}})$ $(T_0/10^4{\rm K})
   =4.2 \times 10^3$ ${\rm K\,cm^{-3}}$ $(n_0/0.2{\rm cm^{-3}})$ $(T_0/10^4{\rm K})$,
 the plasma beta of the ISM is equal to 
 $\beta_0=0.58(n_0/0.2{\rm cm^{-3}})$ $(T_0/10^4{\rm K})$ $(B_0/5\mu{\rm G})^{-2}$. 
In Figure \ref{fig:1}, we plotted cross-cut views along the $z$- (a and c) and 
 the $r$-axes (b and d).
Upper panels (a and b) display density distributions,
 while the lower ones (c and d) display temperature distributions.
The shell formation time at which the SNR changes its structure 
 from the adiabatic Sedov phase \citep{sed59} to the
 pressure-driven snowplow phase is given \citep{cio88} as
\begin{equation}
 t_{\rm sf}(E_0,n_0)\simeq 7.8\times 10^4{\rm yr} 
           \left(\frac{E_0}{5\times 10^{50}{\rm erg}}\right)^{3/14}
           \left(\frac{n_0}{0.2{\rm cm^{-3}}}\right)^{-4/7}.
\label{eqn:1}
\end{equation}  
The first snapshot ($t=7.97\times 10^4{\rm yr}$; solid line) corresponds to the final stage of
 the adiabatic phase or the transition stage to the pressure-driven snow-plow phase.
Figure \ref{fig:1} shows that the differences in the $z$- and $r$-cross-cuts
 are slight at this stage.
A dip in the temperature distribution just after the shock front
 indicates that the radiative cooling begins to play a part at this stage.

The second snapshot at $t=1\times 10^5{\rm yr}$
 shows clearly that a thin shell is formed
 [Fig.\ref{fig:1}(a) and (b)].      
At this stage, Figure \ref{fig:1}(a) and (b) indicates
 that the shell is thin and its widths propagating in $z$- and $r$-directions are
 similar.
The temperature distributions also look similar to each other 
 [Fig.\ref{fig:1}(c) and (d)].
Peak density in the shell reaches maximum at this stage and declines gradually 
 for the shell near the $r$-axis [Fig.\ref{fig:1}(b)],
 while it increases further for the shell propagating near the $z$-axis
 [Fig.\ref{fig:1}(a)] until $t=2.52 \times 10^5{\rm yr}$. 

At the age of $3.99\times 10^5{\rm yr}$ (dashed line), the SNR shows a shape far from spherical symmetry,
 that is, the shock front reaches $R\simeq 62$ pc while $Z\simeq 57$ pc. 
The shell width in the $r$-direction $\Delta R$ becomes much wider than
 that in the $z$-direction $\Delta Z$ (short-dashed line) 
 as $\Delta R\simeq 14$ pc while $\Delta Z\simeq 2$ pc.
The shape of the cavity which contains a hot low-density gas indicates an elongated
 shape along the $z$-axis.
That is, the bubble ($T\gtrsim 10^5{\rm K}$) half radius is equal to
 $Z_{\rm bub}\simeq 55{\rm pc}$ while $R_{\rm bub}\simeq 47{\rm pc}$.
This difference in $z$- and $r$-directions comes from the effect of magnetic pressure which works to confine the gas in the $r$-direction.  
The post-shock thermal pressure (not shown) reaches $\simeq 4\times 10^4{\rm K\,cm^{-3}}$
 on the $z$-axis, while it is equal to $\simeq 2 \times 10^4{\rm K\,cm^{-3}}$
 for the shock propagating perpendicularly to the magnetic field (along the $r$-axis).
In a case in which the magnetic field is parallel to the shock front,
 which is applicable to the shock expanding in the $r$-direction,
 the post-shock to pre-shock pressure ratio is given using the plasma beta
 of the pre-shock gas $\beta_0$ and the Mach number of shock propagation speed 
 ${\cal M}$ [the Rankin-Hugoniot relation, see eq.(\ref{eqn:y})].
Since part of the pressure is attributed to the magnetic one,
 the post-shock thermal pressure of the magnetic shock
 must be lower than that of the non-magnetic one 
 (the ratio of post-shock thermal pressure of the magnetic shock
  to that of the non-magnetic shock is plotted in Fig.\ref{fig:A1}).
With the plasma $\beta$ of the ISM $\beta_0=0.58$ (model A) and 
 the Mach number of the shock speed
 $V_s\simeq 50{\rm km\,s^{-1}}$ of ${\cal M}\equiv V_s/c_s=3.3$ at $t=3.99\times 10^5{\rm yr}$ ($c_s$ denotes the adiabatic thermal speed),
 equation (\ref{eqn:y}) anticipates that
 the thermal pressure of magnetized post-shock gas is 0.36 times smaller than
 that of the non-magnetized one. 
This demonstrates that the post-shock magnetic field plays an essential role
 in determining the shock structure at this stage.
%However, the ratio of post-shock pressure on the $r$-axis to that of $z$-axis
% obtained for $t=1.2 \times 10^5{\rm yr}$ is approximately equal to 0.5. 
%Since equation (\ref{eqn:Vs}) expects the ratio of 0.52
% when the Mach number is as small as  ${\cal M}=5$,
% this may indicate the shock is much decelerated than that expected from the Sedov
% solution owing to the radiative cooling.  

The shell width continues to increase with time in both $z$- and $r$-directions.
Even in the $z$-direction, the shell width increases.
This is understood as follows:
In the case of hydrodynamic shock, the shell is confined under the effect of
 the post-shock pressure and the thermal pressure of the hot interior gas.
Ram pressure working at the front of the shell,
 which is proportional to the shock speed squared,  
 decreases with time, since the expansion of the shock front is decelerated.
At the same time, the thermal pressure in the 
 hot interior gas also decreases owing to the adiabatic expansion.
These thicken the shell thick \citep{che74}.
In the $r$-direction, the magnetic force prevents the shell from being 
compressed and works to support the shell.
The width of the shell increases owing to the decrease in the confining pressures
 and the difference between the shells propagating in $z$- and $r$-directions comes from
 the magnetic force . 

\begin{figure}
{\centering
(a)\hspace*{7cm}(b)\\[-5mm]
\plottwo{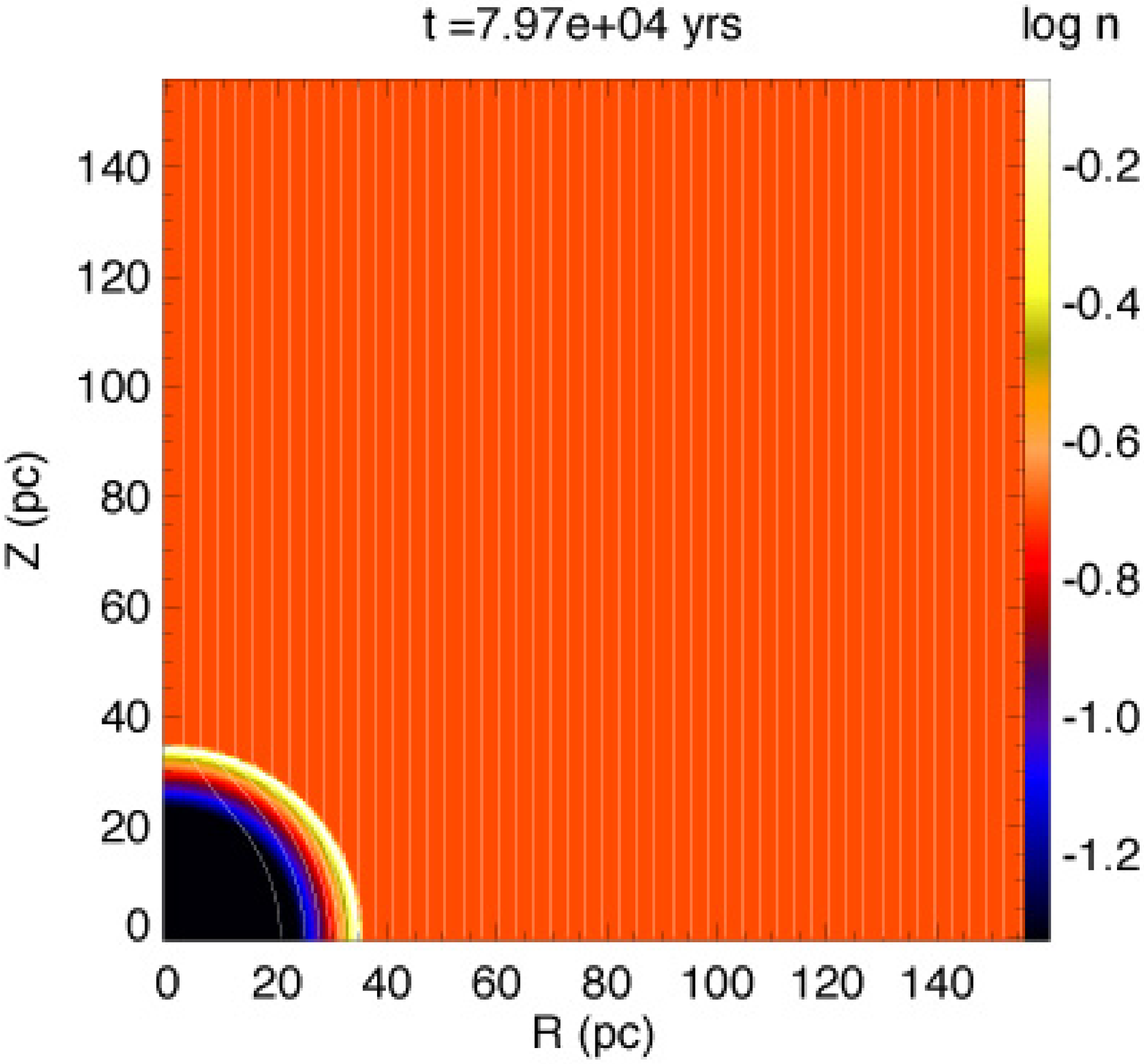}{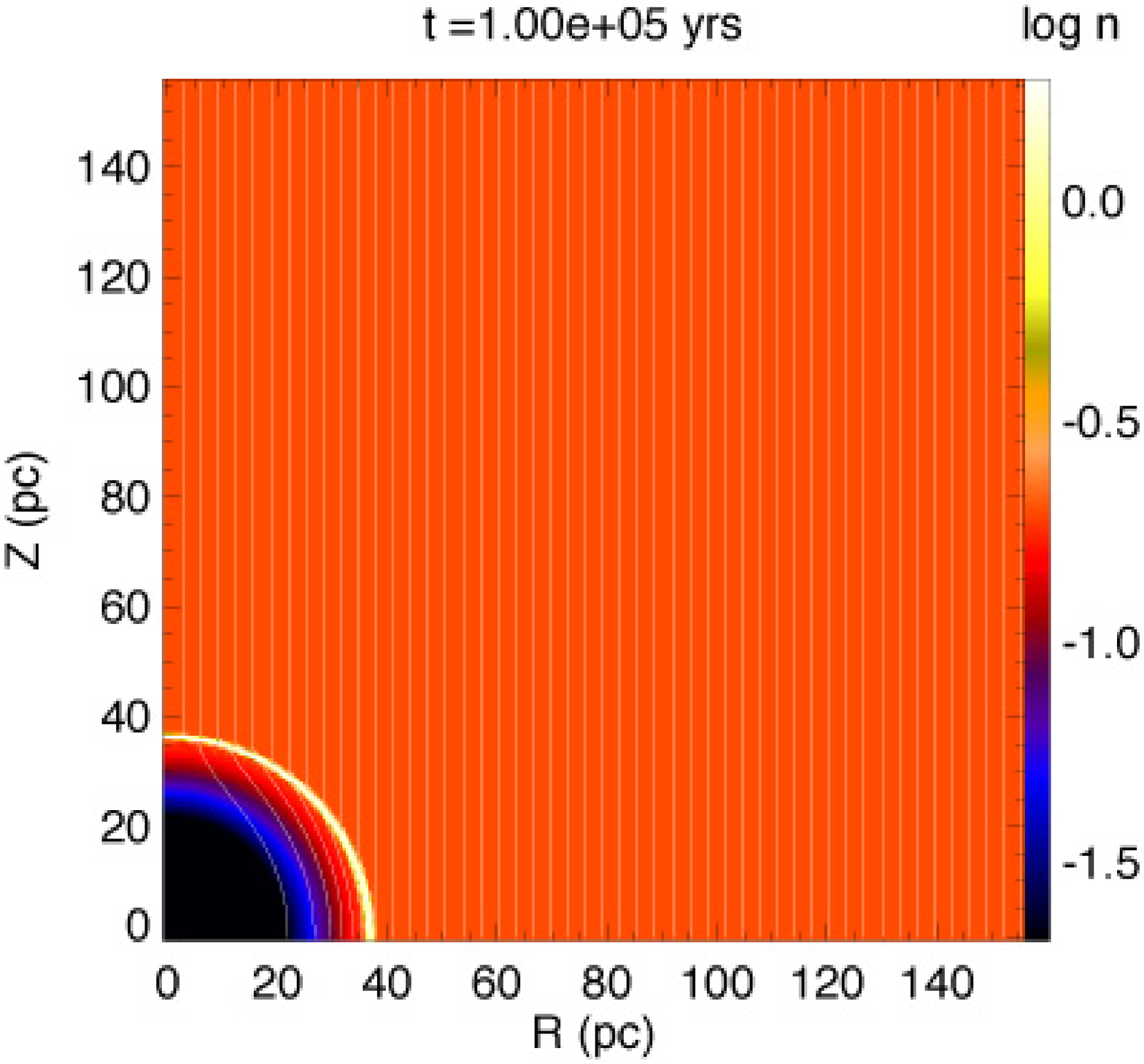}\\[-5mm]
(c)\hspace*{7cm}(d)\\[-5mm]
\plottwo{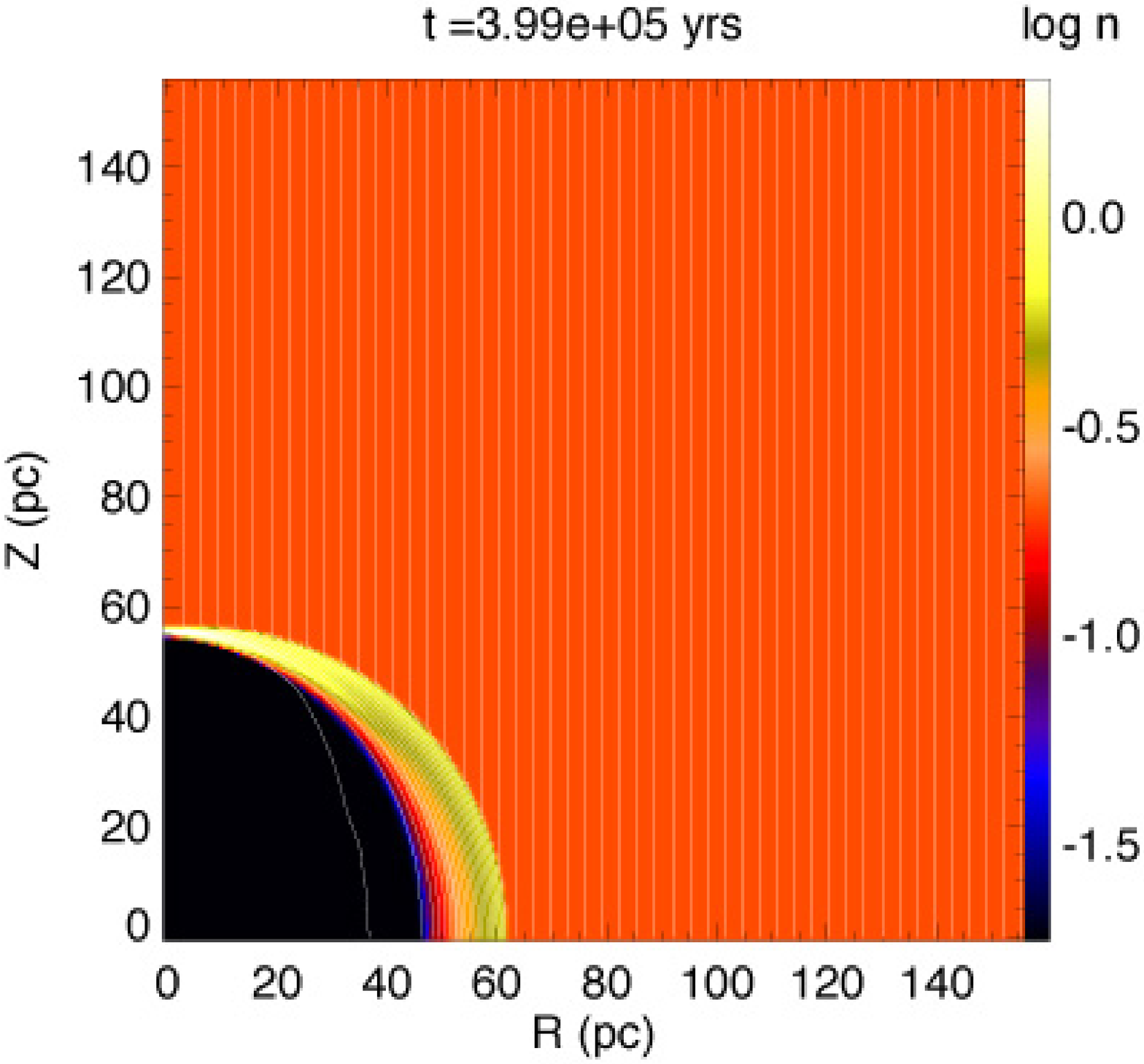}{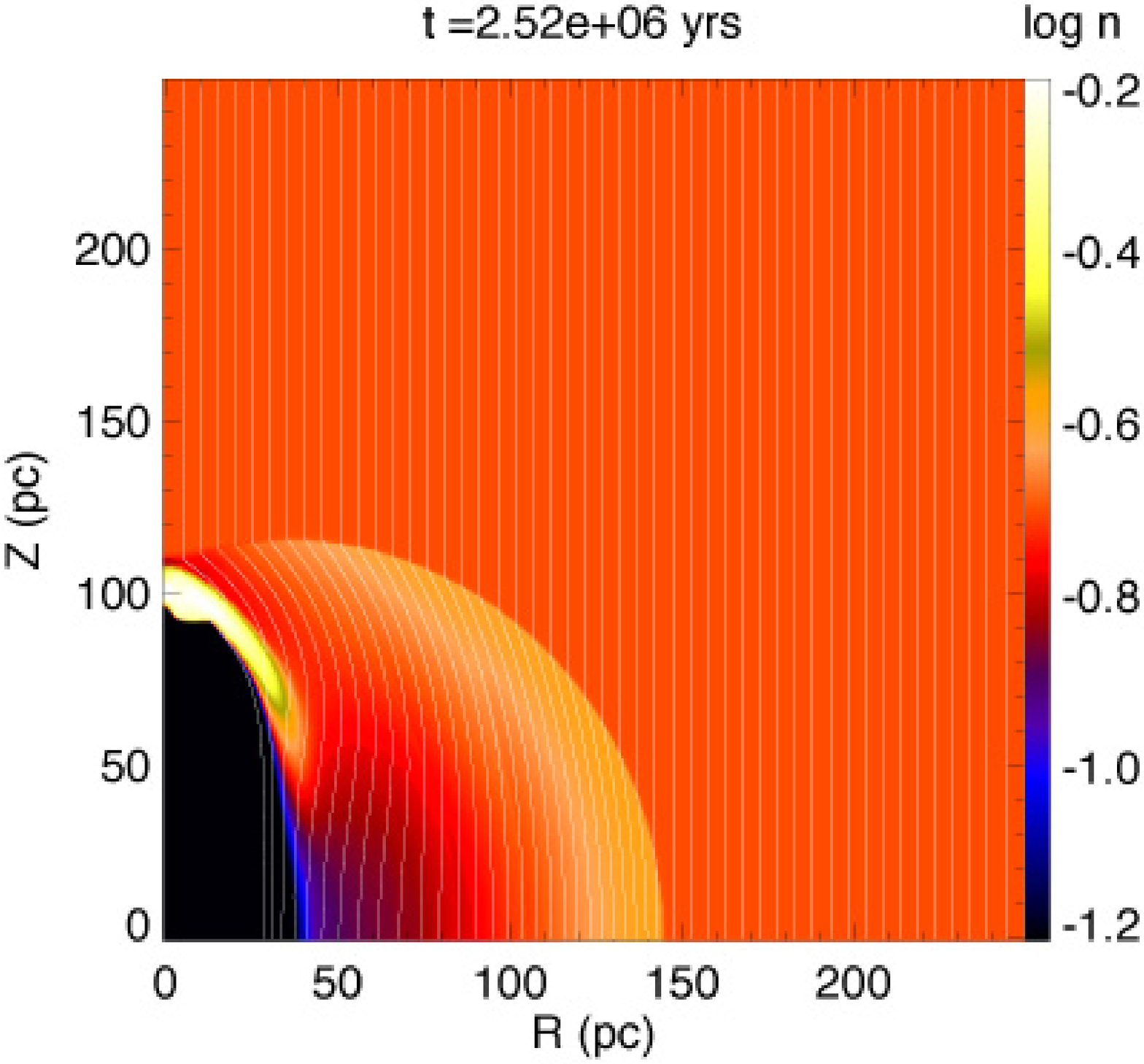}\\[-5mm]
(e)\\[-5mm]
\includegraphics[width=7.5cm,clip]{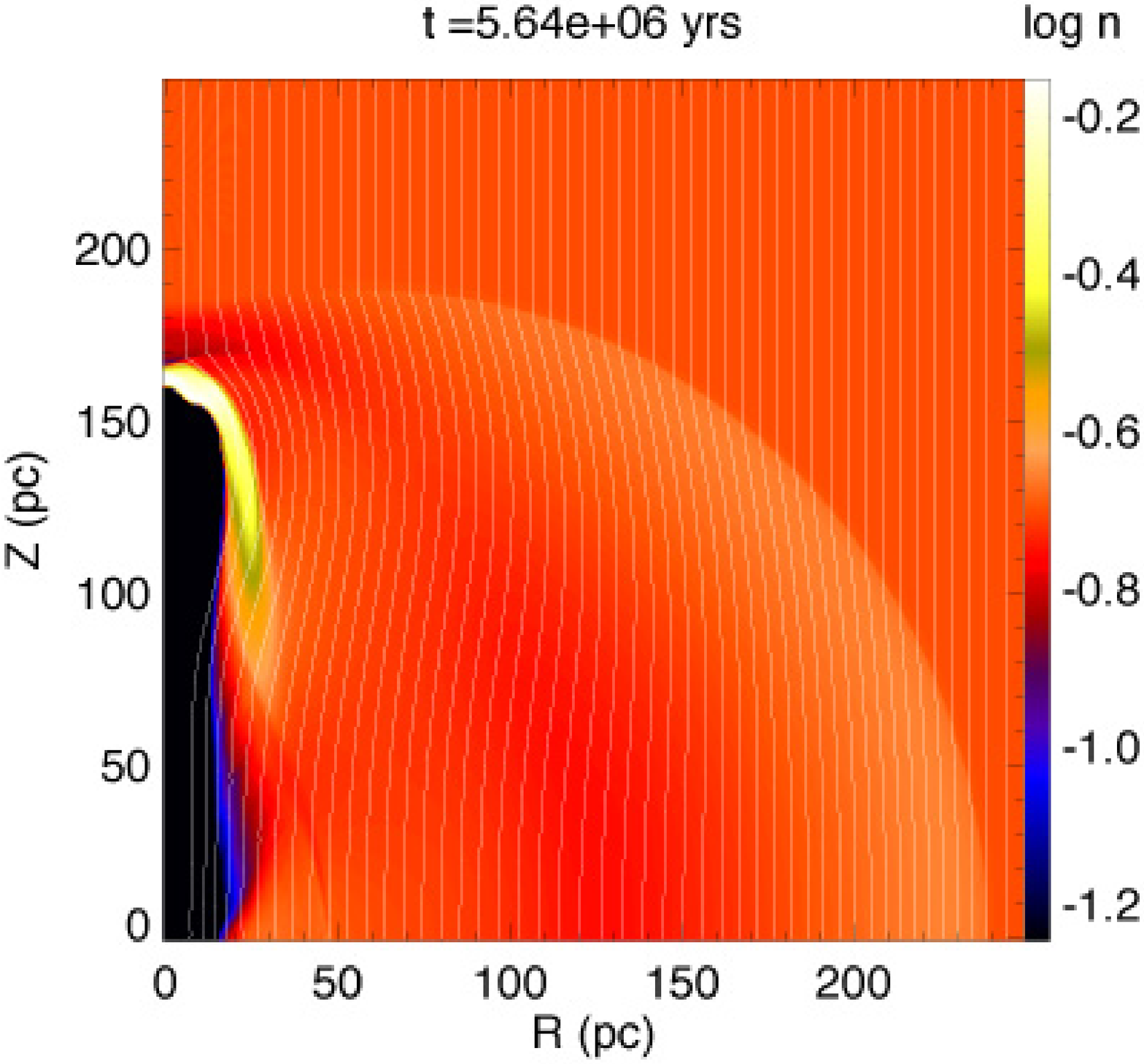}\\
}
\caption{
Structures of the magnetized SNR (model A).
Density distributions (false color) and magnetic field lines (white lines)
 are shown.
Five snapshots are taken at the ages of
 $7.97\times 10^4{\rm yr}$ (a),
 $1\times 10^5{\rm yr}$ (b), 
 $3.99\times 10^5{\rm yr}$ (c),
 $2.52 \times 10^6{\rm yr}$ (d),
 and $5.64 \times 10^6{\rm yr}$ (e).
Panels (a)-(c) cover the region of 150pc $\times$ 150pc, while
 panels (d) and (e) cover 250pc $\times$ 250pc. 
}
\label{fig:2}
\end{figure}

\begin{figure}
{\centering
(a)\hspace*{7cm}(b)\\[-5mm]
\plottwo{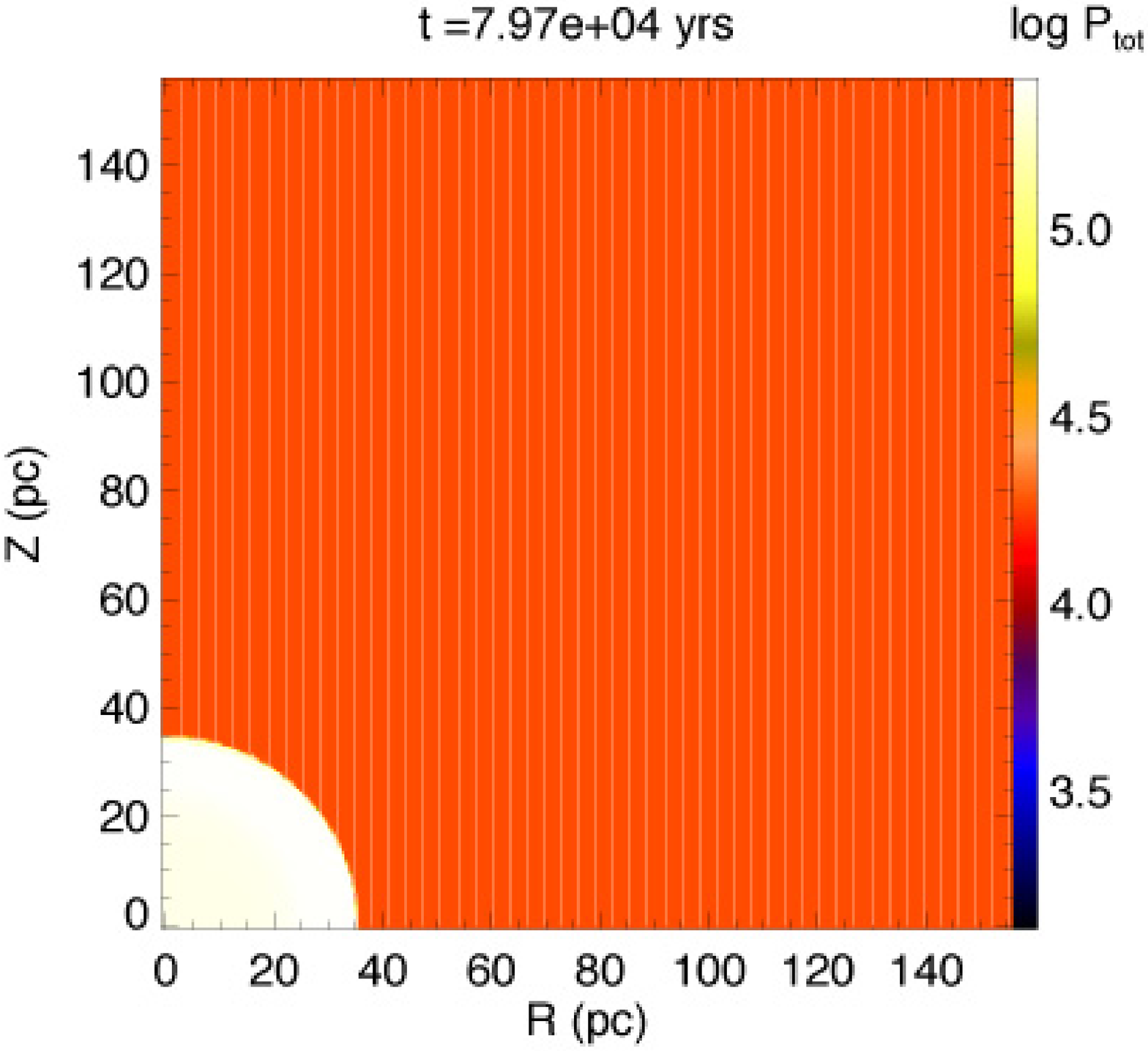}{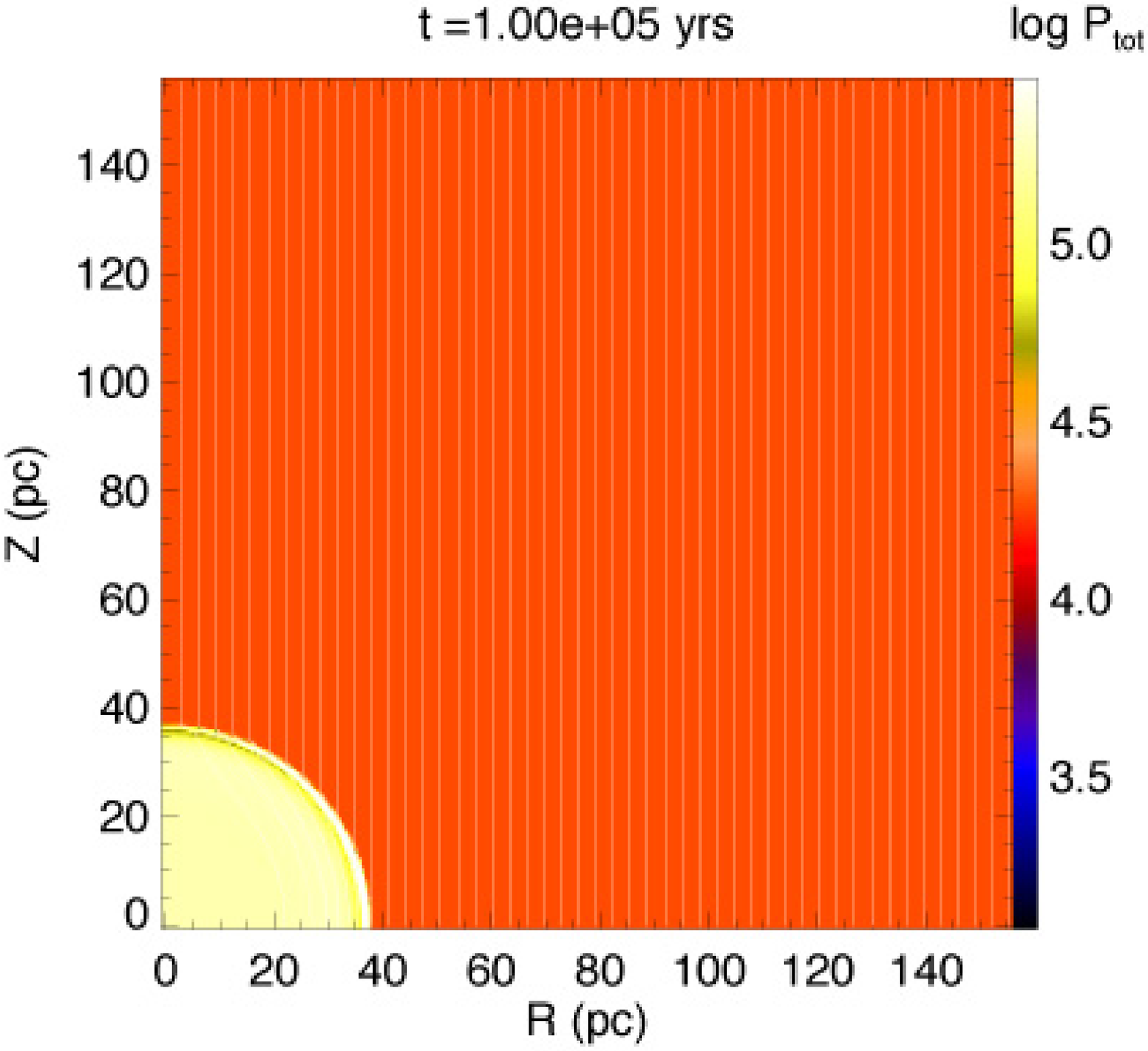}\\[-5mm]
(c)\hspace*{7cm}(d)\\[-5mm]
\plottwo{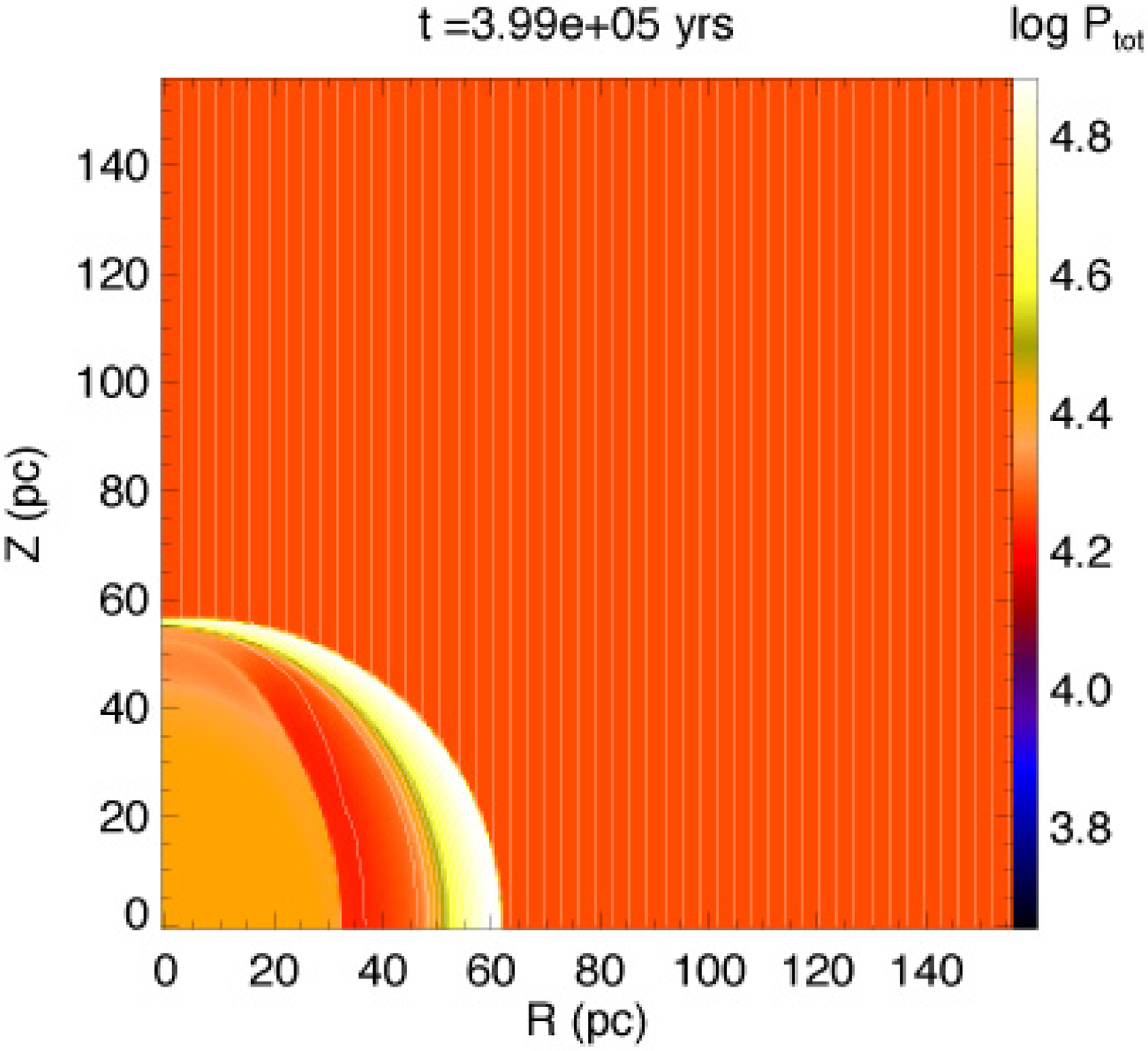}{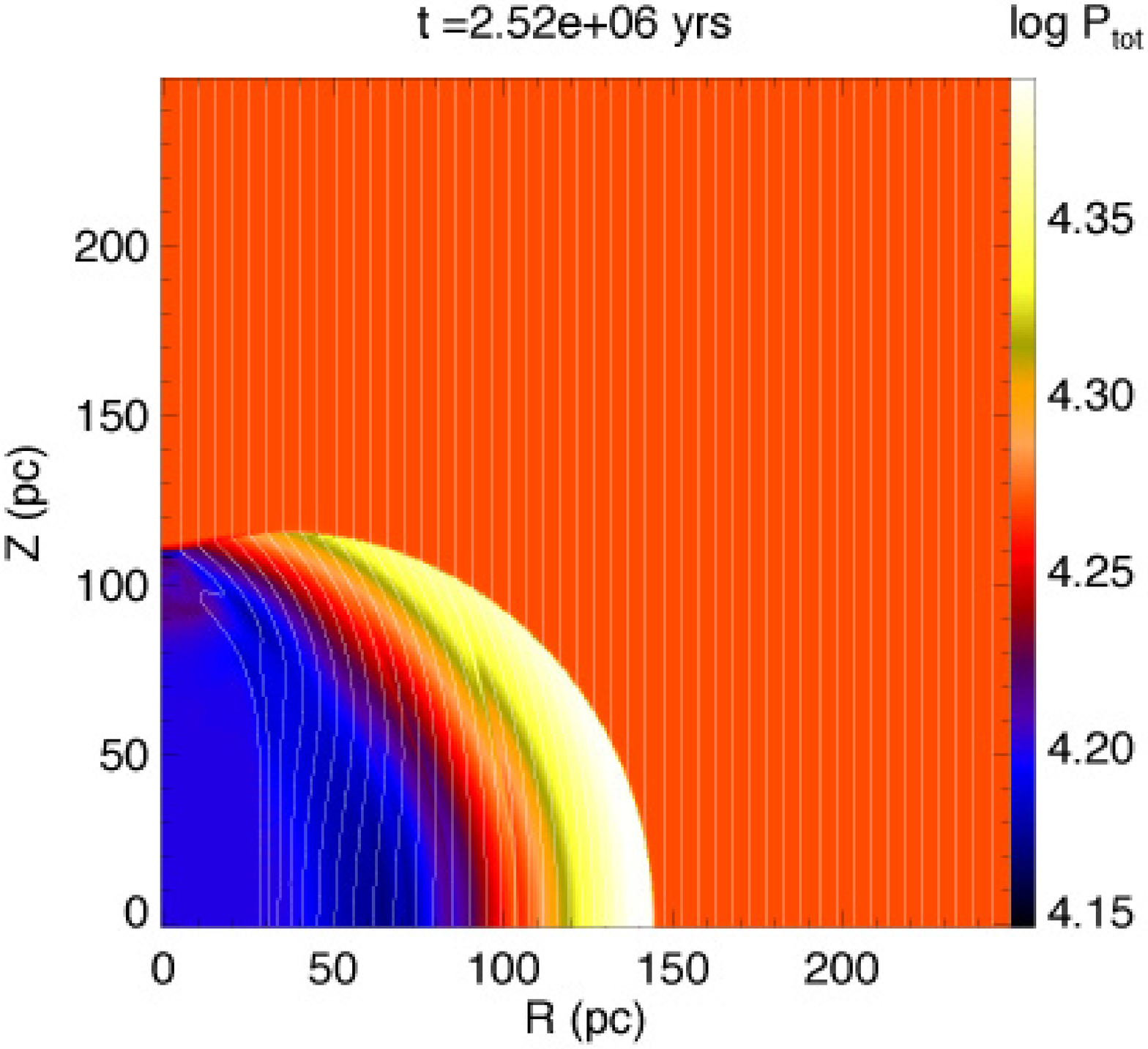}\\[-5mm]
(e)\\[-5mm]
\includegraphics[width=7.5cm,clip]{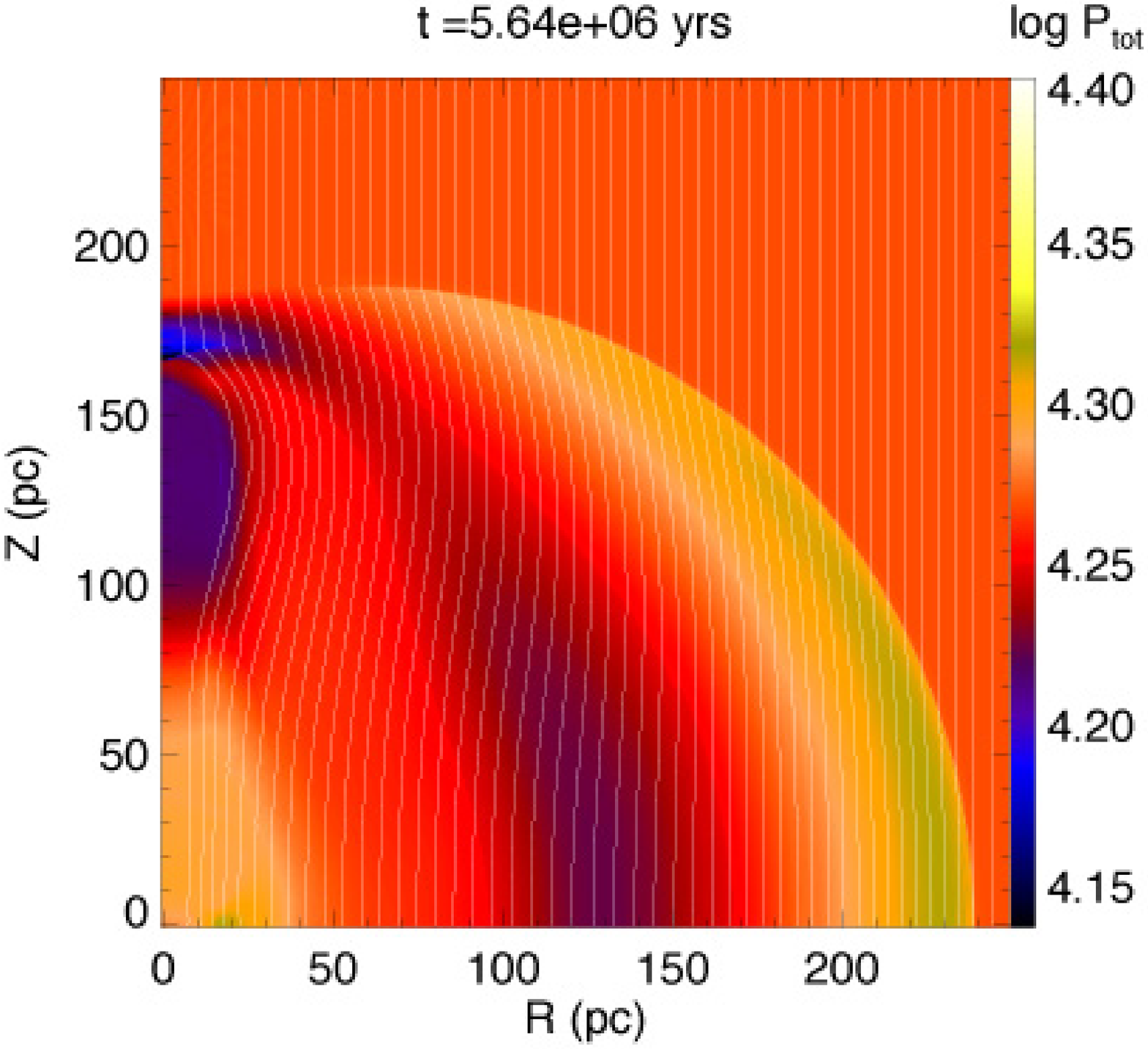}\\
}
\caption{
Total pressure $p+\ve{B}^2/8\pi$ distribution and the magnetic field lines for model A. 
Five snapshots are taken at the ages of
 $7.97\times 10^4{\rm yr}$ (a),
 $1\times 10^5{\rm yr}$ (b), 
 $3.99\times 10^5{\rm yr}$ (c),
 $2.52 \times 10^6{\rm yr}$ (d),
 and $5.64 \times 10^6{\rm yr}$ (e).
Panels (a)-(c) cover the region of 150pc $\times$ 150pc, while
 panels (d) and (e) cover 250pc $\times$ 250pc.  
}
\label{fig:3}
\end{figure}

In Figures \ref{fig:2} and \ref{fig:3},
 we plotted density distributions, total (thermal + magnetic)
 pressure distributions and magnetic field lines.
For each snapshot, the cross-cut views are shown in Figure \ref{fig:1}.
Figures \ref{fig:2}(a) and \ref{fig:3}(a) show
 the structure of SNRs at the final stage of the adiabatic phase ($t=7.97\times 10^4{\rm yr}$).
The interior pressure (mainly thermal) predominates over the ISM pressure and
 the SNR holds the spherical symmetry.
The shell thickness is approximately equal to 10\% of the radius,
 which decreases to form a cooled shell
 [Figs. \ref{fig:2}(b) and \ref{fig:3}(b); $t=1\times 10^5{\rm yr}$].
The SNR looks spherically symmetric in the adiabatic phase, even with the magnetic field.
Compared with panels (c) of $t=3.99\times 10^5{\rm yr}$,
it is shown that the shell width increases with time 
in the pressure-driven snow-plow phase after the cool shell formation, especially in the $r$-direction.

As is shown in Figure \ref{fig:1},
 the distance of the shock front from the explosion site reaches $R\simeq 62$pc
 in the $r$-direction, while $Z\simeq 57$pc in the $z$-direction
 (panels c of $t=3.99\times 10^5{\rm yr}$).
This means that
 the shock wave propagates faster in the direction perpendicular to the magnetic field
 than in the parallel direction.
Since the magnetic field lines are bent reaching the shock front
 after passing the shock front, the shock running outward is a fast shock.  
The phase velocity of the fast wave increases in the direction perpendicular
 to the magnetic field. 
It is natural that the fast shock propagates faster in the $r$-direction
 than in the $z$-direction.
Thermal pressure in the hot low-density cavity is twice as high as
 the total pressure $p+\ve{B}^2/8\pi$ of the ISM at this phase.
However, it becomes lower than the total pressure of the ISM after $t\gtrsim 5\times 10^5{\rm yr}$.
Around this equilibrium epoch, the hot cavity begins to contract in the $r$-direction, 
while in the $z$-direction, the inner boundary of the shell continues to expand.  

Figures \ref{fig:2}(d) and \ref{fig:3}(d)
 show the structure at $2.52\times 10^6{\rm yr}$, at which
 the snapshot is also made in Figure \ref{fig:1} (dash-dotted line). 
As indicated by a magnetic field line running nearest to the $z$-axis in these figures,
 the inner boundary of the shell contracts radially between $3.99\times 10^5{\rm yr}$ 
 ($R\simeq 47$ pc) and $2.52\times 10^6{\rm yr}$ ($R\simeq 28$ pc), 
while the outward-facing shock front continues to expand
 from $R\simeq 62$pc at $t=3.99\times 10^5{\rm yr}$
 to $R\simeq 145$pc at $t=2.52\times 10^6{\rm yr}$.
The portion traced by the above nearest magnetic field line corresponds to
 the temperature jump between a gas in the shell $T\lesssim 10^4$K
 and a hot gas in the cavity $T\gtrsim 10^5$K. 
As for the expansion in the $z$-direction, the temperature transition located   
 near $Z\simeq 55$ pc ($3.99\times 10^5{\rm yr}$) expands to $Z\simeq 95$ pc ($2.52\times 10^6{\rm yr}$)
 shown in Figure \ref{fig:1}(c), 
 while the cavity contracts in the $r$-direction 
 as the radial size of the cavity decreases from $R\simeq 47$ pc at $t=3.99\times 10^5{\rm yr}$
 to $R\simeq 28$ pc at $t=2.52\times 10^6{\rm yr}$ shown in Figure \ref{fig:1}(d). 
This indicates that the expansion of the hot gas is prevented
 by the effect of the magnetic field in the direction perpendicular to
 the magnetic field.
As for the shock front, panels (d) show that
 the angle measured from the shock normal vector to the pre-shock magnetic field line is positive for
 $r\gtrsim 40$ pc.
In other words, the magnetic field component parallel to the shock front
 does not change its sign passing through the shock front.
This is characteristic of the fast shock, 
while in the region near the $z$-axis ($r\lesssim 40$ pc), the parallel components have different
 signs comparing the pre- and post-shock magnetic fields. 
Thus, this part of the shock is the intermediate shock \citep{jef64}.  

The structure at $t=5.64\times 10^6{\rm yr}$ is shown in Figures \ref{fig:1} (long dashed line),
 \ref{fig:2}(e) and \ref{fig:3}(e).   
The expansion of the front continues.
In the thick diffuse shell propagating in the $r$-direction,
 the thermal pressure has an excess of only $\sim$ 10\% over the ISM and
 the excess total pressure in the shell is at most $\sim$ 20\% of the ISM,
 even including the magnetic pressure. 
The shock is weak.
The expansion in the $r$-direction is steadily decelerated till $t=5.64\times 10^6{\rm yr}$.
Although the deceleration along the $z$-axis continues for $t\lesssim 10^6{\rm yr}$,
 the expansion speed becomes constant $V\sim 20{\rm km\,s^{-1}}$ after $t\gtrsim 10^6{\rm yr}$.
The shell propagating along the $z$-axis is not decelerated for $t\gtrsim 10^6{\rm yr}$.
This may seem strange.
However, this can be explained as follows:
The mechanical equilibrium in the $r$-direction requires that 
 the thermal pressure in the hot low-density cavity $p_{\rm cav}$,
 where the magnetic field is weak,
 is in pressure equilibrium with the total pressure of the ISM
 as $p_{\rm cav}\simeq p_0+B_0^2/8\pi$
 in the late phase of SNRs.
This leads to a thermal pressure imbalance of $p_{\rm cav} > p_0$. 
This difference of $B_0^2/8\pi$ pushes the shell near the $z$-axis,
 where no magnetic force works.
The equation of motion for the part expanding along the $z$-axis is written
\begin{equation}
 \frac{d\sigma V}{dt}=\frac{B_0^2}{8\pi},
\end{equation}
where $\sigma$ represents the column density of $\sigma=\int\rho dz$.
Constant force per unit area drives the mass-accumulating shell to move with a
 constant velocity as $\sigma\propto Z \propto t$.

\subsection{Anisotropy Driven by the Magnetic Field}

\begin{figure}
\noindent
{\centering
(a)\hspace*{7.7cm}(b)\\[-5mm]
\plottwo{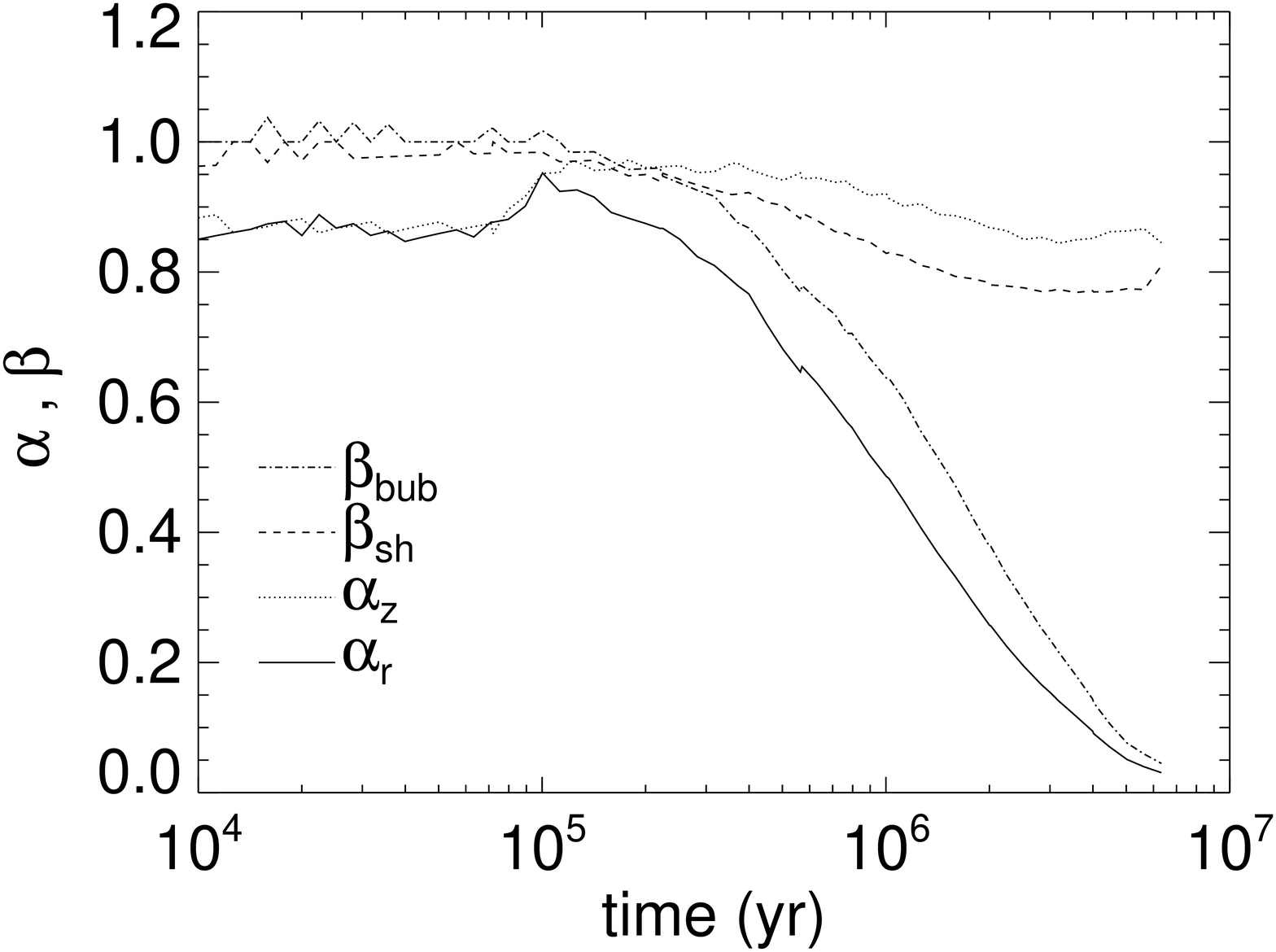}{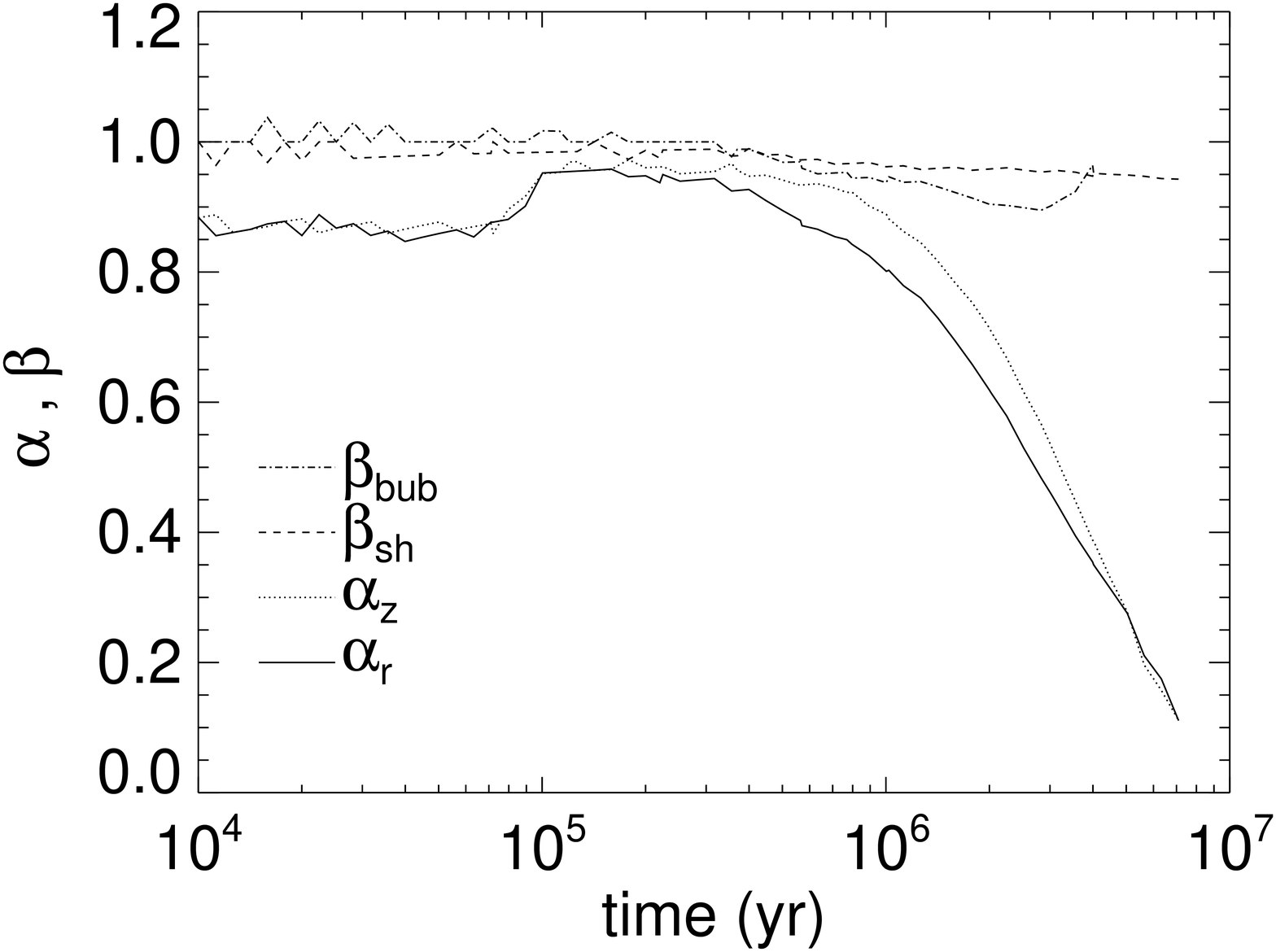}}
\caption{
The ratios of the bubble radii to the shell radii as
 $\alpha_z\equiv Z_{\rm bub}/Z_{\rm sh}$
 and $\alpha_r\equiv R_{\rm bub}/R_{\rm sh}$.
We also plot the time variations of asymmetric factors 
$\beta_{\rm sh}\equiv Z_{\rm sh}/R_{\rm sh}$
 and $\beta_{\rm bub}\equiv R_{\rm bub}/Z_{\rm bub}$.
Shell radii $Z_{\rm sh}$ and $R_{\rm sh}$ are calculated 
 as the distance of the shell from the explosion site along $z$- and $r$-axes,
 respectively.
Bubble radii $Z_{\rm bub}$ and $R_{\rm bub}$ are
 the sizes of the hot interior cavity measured along the $z$- and $r$-axes.
Models A (a: $B_0=5\mu{\rm G}$) and C (b: $B_0=1\mu{\rm G}$) are plotted.
\label{fig:4}}
\end{figure}

To see the departure from the spherically symmetric SNR more quantitatively,
 we define the ellipticity of the SNR shell as
\begin{equation}
 \beta_{\rm sh}\equiv Z_{\rm sh}/R_{\rm sh},
\end{equation}
where $Z_{\rm sh}$ and $R_{\rm sh}$ are the distances from the explosion site along the $z$- and
 $r$-axes and the distance is measured for the point at which the density changes from that of
 the ISM in the cross-cuts by more than 1\% (Fig.\ref{fig:1}).
As for the bubble, its ellipticity is calculated as
\begin{equation}
 \beta_{\rm bub}\equiv R_{\rm bub}/Z_{\rm bub},
\end{equation}
where $Z_{\rm bub}$ and $R_{\rm bub}$ are the distances from the surface of the bubble to the
 explosion site.
The surface of the bubble is defined using the temperature isosurface of $T=10^5$K and
 the iso-density surface with the same density as the ISM $n_0$ as
\begin{equation}
 R_{\rm bub}\equiv \min(R_{T=10^5{\rm K}}, R_{n=n_0}),
\end{equation}
\begin{equation}
 Z_{\rm bub}\equiv \min(Z_{T=10^5{\rm K}}, Z_{n=n_0}),
\end{equation}
where the former temperature isosurface indicates the surface in the pressure-driven
 phase ($t\gtrsim 10^5{\rm yr}$) while the latter density isosurface does that in the adiabatic stage.
We found the minimum of these two gives the bubble surface appropriately. 
It should be noted that these two factors are chosen to be smaller than unity for ordinary
 magnetized SNRs ($Z_{\rm sh} < R_{\rm sh}$ while $R_{\rm bub} < Z_{\rm bub}$).
Figure \ref{fig:4} plots ellipticities $\beta_{\rm sh}$ and $\beta_{\rm bub}$.
After $t\gtrsim 10^5{\rm yr}$, $\beta_{\rm sh}$ begins to decrease and reaches $\simeq 0.8$
 after $t \gtrsim 2\times 10^6{\rm yr}$.
This shows that the maximum difference between $R_{\rm sh}$ and $Z_{\rm sh}$ is 20\% and
 the SNR is observed as elongated in the $r$-direction (an oblate spheroidal shape),
 if we take the shock front.
Asymmetry in the {\em bubble} shape also appears  $t\gtrsim 10^5{\rm yr}$ and grows steadily.
The bubble shrinks in the $r$-direction to form a very thin cigar-like structure
 $\beta_{\rm bub}\equiv R_{\rm bub}/Z_{\rm bub}\ll 1$.
The hot cavity in the pressure-driven expansion phase forms a prolate spheroid.

Figure \ref{fig:4} also plots the ratio of the bubble radius to the shell radius
 in respective directions as
\begin{equation}
 \alpha_z=Z_{\rm bub}/Z_{\rm sh},
\end{equation}
and
\begin{equation}
 \alpha_r=R_{\rm bub}/R_{\rm sh}.
\end{equation}
Since in the Sedov phase the shell size is approximately equal to one-twelfth of the
 shock radius $\Delta R \simeq R/12$ \citep{sak96}, $\alpha_z$ and $\alpha_r$ keep constant $\simeq 0.9$
 in the adiabatic phase $t\lesssim 7\times 10^4{\rm yr}$.
As a cool shell forms, the difference between the shock radius and that of the hot cavity
 decreases.
The quantities $\alpha_z$ and $\alpha_r$ reach $\simeq 0.95$ at the age of $t\simeq 10^5{\rm yr}$.
The shell width near the $z$-axis ($\Delta Z=Z_{\rm sh}-Z_{\rm bub}$) continues to be thin
 for $t\lesssim 4\times 10^5{\rm yr}$ and increases gradually after $t\gtrsim 4\times 10^5{\rm yr}$.
On the other hand, the shell width in the $r$-direction ($\Delta R=R_{\rm sh}-R_{\rm bub}$)
 increases steadily after $t\gtrsim 10^5{\rm yr}$ ($\alpha_r$ decreases monotonically).
Figure \ref{fig:4} shows us that asymmetry from the spherically symmetric expansion
 begins just after the effective radiative cooling forms a thin shell.
The diffuse shell formation owing to the magnetic field is triggered
 just after this shell formation.

\subsection{Model with $B_0=1\mu{\rm G}$} 

\begin{figure}
\noindent
{\centering
(a)\hspace*{7cm}(b)\\
\plottwo{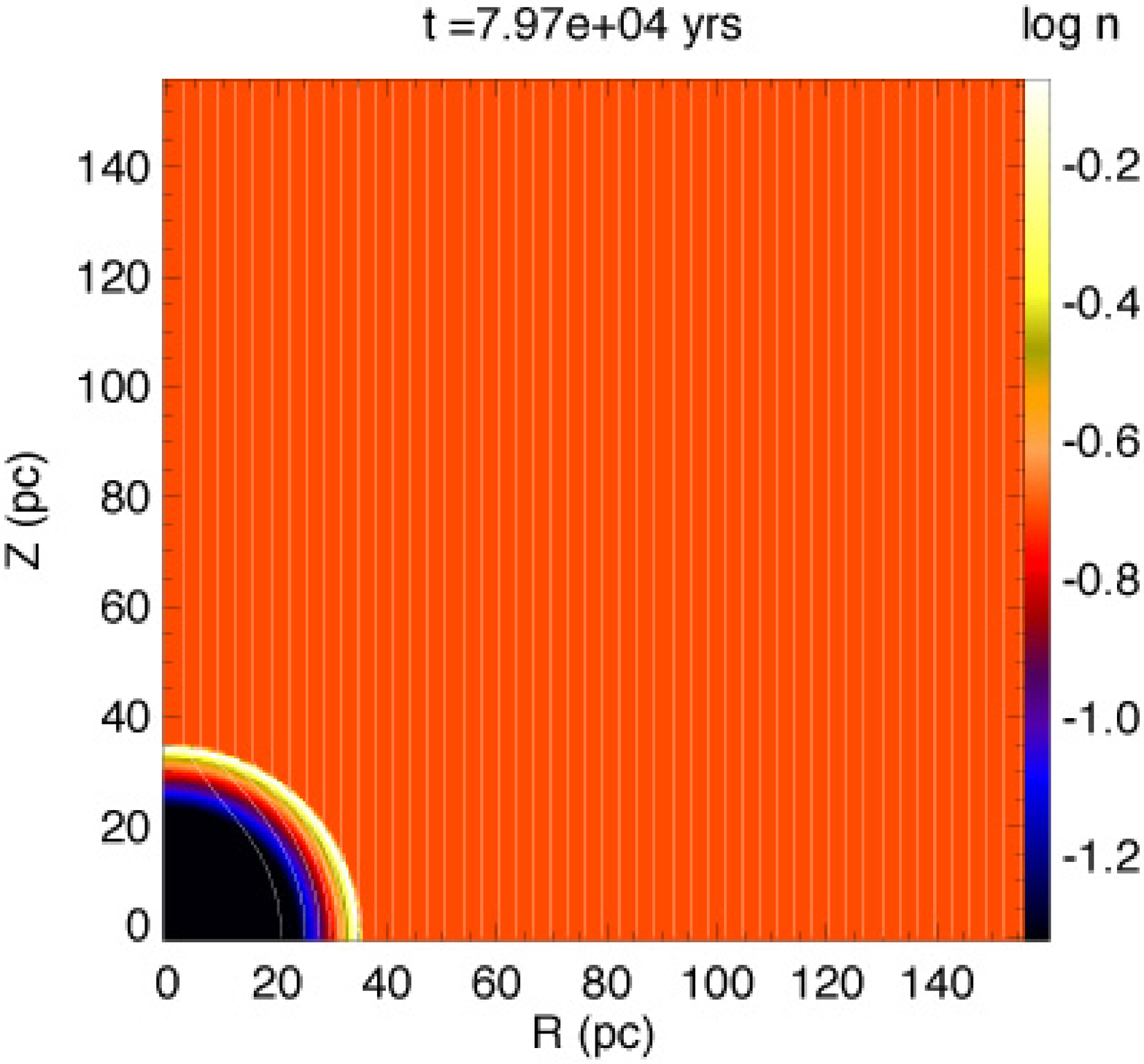}{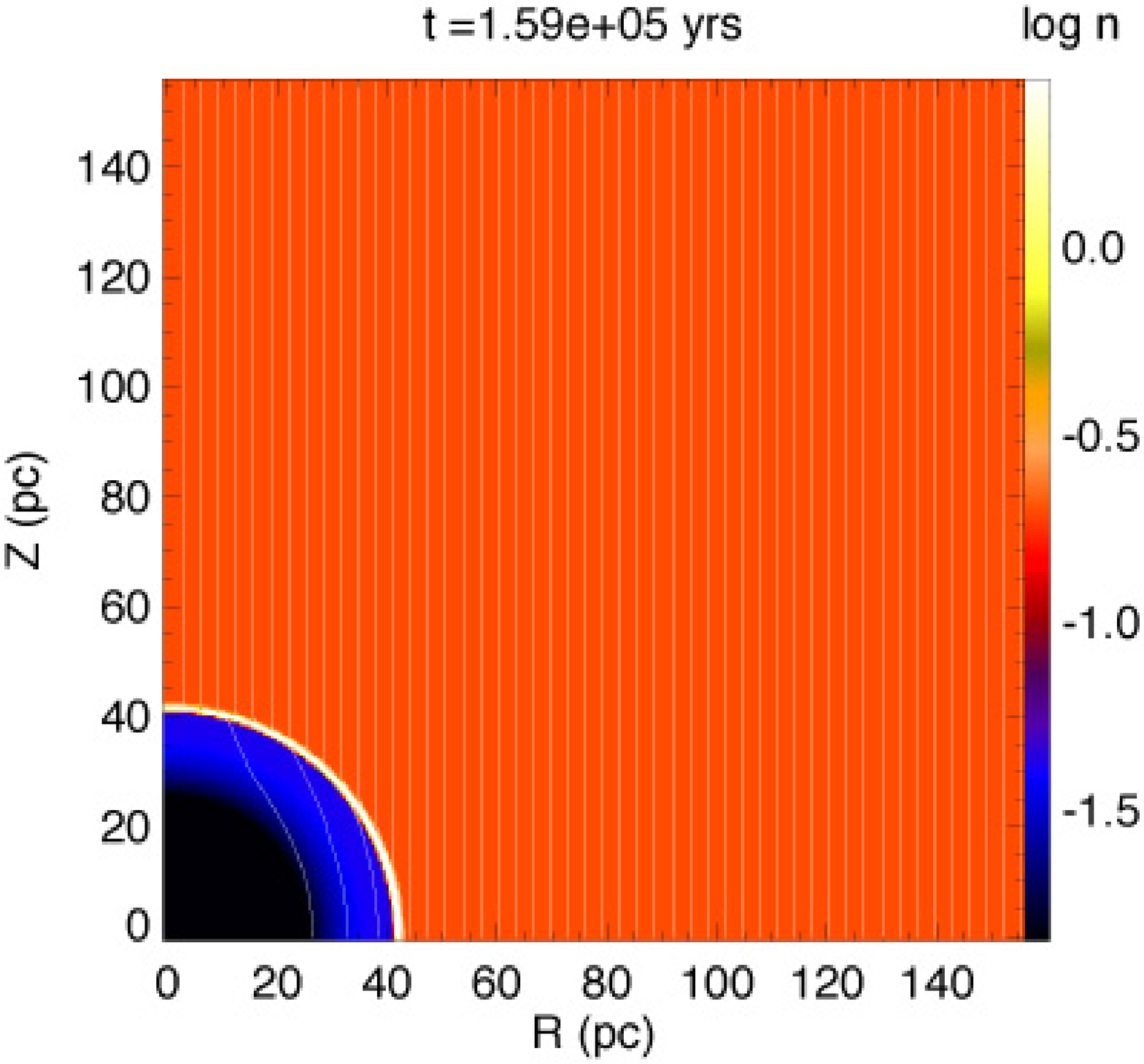}\\
(c)\hspace*{7cm}(d)\\
\plottwo{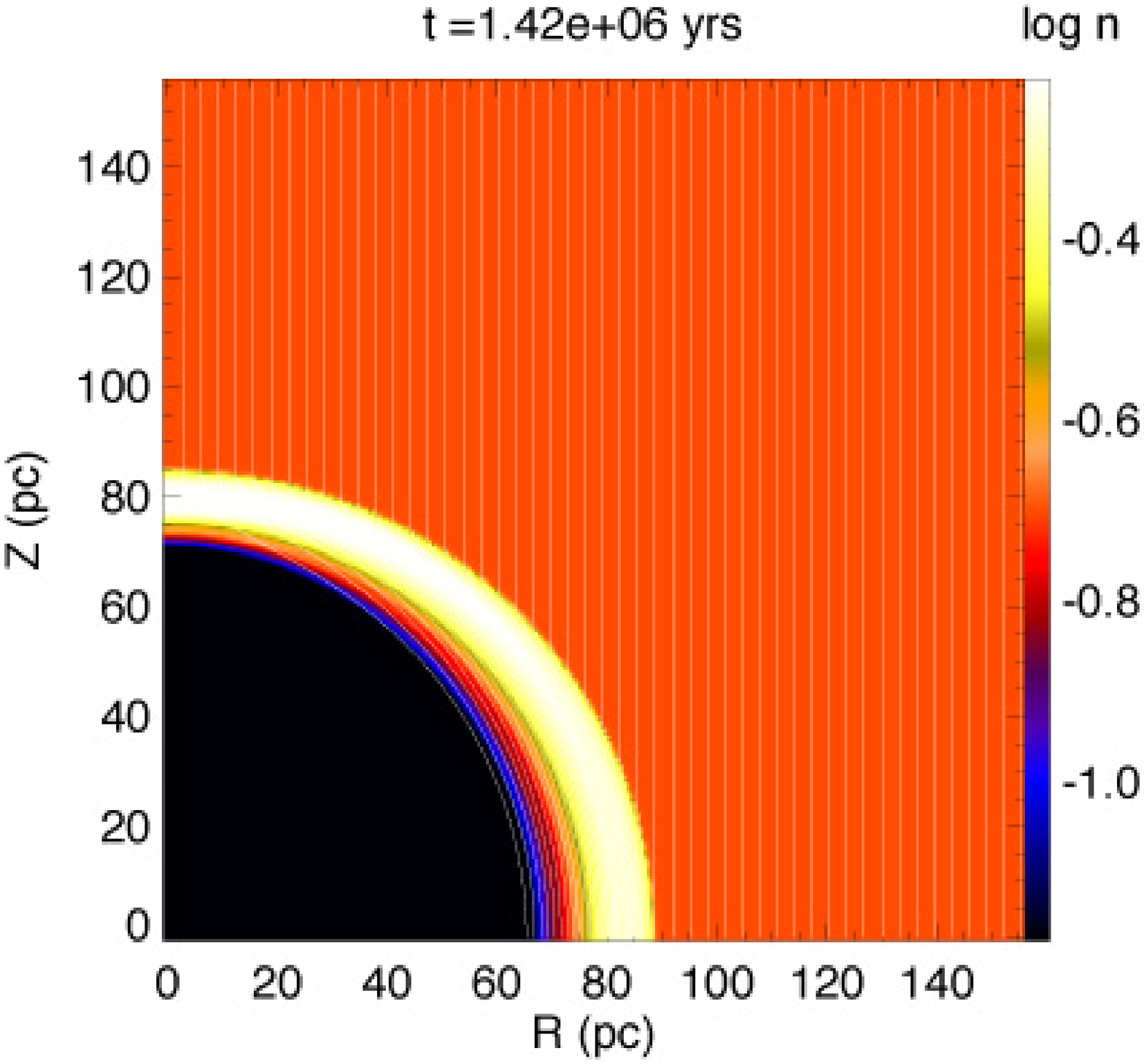}{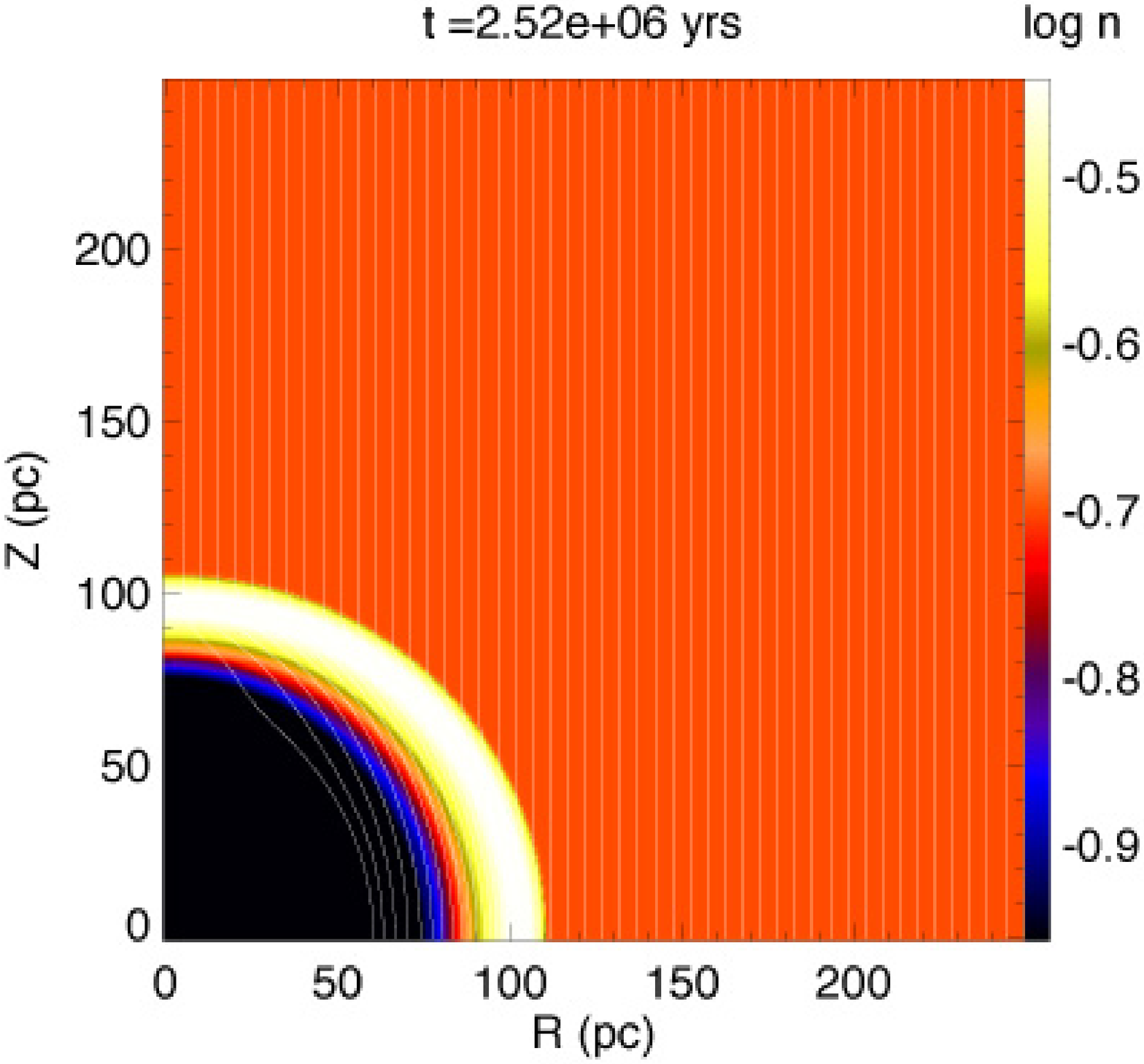}\\
}
\caption{
Structures of the SNR expanding in the weak magnetic field model C.
Density distributions (false color) and magnetic field lines (white lines)
 are shown.
Four snapshots are taken at the ages of
 $7.97\times 10^4{\rm yr}$ (a), 
 $1.59\times 10^5{\rm yr}$ (b), 
 $1.42\times 10^6{\rm yr}$ (c), 
 and $2.52 \times 10^6{\rm yr}$ (d).
\label{fig:5}}
\end{figure}

In this section, we describe model C with a weak magnetic field of $B_0=1\mu{\rm G}$.
Other parameters are the same as model A.
Since the magnetic pressure of the ISM is equal to
 $B_0^2/8\pi=4\times 10^{-14}{\rm dyn\,cm^{-2}}(B_0/1\mu {\rm G})^2$,
 the plasma $\beta$ of the ISM is as large as $\beta_0=14.5$. 
Before and at the thin shell formation stage [Fig. \ref{fig:5}(a) and (b)],
 the SNR is spherically symmetric and looks very similar to model A
 (first two panels of Figs. \ref{fig:2} and \ref{fig:3}). 
However, although the structure of the magnetized SNR in $B_0=5\mu{\rm G}$
 (model A) is far from being the spherically symmetric in the pressure-driven expansion stage 
 [last three panels of Figs. \ref{fig:2} and \ref{fig:3}],
 the SNR in a weak magnetic field (model C) remains its shape spherically symmetric 
 [Fig. \ref{fig:5}(c) and (d)].
For example, at $t=1.42\times 10^6{\rm yr}$,
 the shell extends in the $r$-direction from 65 to 89 pc
 and in the $z$-direction from 70 to 85 pc. 

In Figure 4 (b), we plot the ratio of bubble-to-shock radii, and major-to-minor axis ratio
 of the shock front and the bubble for model C.
Compared with Figure 4 (a) of model A,
 although the ratio of bubble-to-shock radii is as small as 
 $\alpha_r\simeq 0.5$ at the age of $10^6{\rm yr}$ for model A,
 it is $\alpha_r\simeq 0.8$ for model C.
This means that the thickness of the shell, which propagates in the direction perpendicular to the magnetic field ($r$-direction), is much reduced in model C.
The bubble begins to contract after $1.42 \times 10^6{\rm yr}$, which is much longer than
 that of model A $t\simeq 4\times 10^5{\rm yr}$.
The difference comes from the effect of the magnetic force working to compress the bubble.
In both models A and C, $\alpha_r$ decreases with time 
 and reaches 0.1 in $t\simeq 4\times 10^6{\rm yr}$ (model A)
 and $t\simeq 7\times 10^6{\rm yr}$ (model C).

As for the same ratio for the parallel ($z$-)direction,  
 both models have similar values as $\alpha_z\simeq 0.93$ for model A and
 $\alpha_z\simeq 0.88$ for model C at the age of $10^6{\rm yr}$.
However, although model A remains at $\alpha_z\simeq 0.85$ for the final phase
 $t\gtrsim 5\times 10^6{\rm yr}$, $\alpha_z$ continues to decrease in model C and
 $\alpha_z\simeq 0.1$ for $t\sim 7\times 10^6{\rm yr}$.
This means the shell propagating along the $z$-axis is confined to being thin in model A,
 while the shell thickness increases with time in model C.
In the last paragraph of section 3.1, we already mentioned that the expansion 
 of the shell propagating in the $z$-direction
 is driven by the thermal pressure in the bubble, which is sustained by the radial
 compression due to the magnetic force.
This seems to have the effect of keeping the shell that is propagating in the $z$-direction thin.

\subsection{Effect of the Density}

In this section, the models with high ISM density $n_0=1{\rm cm^{-3}}$ are shown.
Model E has the same parameters as model A except for the interstellar density.   
The heating rate is taken to be 5 times larger than model A to keep $T_0=10^4{\rm K}$. 

\begin{figure}
\noindent
{\centering
(a)\hspace*{7cm}(b)\\
\plottwo{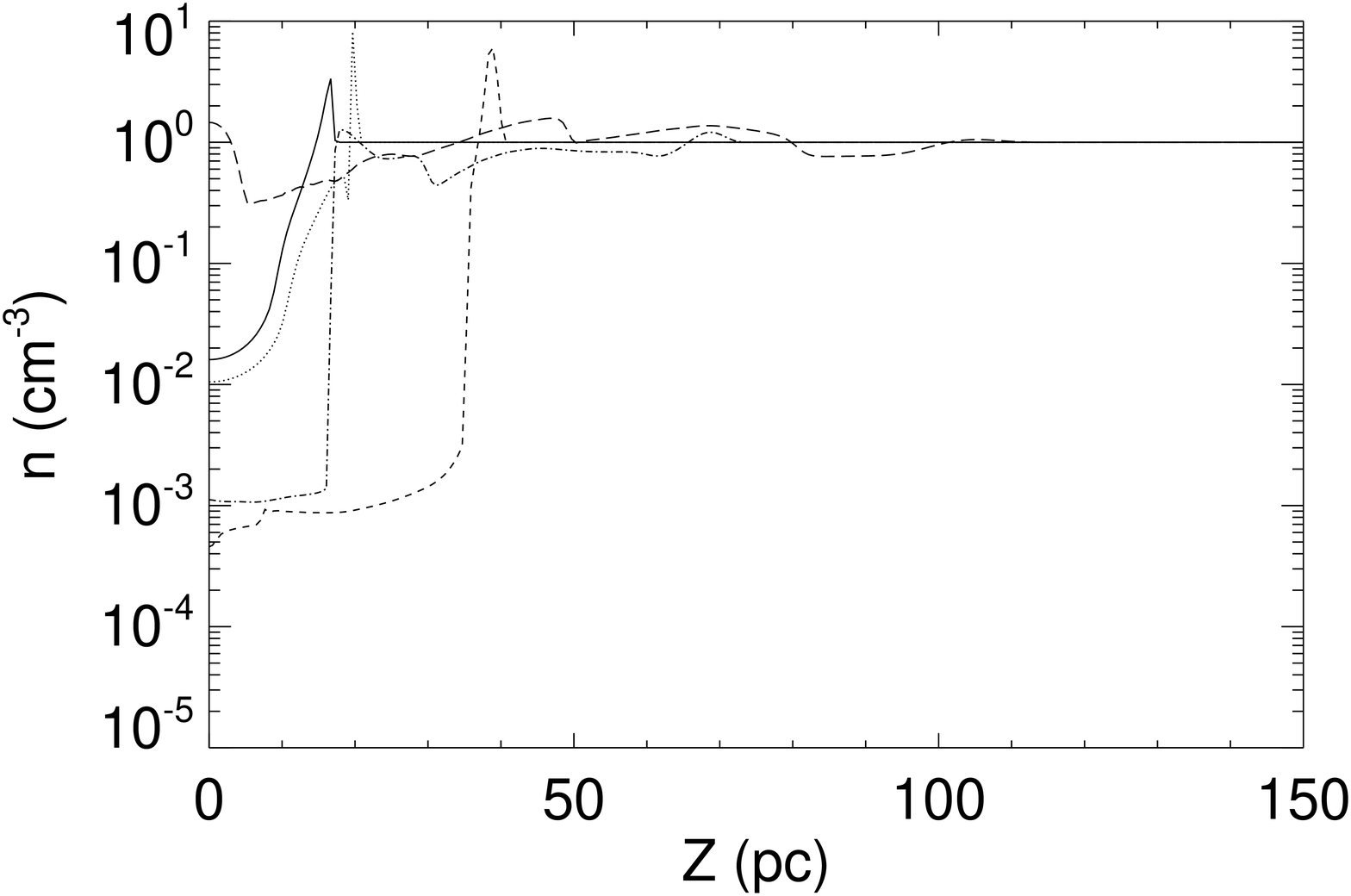}{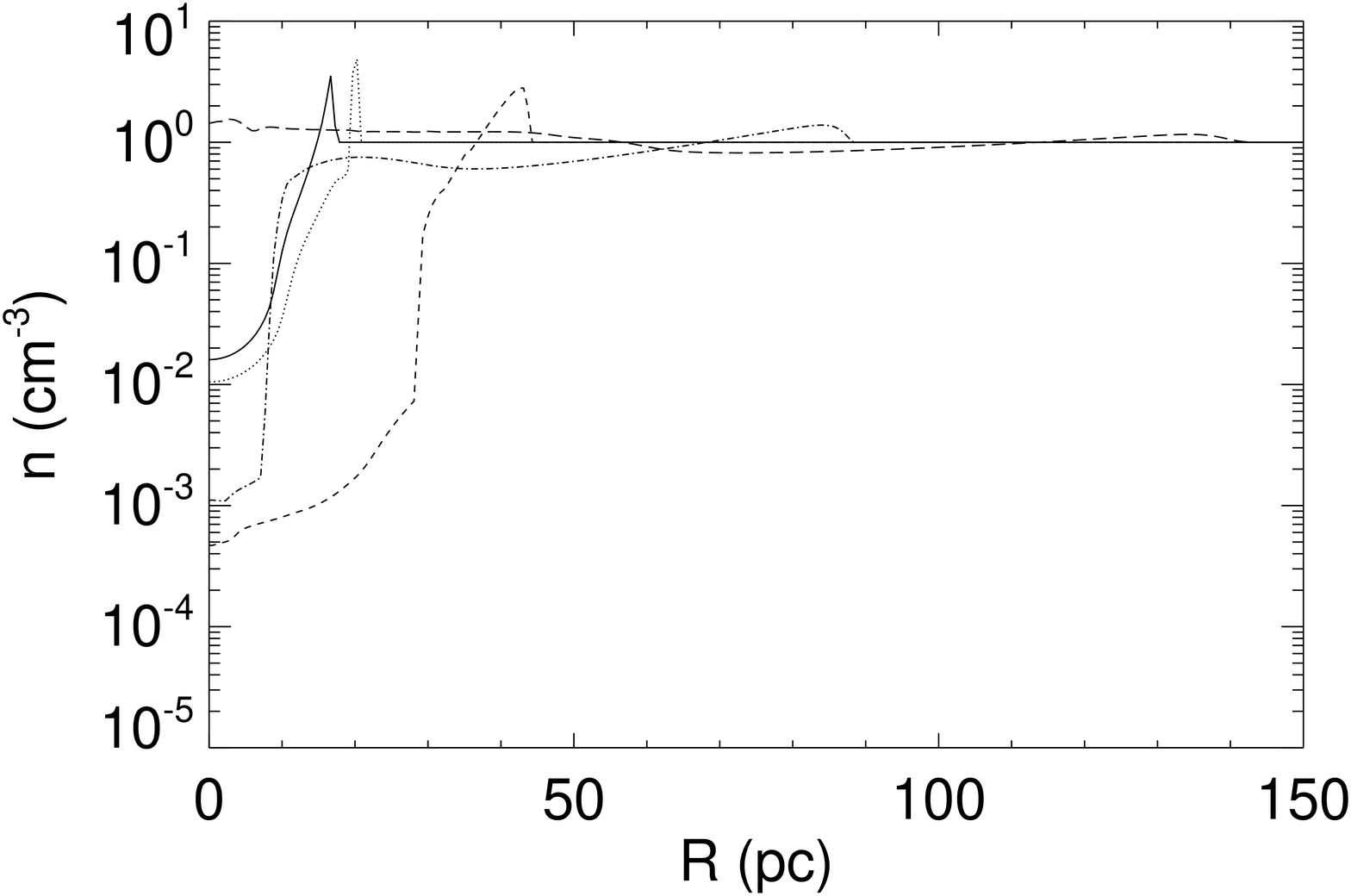}\\
(c)\hspace*{7cm}(d)\\
\plottwo{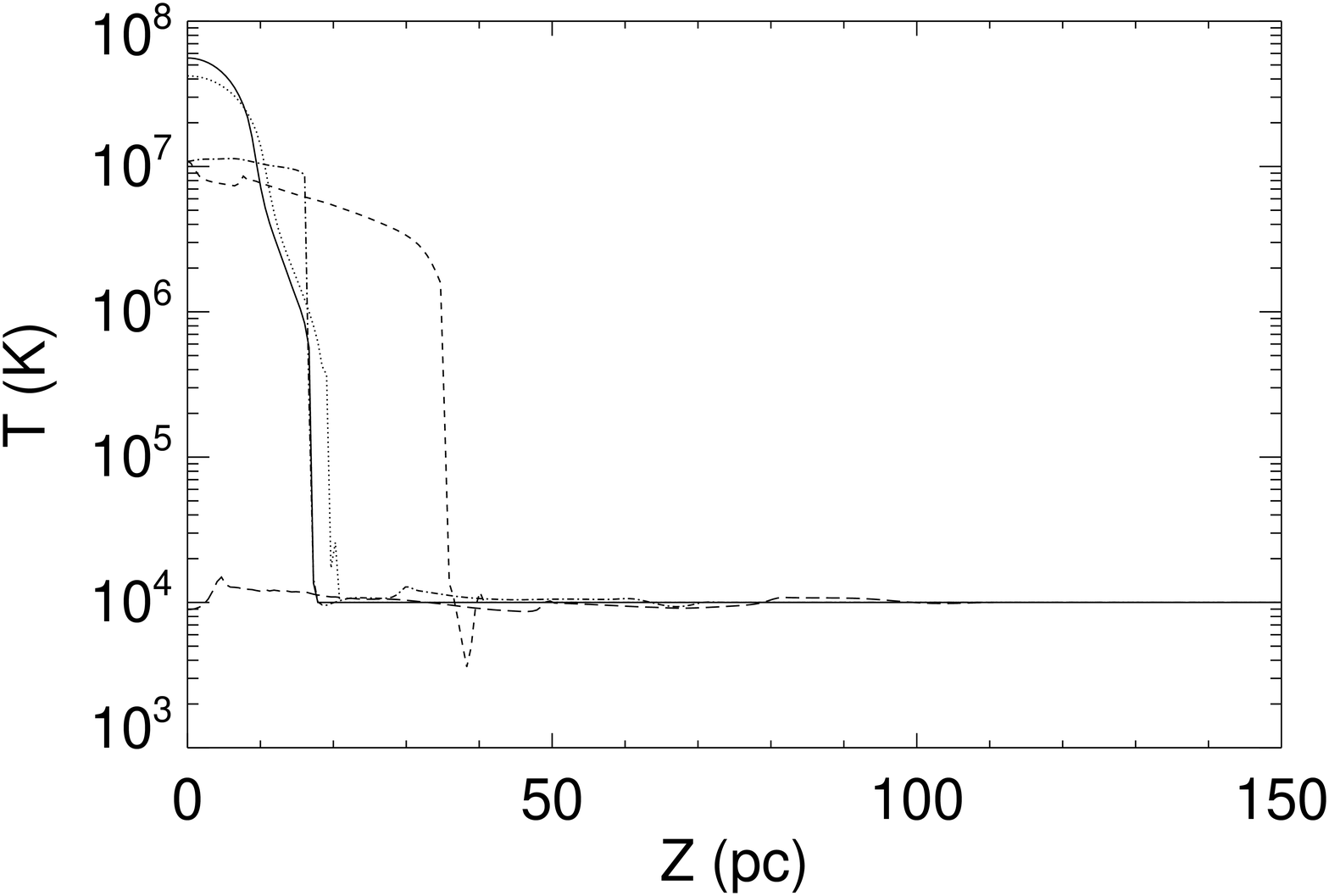}{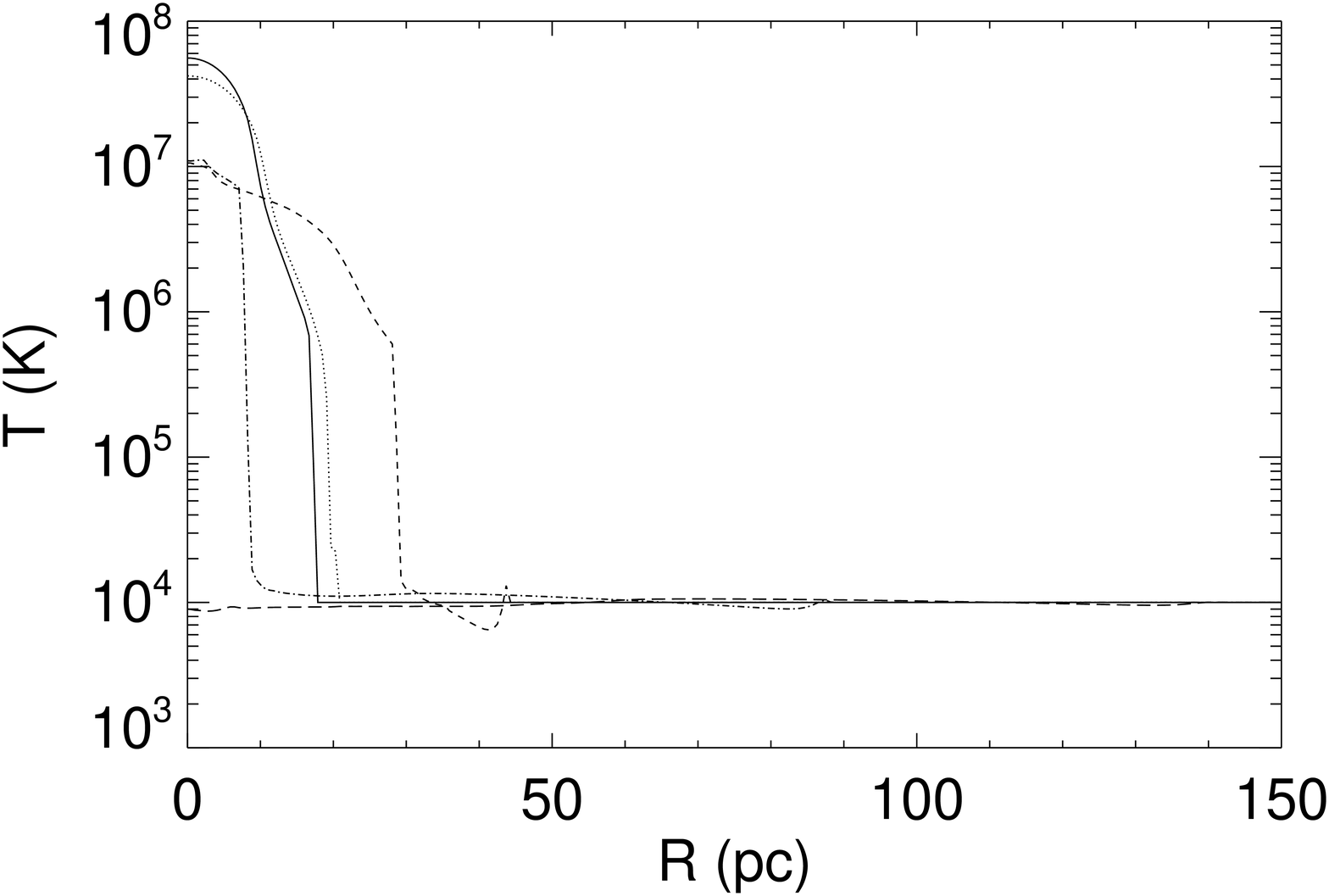}\\
}
\caption{
The same as Fig.\ref{fig:1} but for SNR in $n_0=1{\rm cm^{-3}}$ (model E).
Structures of the SNR expanding in the dense ISM with $n_0=1{\rm cm^{-3}}$.
Five snapshots are taken at the ages of
 $2.83\times 10^4{\rm yr}$ (solid line),
 $5.03\times 10^4{\rm yr}$ (dotted line), 
 $5.03\times 10^5{\rm yr}$ (dashed line), 
 $2.52\times 10^6{\rm yr}$ (dash-dotted line),
 and $5.64 \times 10^6{\rm yr}$ (long dashed line).
\label{fig:6}}
\end{figure}

Equation (\ref{eqn:1}) gives the timescale to enter the pressure-driven snow-plow phase
 from the Sedov stage as $t_{\rm sf}\simeq 3.1 \times 10^4 {\rm yr} 
           \left(E_0/5\times 10^{50}{\rm erg}\right)^{3/14}
           \left(n_0/1{\rm cm^{-3}}\right)^{-4/7}$, which is three times shorter
 than the models with low density $n_0=0.2{\rm cm}^{-3}$. 
We plotted the snapshot of the adiabatic stage at $t=2.83\times 10^4{\rm yr}$
 in Figure \ref{fig:6} (solid line).
The next snapshot is at $t=5.03 \times 10^4{\rm yr}$ (dotted line),
 at which the density of the shell 
 propagating in the $r$-direction takes the maximum.
This shows the thin shell phase of SNR.

At $t=5.03 \times 10^5{\rm yr}$ (dashed line),
 $R_{\rm bub}$ takes the maximum $R_{\rm bub}=29{\rm pc}$
 and the bubble begins to contract in the $r$-direction.
This timescale is almost the same as that of model A ($t=4\times 10^5{\rm yr}$),
 even if the adiabatic to pressure-driven transition timescale $t_{\rm sf}$
 is three times shorter than low-density model A.
At this stage, the asymmetry between $z$- and $r$-directions has developed.
That is, the shell width of the $r$-direction
 $\Delta R=R_{\rm sh}-R_{\rm bub}\simeq 15{\rm pc}$ and that of the $z$-direction
 $\Delta Z=Z_{\rm sh}-Z_{\rm bub}\simeq 5{\rm pc}$ differ.
And as a consequence, the shell propagating in the $z$-direction is thin, while
 that propagating in the $r$-direction is thick.

The shock wave propagates outwardly in both $z$- and $r$-directions.
The next snapshot is at $t=2.52\times 10^6{\rm yr}$ (dash-dotted line). 
The bubble has almost been crushed $R_{\rm bub}\lesssim 10{\rm pc}$.
In $\simeq 3\times 10^6{\rm yr}$ the bubble is completely crushed.
The contraction time-scale is much shorter than that of the low-density models
 ($\gtrsim 6\times 10^6{\rm yr}$ for model A).

%In high-density model E, the scale of the whole structure is $\simeq$1.8 times smaller than that of low-density Model A because the timescale $t_{\rm sf} to enter the pressure-driven snow-plow phase from the Sedov stage is three times shorter than the models with low density $n_0=0.2{\rm cm}^{-3}$.
%The bubble contracts $\simeq$3 times earlier to form the thin shell, and to contract.
%On the other hand, the Lorentz force of both Model E and Model A to shrink the bubble is the same order, $\simeq 10^{-29}{\rm dyn\,cm^{-3}}$ at the thin shell phase, and the relation of it doesn't change after the radiative cooling phase.

\subsection{Effect of the Heating Rate}

\begin{figure}
\noindent
{\centering
(a)\hspace*{7cm}(b)\\
\plottwo{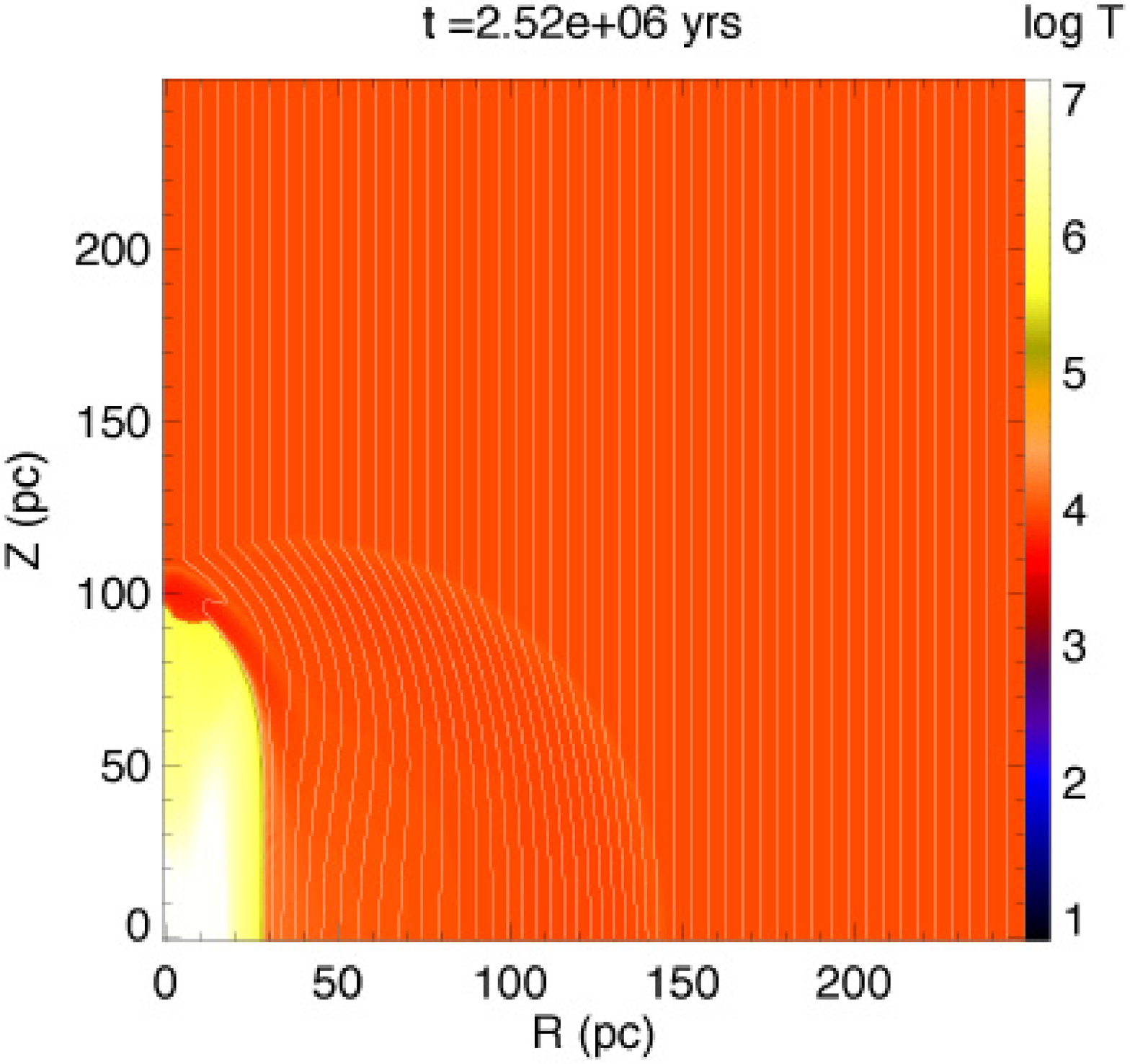}{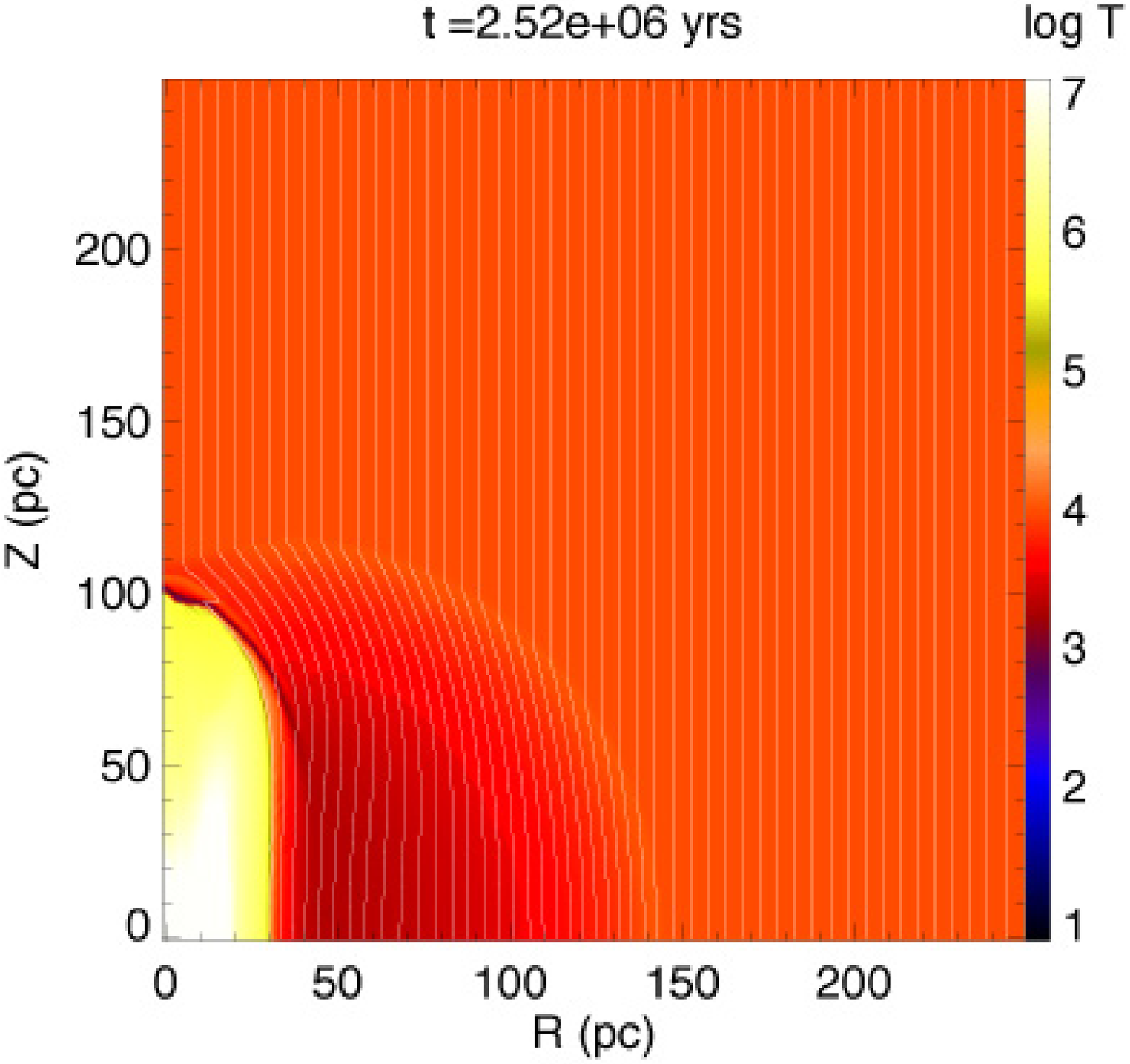}\\
(c)\hspace*{7cm}(d)\\
\plottwo{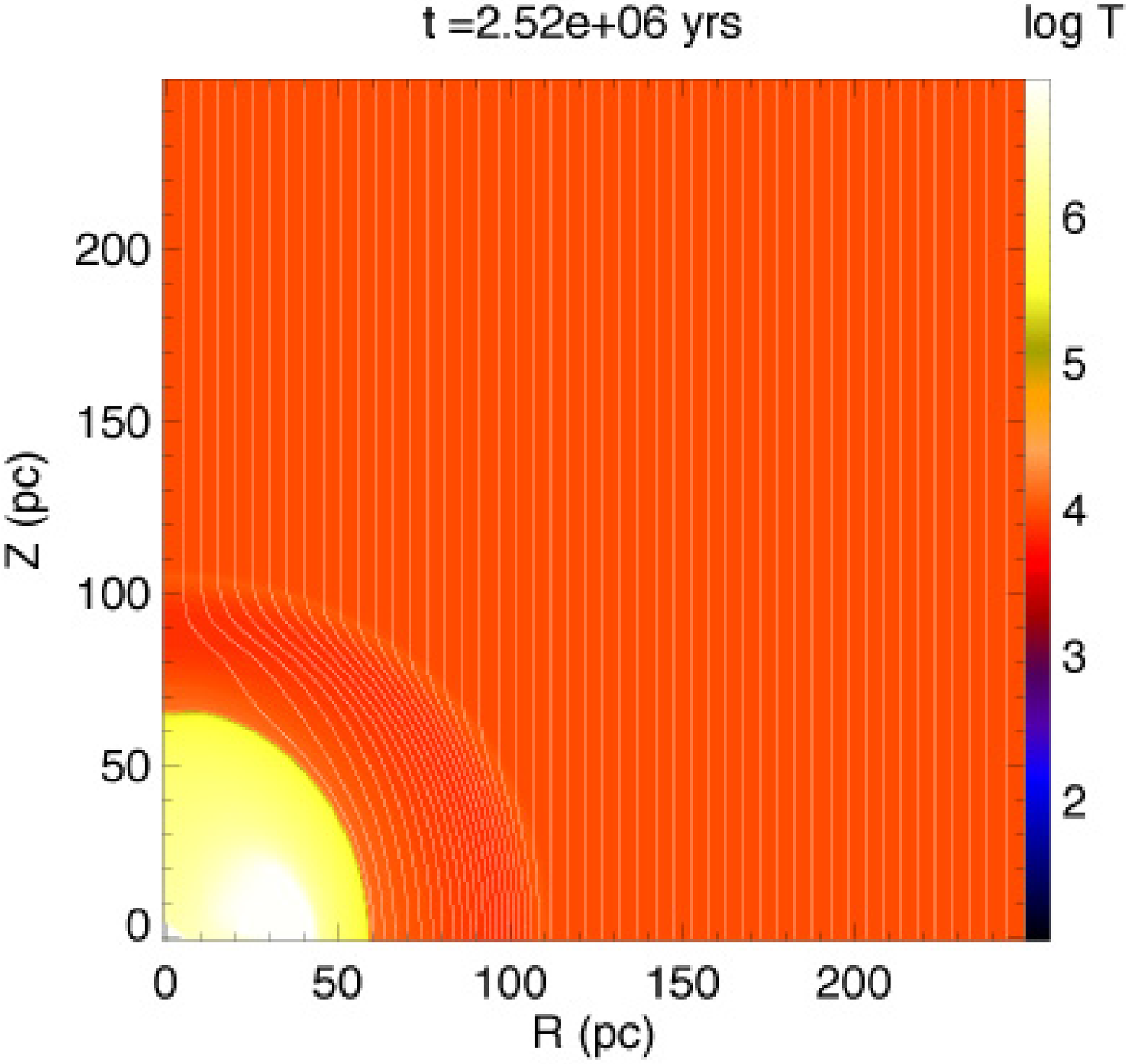}{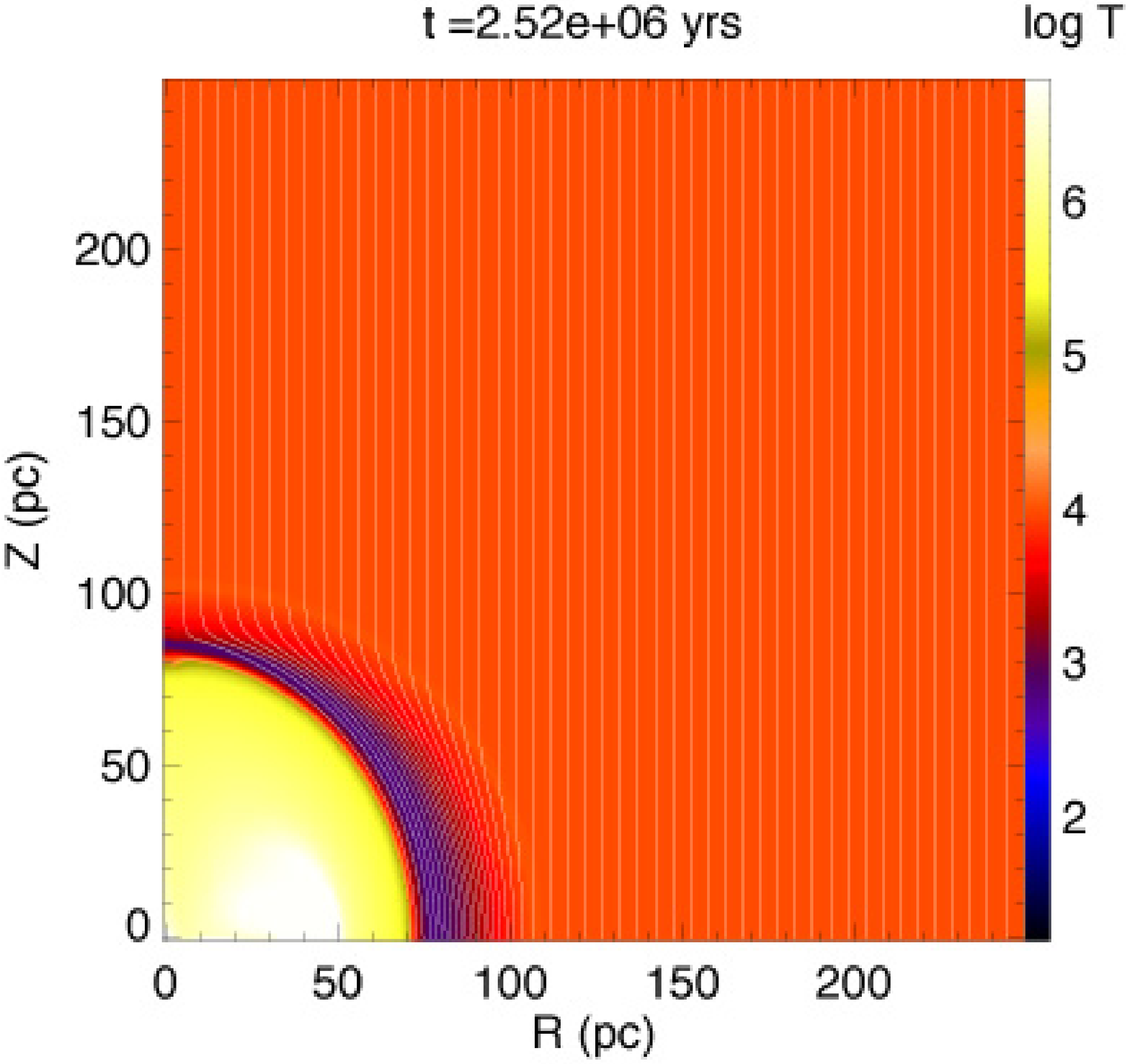}\\}
\caption{Comparison between models with strong heating rates and
 inefficient heating rates.
Temperature distributions are plotted
 for models A (panel a), AW (panel b), C (panel c), and CW (panel d) 
at the age of $2.52 \times 10^6{\rm yr}$.
\label{fig:7}}
\end{figure}

In this section, we compare models with the same parameters except for the 
 heating rate.
In models AW-DW, we take $\Gamma_0=5\times 10^{-27}{\rm erg\,s^{-1}}$,
 which is $\simeq 1/10$ smaller than previous models A-D. 
In Figure \ref{fig:7}, we plot snapshots at $t=2.52\times 10^6{\rm yr}$
 of the temperature distribution of 
 models A (a) and AW (b) with $B_0=5\mu {\rm G}$ and
 models C (c) and CW (d) with $B_0=1\mu {\rm G}$.
First, it is shown that the structures are very similar for models A and AW,
 even if $\Gamma_0$ differs $\simeq 10$ times.  
Although the height of the bubble  $Z_{\rm bub}$ differs several \%,
 the radius of the bubble $R_{\rm bub}$ differs only slightly.
On the other hand, the sizes of the bubbles of models C and CW are different,
 as the bubble size differs $\simeq 20\%$ for both  $R_{\rm bub}$ and $Z_{\rm bub}$. 

To closely see model A, as for the structure along the $r$-axis,
 the shell ($30 {\rm pc} < r < 140 {\rm pc}$) in model A has
 a temperature of $T\simeq 10^4{\rm K}$,
 while that for model AW is as low as $T\simeq 10^3{\rm K}$ owing to inefficient heating.
As a result, the thermal pressure of model A
 [$p_{\rm th}\simeq 4\times 10^3{\rm cm^{-3}\,K}$]
 is much higher than that of model AW, and decreases
 from the post-shock value of $p_{\rm th}\simeq 4\times 10^3{\rm cm^{-3}\,K}$
 to $p_{\rm th}\simeq 3\times 10^2{\rm cm^{-3}\,K}$ near the bubble boundary.
Although the thermal pressure differs one order of magnitude,
 the expansion of the bubble is unchanged.
This is owing to the fact that magnetic pressure is dominant in the shell of these models.

As for the structure along the $z$-axis,
 the temperature of the shell of model AW ($T\lesssim 10^3{\rm K}$) is much lower than 
 that of model A ($T\simeq 6\times 10^3{\rm K}$).
However, the thermal pressures in the shell look similar, as follows: 
$p_{\rm th}\simeq 3\times 10^3{\rm cm^{-3}\,K}$ (model A) and
$p_{\rm th}\simeq 2\times 10^3{\rm cm^{-3}\,K}$ (model AW).
Further, those in the bubble are equal to 
 $p_{\rm th}\simeq 8\times 10^3{\rm cm^{-3}\,K}$ (model A) and
 $p_{\rm th}\simeq 6\times 10^3{\rm cm^{-3}\,K}$ (model AW).
Since the thermal pressure in the bubble is driving the shell in the $z$-direction, 
 this similarity explains the fact that the expansion is very similar
 even if $\Gamma_0$ changes 10 times. 
As a result, it is concluded that the evolution does not depend on the heating rate
 for models with strong magnetic fields of $B_0\gtrsim 3\mu{\rm G}$.

\begin{figure}
\noindent
{\centering
(a)\hspace*{7cm}(b)\\
\plottwo{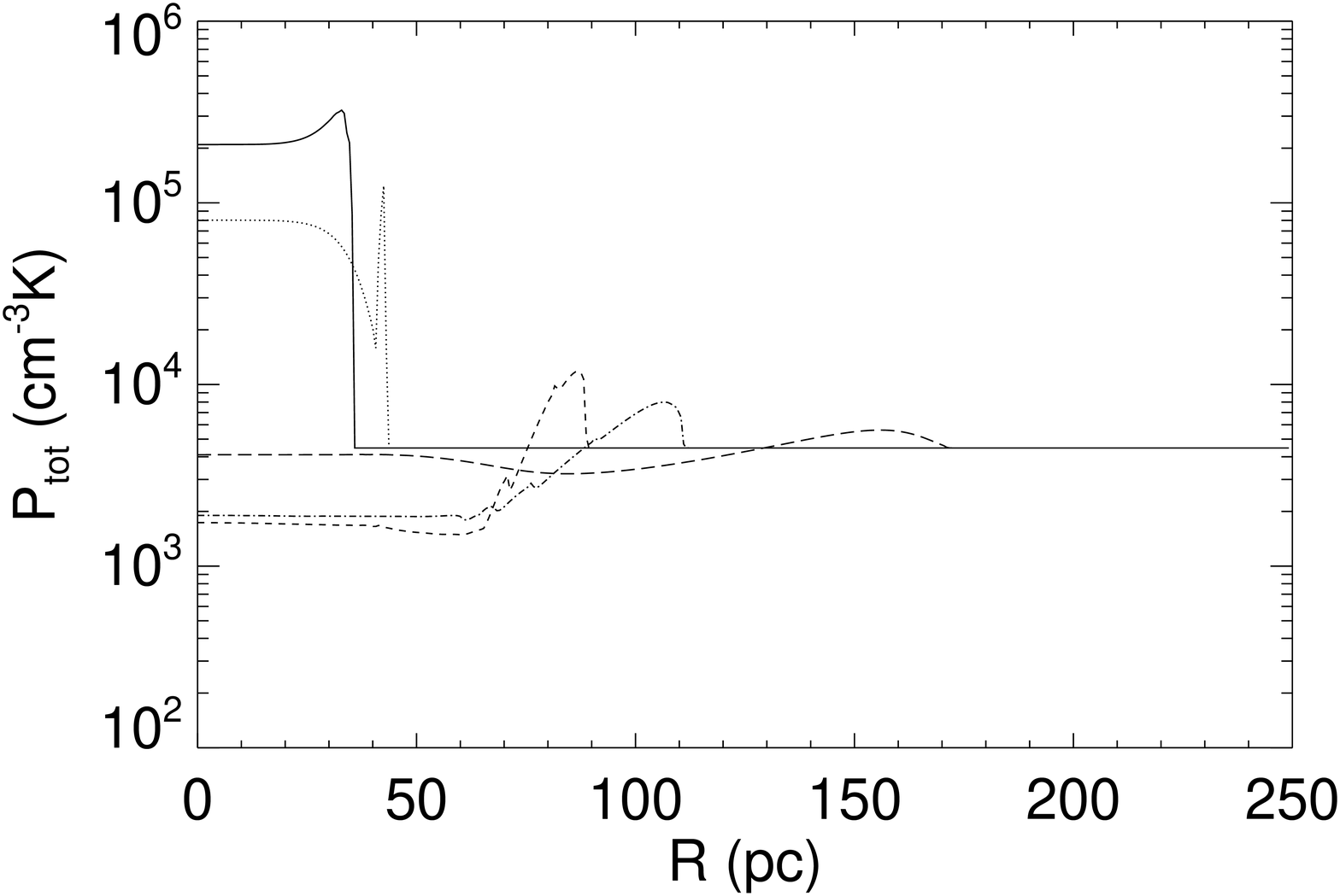}{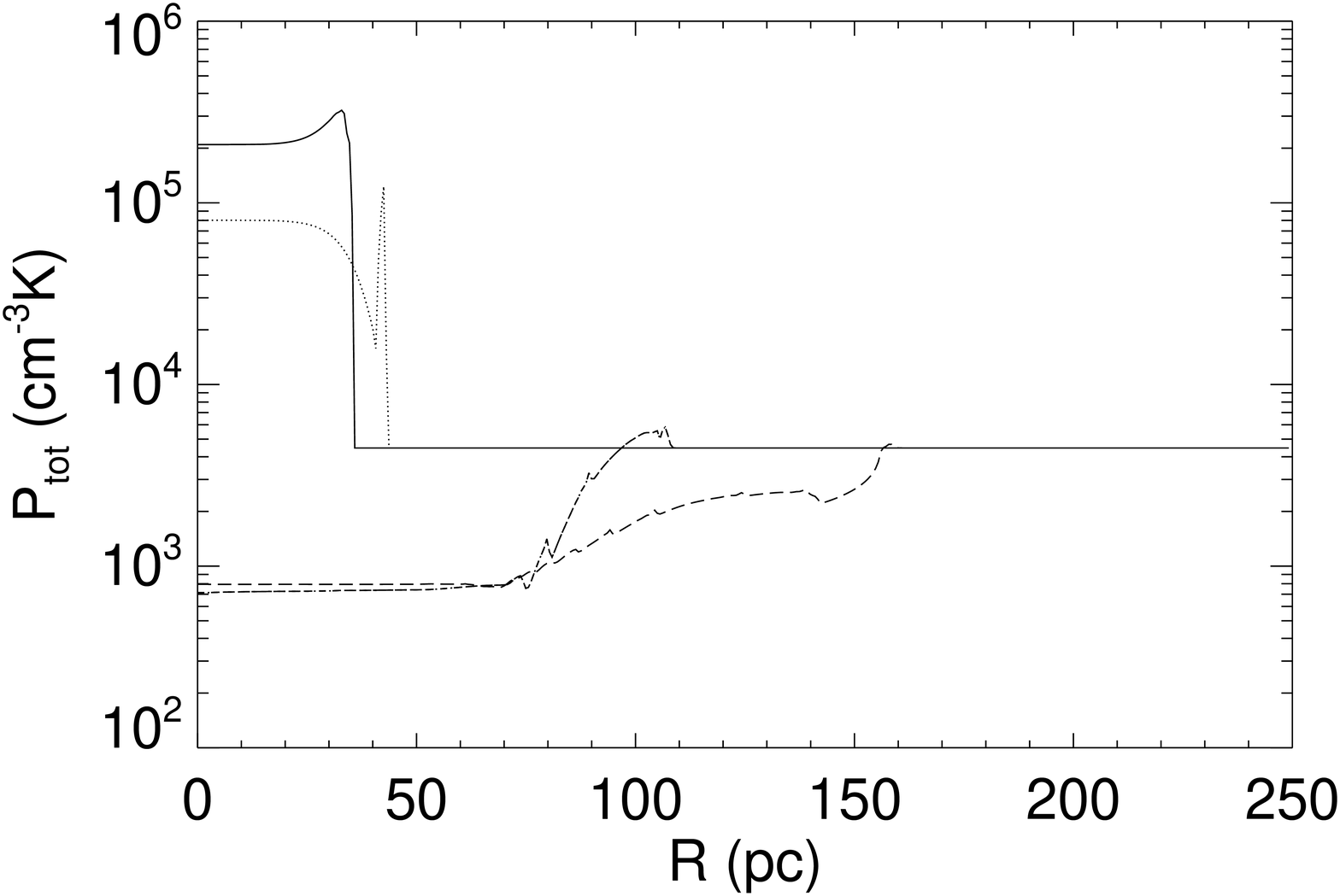}\\
(c)\hspace*{7cm}(d)\\
\plottwo{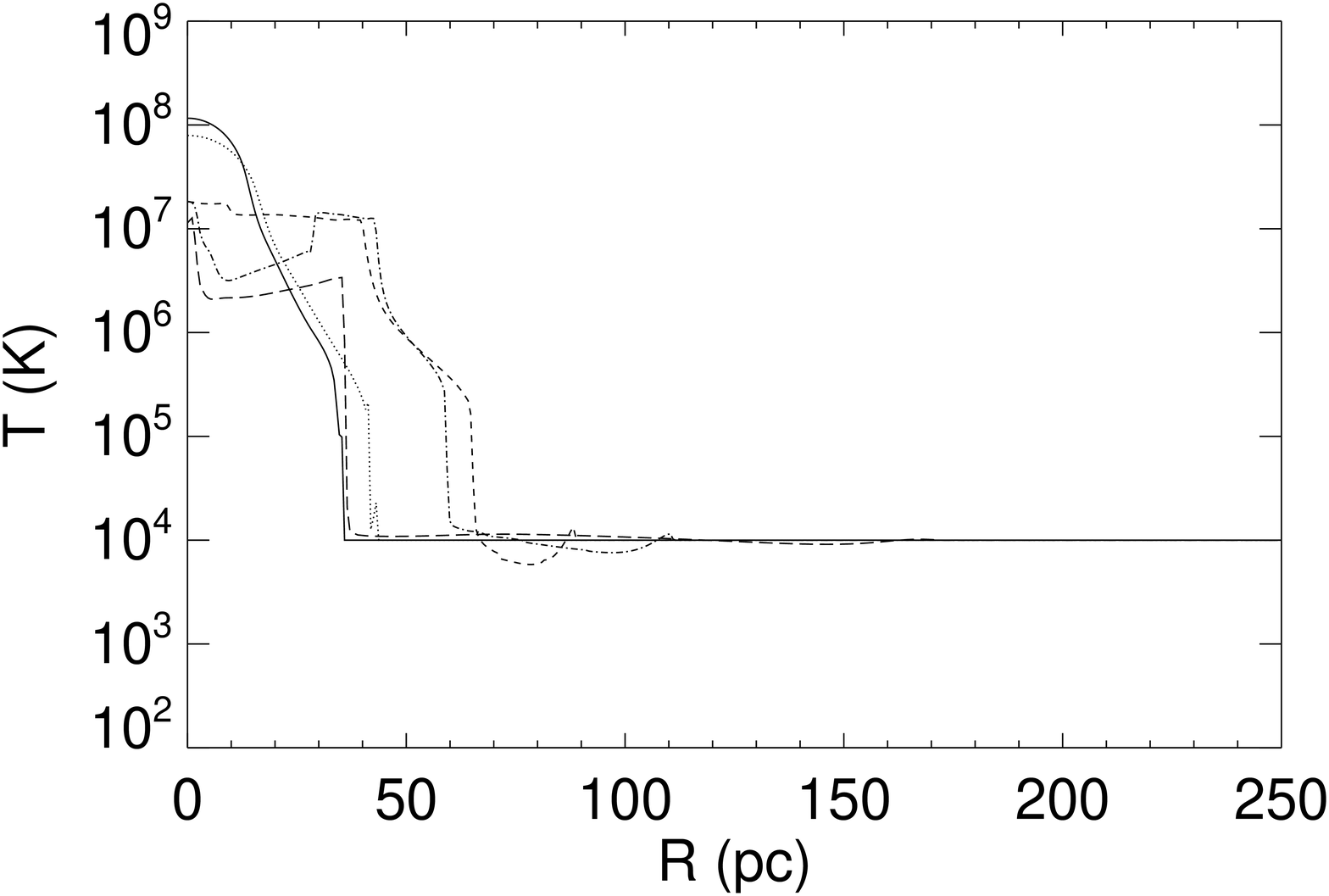}{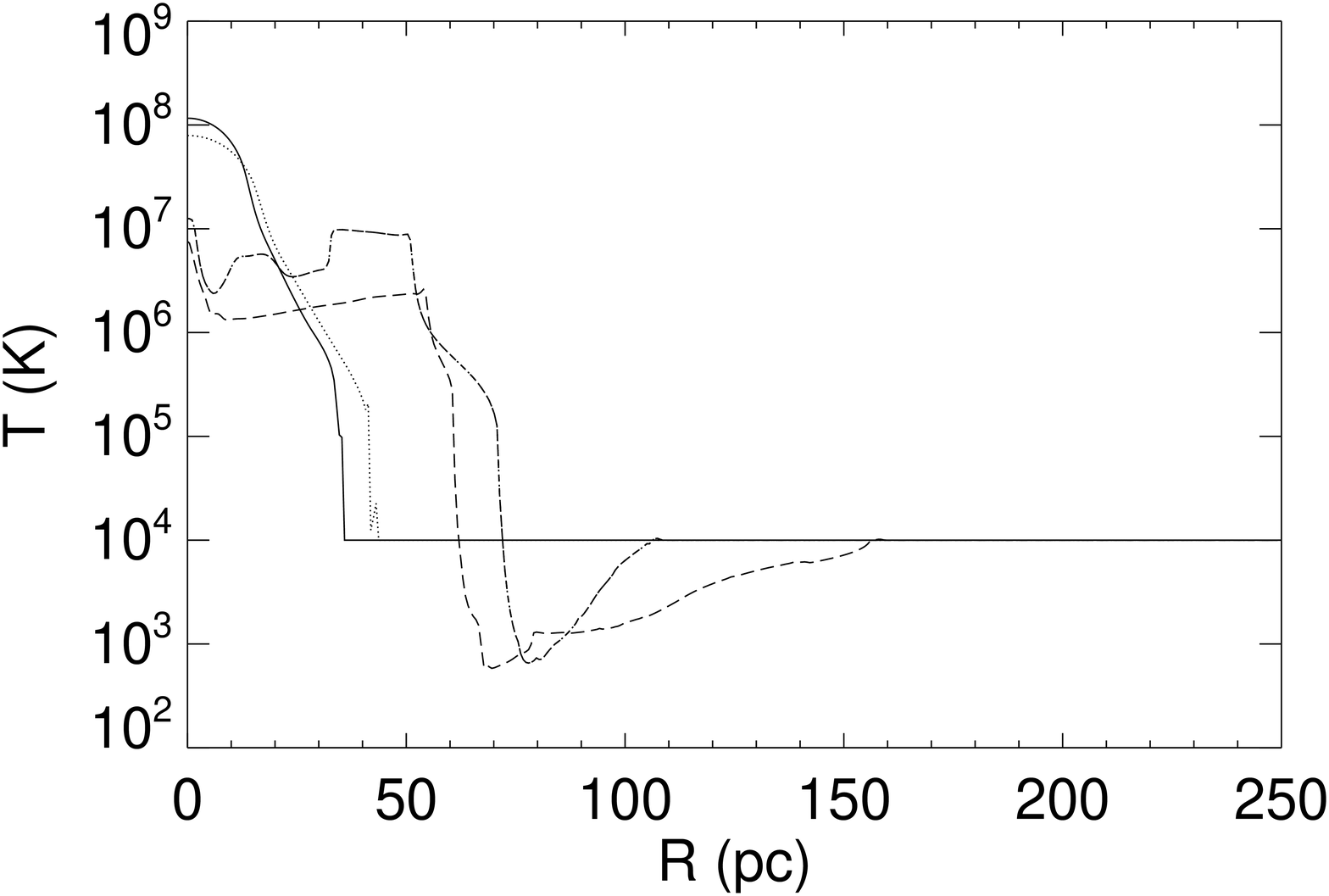}} 
\caption{
Comparison between models C and CW.
Thermal pressure distribution along $r$-axis of model C (a) and that of 
 model CW (b) are plotted.
Temperature distribution along $r$-axis of model C (c) and that of 
 model CW (d) are plotted. 
Snapshots are taken at the ages of
 $7.97\times 10^4{\rm yr}$ (solid line),
 $1.59\times 10^5{\rm yr}$ (dotted line), 
 $1.42\times 10^6{\rm yr}$ (dashed line), 
 $2.52\times 10^6{\rm yr}$ (dash-dotted line),
 and $5.64 \times 10^6{\rm yr}$ (long dashed line).
\label{fig:8}}
\end{figure}

In Figure \ref{fig:8}, we plot the thermal pressure (a)-(b)
  and temperature distributions (c)-(d) of models C and CW.
Before and at the shell formation stage (solid and dotted lines)
 the two evolutions are the same,
 since the heating rate has no effect on evolution at this stage.
After the shell formation (dashed, dash-dotted, and long dashed lines),
 these two models show different evolutions.
At $t=2.52\times 10^6{\rm yr}$,
 the bubble occupies $R\lesssim 60 {\rm pc}$ in model C while $R\lesssim 70 {\rm pc}$ in model CW, measured
 from the radius with a steep temperature gradient.
At $t=5.64\times 10^6{\rm yr}$,
 these are $R\lesssim 35 {\rm pc}$ in model C and $R\lesssim 60 {\rm pc}$ in model CW
 (see also Fig.\ref{fig:9}).
This shows that the bubble shrinks to a greater degree in models with strong heating rates.

Structures in the shell are significantly different: 
 the thermal pressure and the temperature in model CW are much lower than those
 in model C.
The low heating rate leads the shell to be cool and under low pressure.
High pressure in the shell hinders the expansion of the bubble in model C and
 thus the bubble size is reduced.   
The fact that the pressure in the shell controls the bubble radius means
 the contraction of the bubble is partly owing to the gas pressure in the shell. As a result, we conclude that for models with weak magnetic field 
$B_0\lesssim 1\mu{\rm G}$, the shell thickness increases and bubble size decreases as the heating rate increases.

Comparing models E and EW with $n_0=1{\rm cm^{-3}}$, 
even in the models with $B_0$=5$\mu$G, 
the size of the bubble depends on the heating rate as $R_{\rm bub}$=26.3 pc (model E), 29.9 pc (model EW) at $t=7.97\times10^5$yr. 
The radius of the bubble differs $\simeq 14\%$ for $R_{\rm bub}$ between models E and EW, and the total pressure along the $r$-axis in model E is twice as high as that in model EW at $t=7.97\times10^5$yr. 
The difference seems to come from a large difference in $\Gamma_0$ ($\Gamma_0$=$2.5\times 10^{-25}{\rm erg~s^{-1}}$ for model E and $5\times 10^{-27}{\rm erg~s^{-1}}$ for model EW). 
In high-density models, the effect of $\Gamma_0$ exceeds 
that of the magnetic field,
 and the evolution of the bubble is different in models E and EW, in contrast to models A and AW.

\subsection{Expansion Law}
\begin{figure}
\noindent
{\centering
(a)\hspace*{7cm}(b)\\
\plottwo{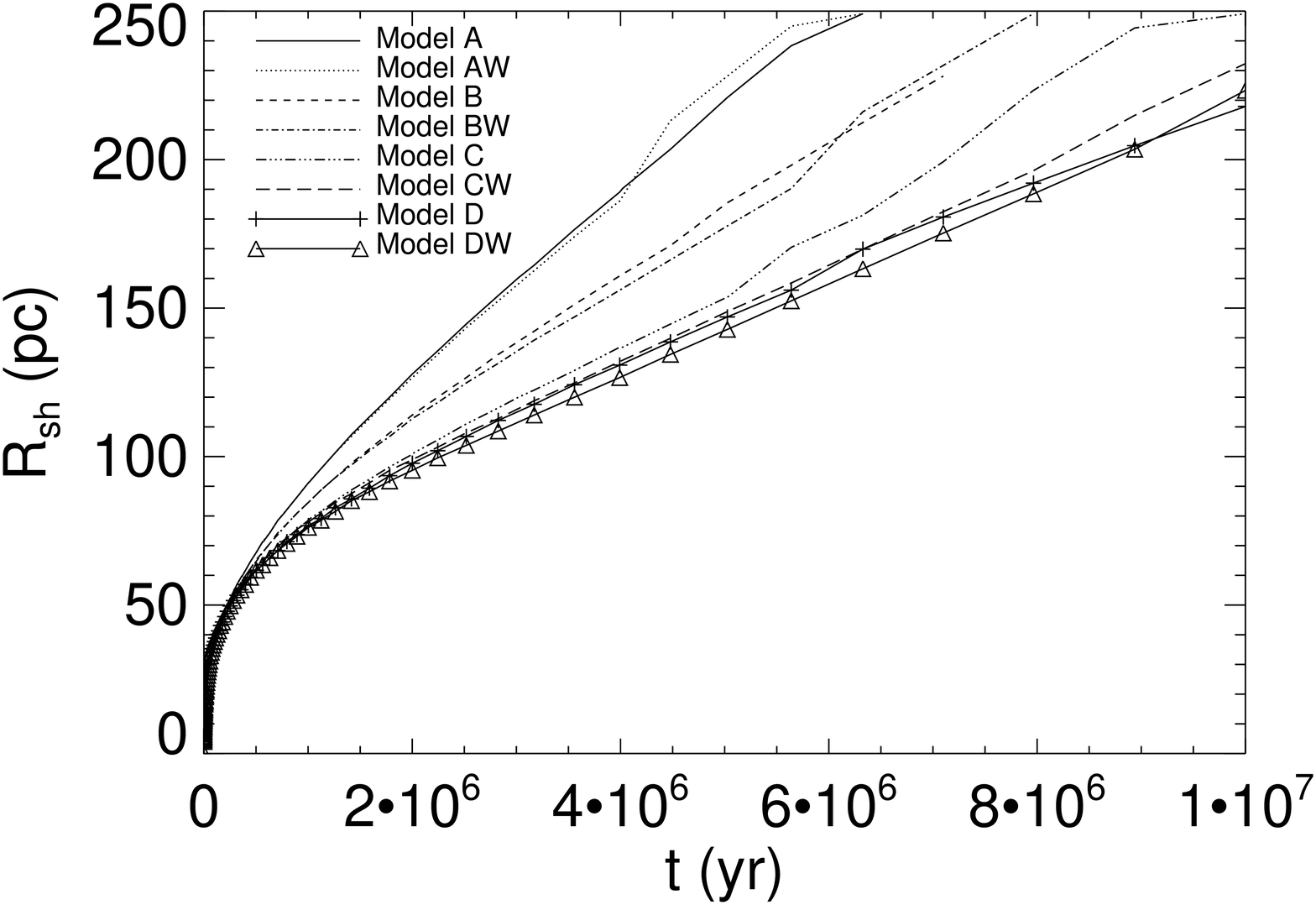}{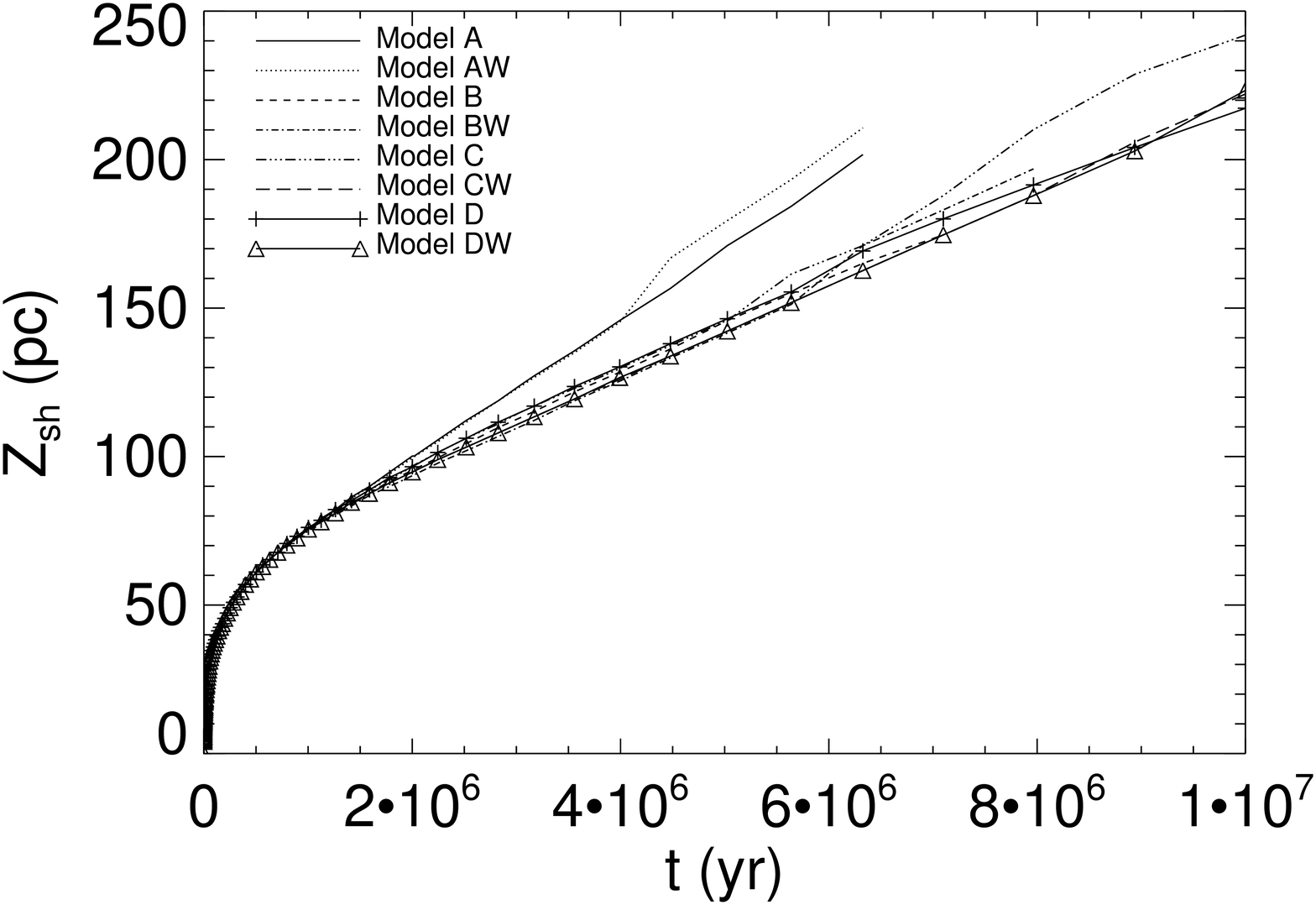}\\
(c)\hspace*{7cm}(d)\\
\plottwo{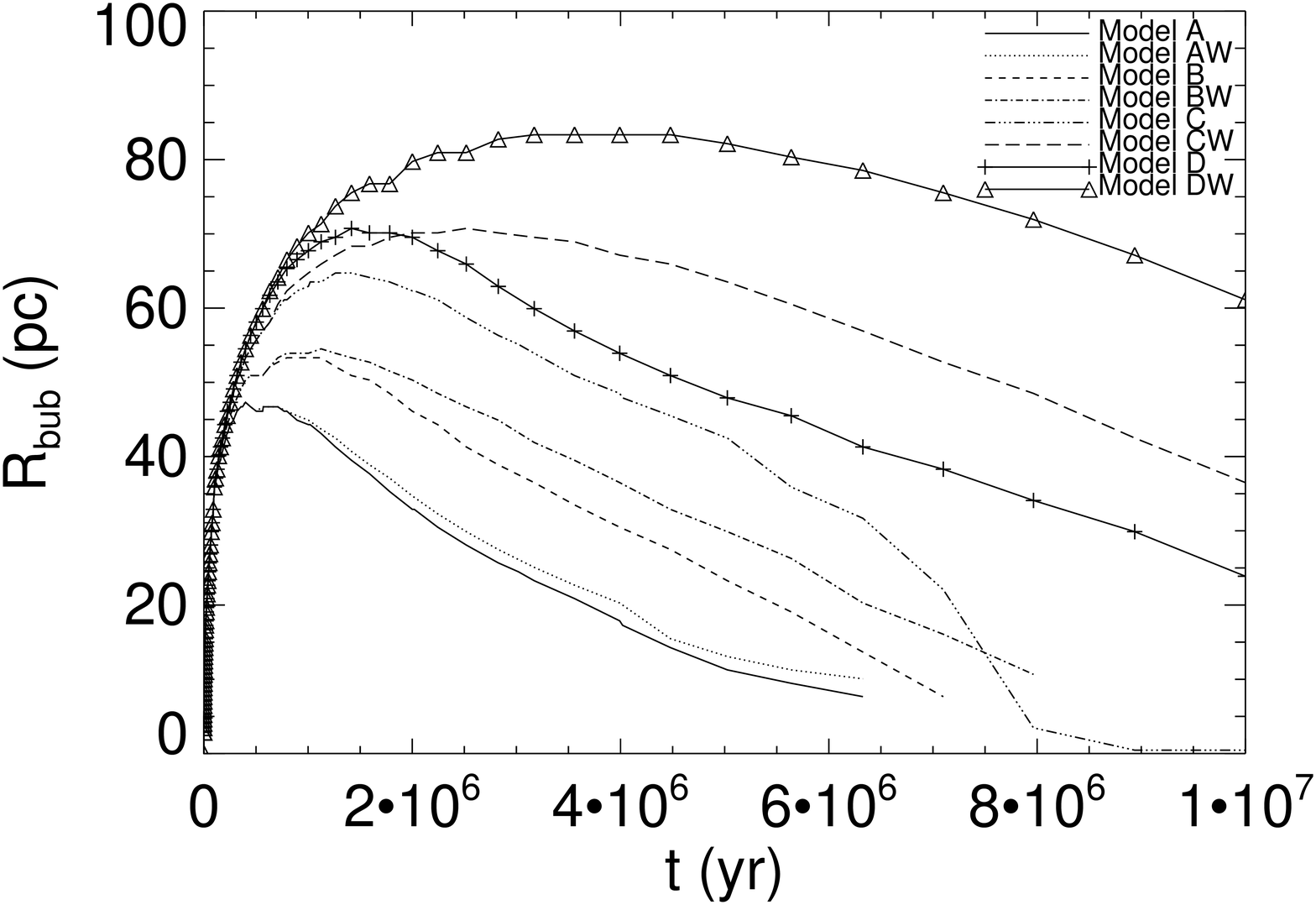}{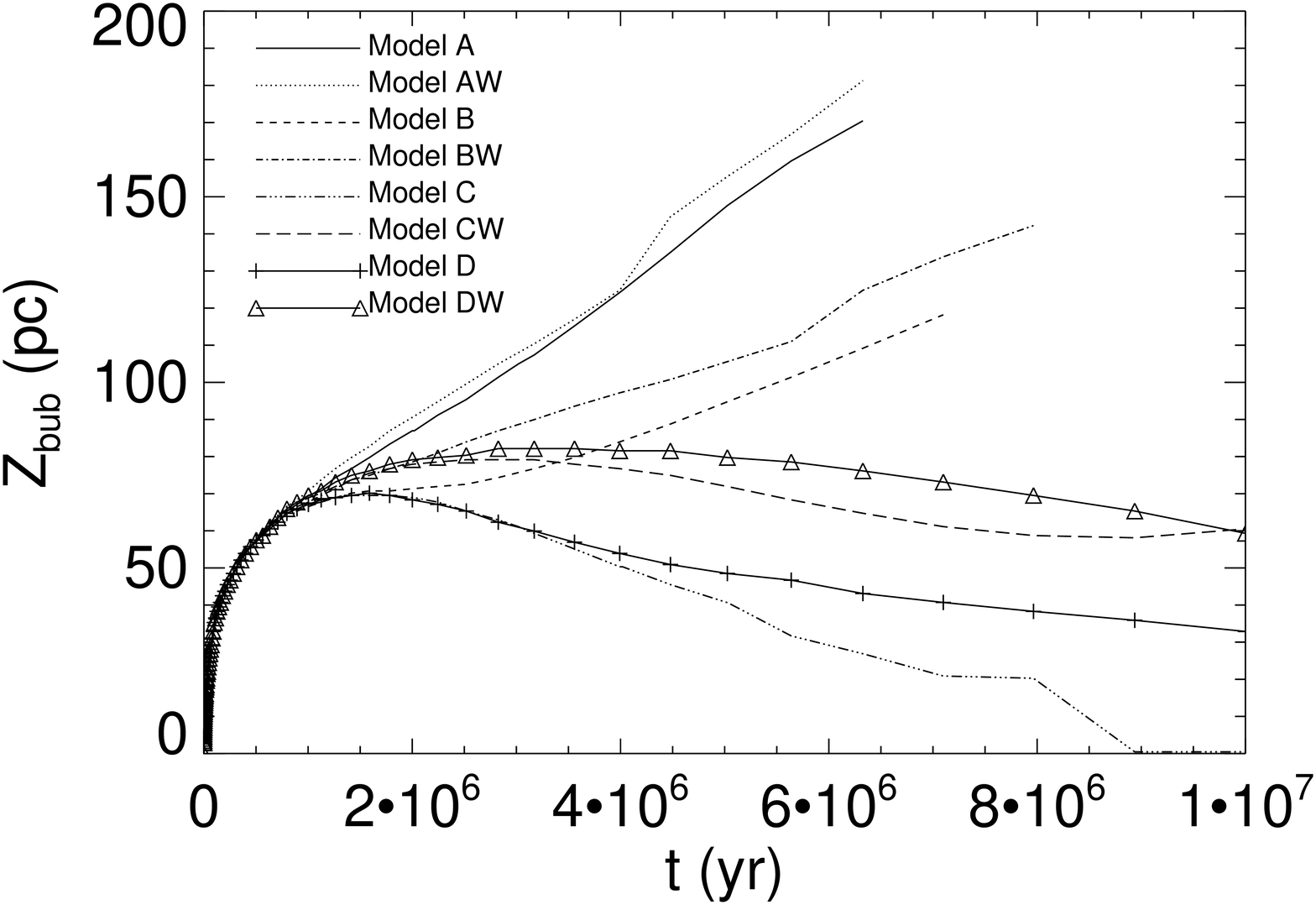}\\
}
\caption{
Expansion laws of the shell size, $R_{\rm sh}$ (a) and $Z_{\rm sh}$ (b) as well as 
 the size of the bubble, $R_{\rm bub}$ (c) and $Z_{\rm bub}$ (d) are plotted.
Models A-D and AW-DW are shown.
The figure shows that the expansion of the shell is affected by the
strength of the magnetic field $B_0$, especially in $R_{\rm sh}$.
A stronger magnetic field induces faster expansion, 
while, the bubble size (especially $R_{\rm bub}$) is affected
 by both the strength of the magnetic field and the heating rate.
\label{fig:9}}
\end{figure}

In Figure \ref{fig:9}, we plot the expansion law of the shell  $Z_{\rm sh}$ 
 and $R_{\rm sh}$ for respective directions of $z$- and $r$-axes as well as
 that of the bubble $Z_{\rm bub}$ and $R_{\rm bub}$ for various models.
Figure \ref{fig:9}(a)
 shows that the expansion of the shock front $R_{\rm sh}$ is affected by the 
 magnetic strength, if $n_0$ and $E_0$ are chosen to be the same.
That is, the expansion is promoted with the strength of the magnetic field.
This is related to the fact that the phase speed of the magneto-sonic wave propagating
 in the direction perpendicular to the magnetic field increases with the magnetic strength. 
However, the expansion of $R_{\rm sh}$ is independent of the value of the heating rate.

As for the shell expanding along the $z$-axis [Fig.\ref{fig:9}(b)],
 the models with $B_0\gtrsim 3\mu{\rm G}$ show expansions different from those
 of from the models with weaker
 magnetic fields $B_0\lesssim 3\mu{\rm G}$.
The expansion of $B_0\gtrsim 3\mu{\rm G}$ is promoted by the strength of the magnetic field, while the expansion speed in $Z_{\rm sh}$ for $B\lesssim 3\mu{\rm G}$ finally approaches the adiabatic thermal speed $c_s$. 
It has been pointed out that the late-phase expansion of $Z_{\rm sh}$ with constant speed 
 comes from the excess thermal pressure inside the bubble, which pushes the shell in the $z$-direction ($\S$\ref{sec:model A}).
Since the excess thermal pressure is well approximated
 by the ambient magnetic pressure as
 $B_0^2/8\pi\simeq 10^{-12}{\rm dyn\,cm^{-12}}(B_0/5\mu{\rm G})^2$,
 this excess becomes greater than the ambient thermal pressure
 $p_0\sim 5.8 \times 10^{-13}{\rm dyn\,cm^{-12}}(n_0/0.2{\rm cm^{-3}})(T_0/10^4{\rm K})$
 only for $B_0 \gtrsim 3.8 \mu {\rm G}$.
This seems to explain the numerical result that the difference in $Z_{\rm sh}$
 disappears for $B\lesssim 3\mu{\rm G}$. 

Figure \ref{fig:9}(c) and (d) shows the bubble sizes measured
 along the $r$- and $z$-axes, respectively.
Figure \ref{fig:9}(c) shows the expansion of $R_{\rm bub}$.
This shows that 
 the magnetic field suppresses the expansion of the bubble radius $R_{\rm bub}$, especially in the late phase evolution.
This is a natural consequence of the fact that
 the contraction of the bubble is driven by the magnetic pressure and tension forces.
Besides this, there is a clear difference in the expansions of models with strong heating
 (models A-D) and weak heating (models AW-DW).
Especially in models B(W) - D(W) with $B_0\lesssim 3\mu {\rm G}$,
 models with weak heating have larger radii $R_{\rm bub}$ compared
 with those with strong heating.
Since $R_{\rm sh}$ is not affected with $\Gamma_0$,
 this means that the shell becomes thinner in models with lower heating rates,
 for weak magnetic field models $B_0\lesssim 3\mu {\rm G}$.   
The fact that the shell density is higher in the weakly heated models BW-DW than in the strongly heated B-D is explained by the fact that higher density is necessary to equilibrate heating and
 cooling in the shell, which will be discussed in $\S$\ref{sec:4.2}.  

Figure \ref{fig:9}(d) shows the expansion of $Z_{\rm bub}$.
There are two distinctly different evolutions of the bubble expansion in the $z$-direction.
In models with $B_0\gtrsim 3\mu {\rm G}$, $Z_{\rm bub}$ [models A(W) and B(W)]
 continues to expand.
On the other hand, in models with $B_0\lesssim 1\mu {\rm G}$,
 $Z_{\rm bub}$ expands first but contracts after $(1.5-3) \times 10^6{\rm yr}$ 
 [models C(W) and D(W)].
Although all models C, CW, D, and DW show contraction in the final stage, 
 the models with ineffective heating (CW and DW) have larger bubble sizes
 in the $z$-direction than those with effective heating (models C and D), respectively.
This indicates models C and D have thicker shells than models CW and DW, respectively.
This is similar to the shells expanding in the $r$-direction. 
In models with strong magnetic field,
 $Z_{\rm bub}$ is larger for stronger magnetic field.
This may seem strange, since the magnetic field has no force working in the $z$-direction
 near the $z$-axis.
However, as already seen,
 the bubble is contracting in the $r$-direction owing to the magnetic tension force
 for $t\gtrsim 10^6{\rm yr}$ for models A(W) and B(W) [Fig.\ref{fig:9}(c)].
The thermal pressure in the hot interior, which mainly drives the shell near the $z$-axis,
 increases with increasing magnetic field strength.
This explains the fact that the inner boundary of the shell $Z_{\rm bub}$ expands
 faster for stronger magnetic fields in the models with a non-negligible magnetic field of $B_0 \gtrsim 3\mu {\rm G}$.

\section{Discussion}

\subsection{Porosity of ISM}

\begin{figure}
\noindent
{\centering
(a)\hspace*{8cm}(b)\\
\plottwo{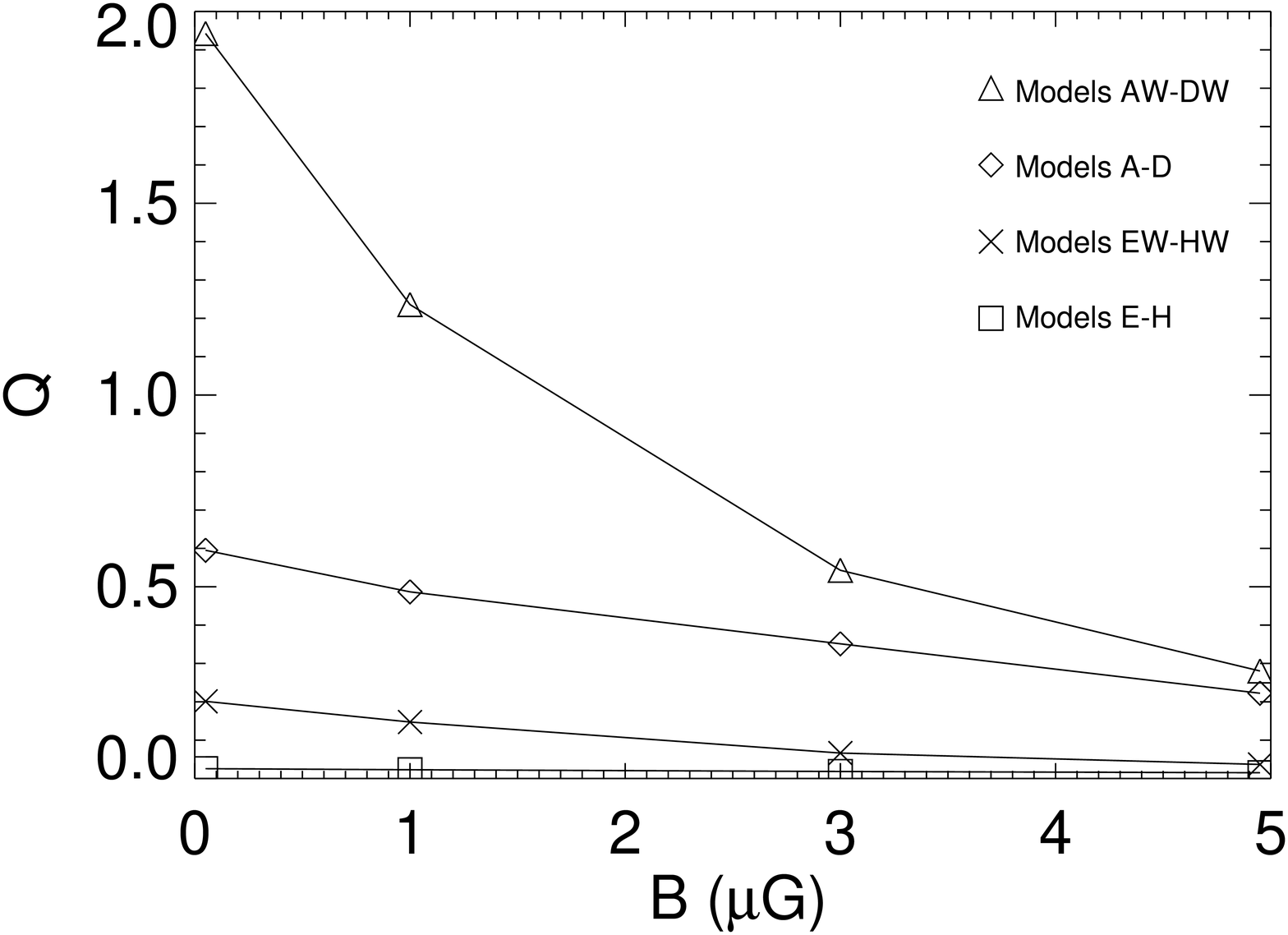}{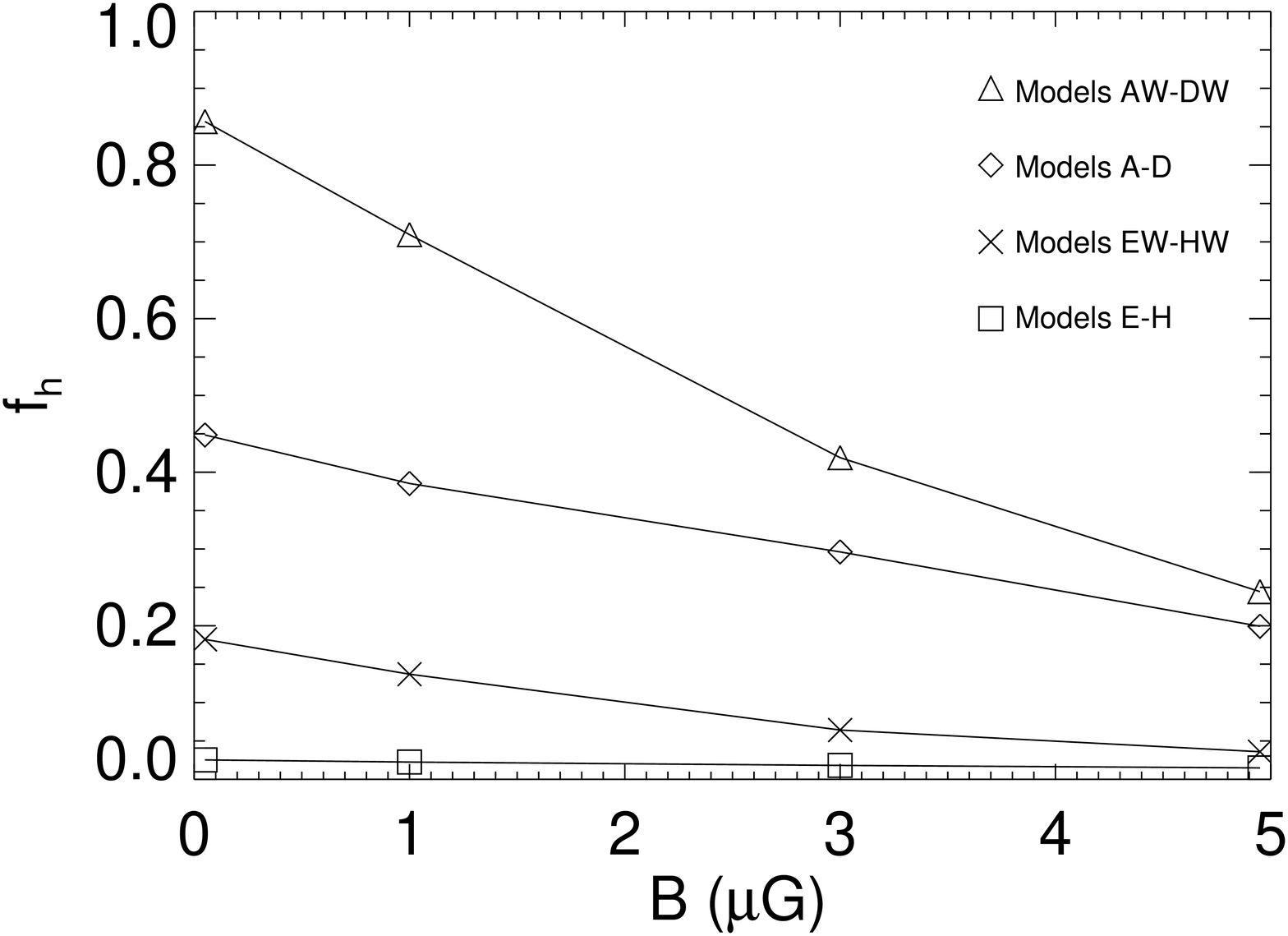}\\
}
\caption{
In panel (a), porosity of the volume occupied with the hot gas is plotted against the strength of the magnetic field. 
We have assumed the supernova rate of $r_{\rm SN}=10^{-13}{\rm yr^{-1}pc^{-3}}$.
 In panel (b), an anticipated volume-filling factor of the hot gas $f_{h}$ is shown. 
Diamonds (models A-D), squares (models E-H), triangles (models AW-DW), and crosses 
(models EW-HW) represent models with different $n_0$ and $\Gamma_0$. 
\label{fig:10}}
\end{figure}

To understand the structure of ISM, the volume fraction of the hot gas $f_{h}$ is a key issue.
The fraction of volumes covered by the hot cavities of SNRs is estimated by a quantity called
 porosity, which is defined as
\begin{equation}
Q \equiv r_{\rm SN}\int_0^{\infty} V(t)dt,
\end{equation} 
where $r_{\rm SN}$ is the supernova rate per volume as $\sim 10^{-13}{\rm yr^{-1}pc^{-3}}$
\citep{mck77} and $V(t)$ represents the volume covered by a hot cavity.
We assume all the SNRs end their lives when the cavities contract and $V(t)=0$. After obtaining the porosity, $f_{h}$ is estimated as $f_{h}=1-\exp(-Q)$.
In Figure \ref{fig:10}, we plot the porosity and the volume fraction of the hot gas against the magnetic field strength.
This shows clearly that the porosity decreases with the magnetic field strength.
Comparing models A-D (diamonds), $Q=0.6r_{-13}$ for non-magnetic ISM
 while $Q=0.22r_{-13}$ for $B_0=5\mu{\rm G}$.
Here $r_{-13}$ represents the SN rate per volume normalized by the standard value,
 $10^{-13}{\rm yr}^{-1}{\rm pc}^{-3}$.
The anticipated volume filling factor of hot gas $f_h=0.45$ for non-magnetic ISM (model D) is
 reduced to $f_h=0.2$ for  $B_0=5\mu{\rm G}$ (model A).
In models with ineffective heating rates (models AW-DW),
 the bubble is compressed more weakly by the thick shell than models A-D.
The porosity of these models is greater than that of models A-D.
In these models, the effect of the magnetic field in reducing porosity is more remarkable,
 that is, $Q=1.95r_{-13}$ for $B_0=0$ (model DW)
 while $Q=0.28r_{-13}$ for $B_0=5\mu{\rm G}$ (model AW).
This anticipates $f_h=0.86$ (model DW) and $f_h=0.24$ (model AW).

Even in models with ISM density of $n_0=1{\rm cm^{-3}}$, 
 the effect of the magnetic field in reducing porosity is evident.
Increasing $B_0$ from $1\mu{\rm G}$ to $5\mu{\rm G}$,
 porosity decreases by a factor of 1.7 from $Q=0.026r_{-13}$ to
 $Q=0.015r_{-13}$.
In the case where the average density of the ISM is high,
 the volume fraction of hot ISM is restricted unless the supernova rate is much higher than that of our Galaxy. 
%To summarize this section,
Consequently,
 it is shown that the volume fraction of the hot gas is restricted as $f_h\sim 0.2$ 
 by the effect of the magnetic field, even if the average ISM density is as low
 as $n_0\simeq 0.2{\rm cm^{-3}}$. 

Observationally, in the solar neighborhood, the ``not strongly absorbing'' 
H{\sc I} gas contains large holes, perhaps ranging up to 400 pc diameter, and 
the large holes occupy 10-20\% of the volume, $f_h\simeq0.1-0.2$ \citep{hei80}. 
For M31, another Sb galaxy, the filling factor which is derived for the H{\sc I} holes, which are found in the range from 7 to 16 kpc from the nucleus, is about 1\%, $f_h\simeq0.01$ \citep{bri86}. 
For M33, an Sc galaxy, the filling factor for H{\sc I} holes, allowing a scaling factor of 2 for non-detected holes, is less than 40\%, $f_h\lesssim0.4$ \citep{deu90}.
H{\sc I} holes arise by OB associations with multiple supernova explosions 
that produce supercavities, 
and are also produced by the independent, random individual supernova explosions \citep{hei87}.
In this section, we considered the latter case, and the porosity and the filling factor of models for $B_0=5\mu{\rm G}$ presumably looks consistent with the observational data.

\subsection{Thermal Equilibrium}
\label{sec:4.2}

To understand how the shell width is determined,
 we consider the gas in thermal equilibrium.
Balance between the radiative cooling $\Lambda(T)n^2$ and heating $\Gamma_0 n$ leads
\begin{equation}
 \frac{\Lambda(T)}{T}=\frac{k\Gamma_0}{p},\label{eqn:equil}
\end{equation}
where $k$ represents the Boltzmann constant.
The cooling function below $T\lesssim 10^4\,{\rm K}$ is an increasing function 
of temperature and we assume $\Lambda=\Lambda_0 (T/10^4\,{\rm K})^q$, 
 ignoring the dependence of the ionization fraction $x$.
Equation (\ref{eqn:equil}) gives the equilibrium pressure proportional to
\begin{eqnarray}
p &\propto& \Gamma_0 T^{1-q}\\ 
  &\propto&{\Gamma_0}^{1/q} n^{(q-1)/q}. 
\end{eqnarray}
If we assume the post-shock gas is cooled and is in a thermal equilibrium
 [eq.(\ref{eqn:equil})] under the post-shock pressure, it is shown that
 the density decreases with the deceleration of the shock velocity as
\begin{equation}
 n \propto \frac{p^{q/(q-1)}}{{\Gamma_0}^{1/(q-1)}}
\propto \frac{n_0^{q/(q-1)} {V_s}^{2q/(q-1)} }{{\Gamma_0}^{1/(q-1)}},\label{eqn:equiln}
\end{equation}
where we used the relation between the shock velocity $V_s$
 and the post-shock gas pressure $p$. 
Since the power $q=3$, the shell density $n$ decreases as the shell expands with time, since the shock velocity $V_s$ decreases with time.

In this discussion we have ignored the magnetic field, which works to prevent compression.
In the thick shell sweeping the magnetic field, dynamics is controlled by the magnetic field as shown in model A. 
In the extreme case, the change in density is controlled under the magnetic field. 
Although equation (\ref{eqn:equil}) does not hold near the shock front,
 the temperature and the thermal pressure reach the equilibrium values expected from equation (\ref{eqn:equil}) near the shell-bubble boundary.
We calculate the equilibrium density under the pressure $p$ in the shell, $\rho_{eq}(r)$ along the $r$-axis. 
The difference from this equilibrium density is given as $\Delta = (\rho(r)-\rho_{eq}(r))/\rho_{eq}(r)$.
The volume average of $\Delta$ decreases with time as 0.91,0.67,0.14, and 0.07 
at $t=1\times10^5, 3.99\times10^5, 2.52\times10^6$, and $5.64\times10^6 \rm{yr}$ respectively in model A.
This shows that, with time, the density of the shell approaches the state of equilibrium expected from equation (\ref{eqn:equil}). 

\acknowledgments
Numerical calculations were carried out by VPP5000 in 
 the Astronomical Data Analysis Center,
 National Astronomical Observatory of Japan.
This work was partially supported by Grants-in-Aid for Scientific Research
 from MEXT, Japan (14540233,16036206).
The authors would like to thank the CANS (Coordinated Astronomical Numerical Software)
 project who provided us with the MHD simulation code.

\clearpage

\appendix
\section{Rankin-Hugoniot Relation}
In the case that the shock front is proceeding in the direction perpendicular to the 
 magnetic field ($x$-direction),
 the mass flux conservation advected with fluid motion is written as  
\begin{equation}
\rho_1 v_{x1}=\rho_2 v_{x2},\label{eqn:A1}
\end{equation}
where subscripts $1$ and $2$ denote the quantities of pre-shock and post-shock,
 respectively, and velocities are measured from the shock front which is moving
 with $V_s$. 
Other momentum, energy, and magnetic flux conservations are written as  
\begin{equation}
p_1+\frac{B_1^2}{8\pi}+\rho_1v_{x1}^2=p_2+\frac{B_2^2}{8\pi}+\rho_2v_{x2}^2,
\label{eqn:A2}
\end{equation}
\begin{equation}
v_{x1}\left(\frac{1}{2}\rho_1v_{x1}^2+\frac{\gamma}{\gamma-1}p_1\right)
 +\frac{v_{x1}B_1^2}{4\pi}=
v_{x2}\left(\frac{1}{2}\rho_2v_{x2}^2+\frac{\gamma}{\gamma-1}p_2\right)
 +\frac{v_{x2}B_2^2}{4\pi},\label{eqn:A3}
\end{equation}
\begin{equation}
v_{x1}B_1=v_{x2}B_2.\label{eqn:A4}
\end{equation}
Using 
\begin{eqnarray}
x&\equiv&\frac{\rho_2}{\rho_1}=\frac{v_{x1}}{v_{x2}}=\frac{B_2}{B_1},\\
y&\equiv&\frac{p_2}{p_1},
\end{eqnarray}
 the above equations (\ref{eqn:A1})-(\ref{eqn:A4}) give following relations \citep{pri82,sak96}
\begin{equation}
\frac{2(2-\gamma)}{\beta_1}x^2 +\gamma\left[\left(\gamma-1\right){{\cal M}}^2
 +\frac{2}{\beta_1}+2\right]x-\gamma\left(\gamma+1\right){{\cal M}}^2=0,
\end{equation}
\begin{equation}
y=\gamma{{\cal M}}^2\left(1-\frac{1}{x}\right)-\frac{x^2-1}{\beta_1}+1,\label{eqn:y}
\end{equation}
where ${\cal M}=V_s/(\gamma p_1/\rho_1)^{1/2}$ and $\beta_1=8\pi p_1/B_1^2$
 represent the Mach number of the shock speed against the pre-shock sound speed
 and the plasma beta for the pre-shock medium, respectively.  

\begin{figure}
{\centering
\plotone{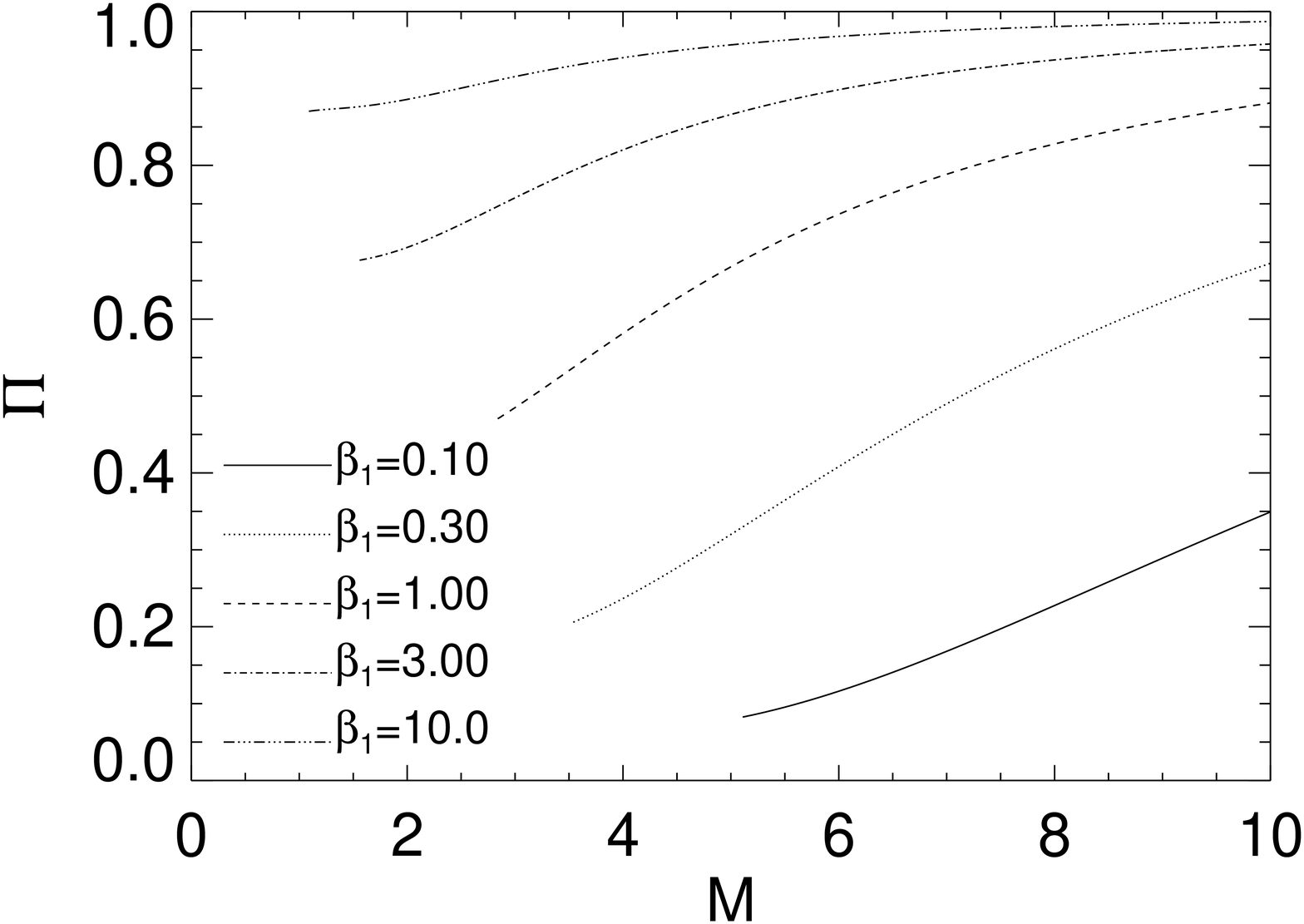}
}
\caption{
Rankin-Hugoniot relation for the MHD shock propagating perpendicular to the magnetic field. 
The post-shock gas pressure normalized by that attained in the non-magnetic shock ($\Pi\equiv y(\beta_1)/y(\beta_1\rightarrow \infty)$)
 is plotted against the shock Mach number (${\cal M}$). 
Each curve corresponds to different plasma $\beta$. 
\label{fig:A1}}
\end{figure}

We plot the ratio of the normalized post-shock pressure $y$
 to that of non-magnetized shock 
 $y(\beta_1\rightarrow\infty)$ as $\Pi\equiv y(\beta_1)/y(\beta_1\rightarrow\infty)$
 against the shock Mach number ${\cal M}$ in Figure \ref{fig:A1}.
Even if ${\cal M}$ is fixed at 10, $\Pi$ increases with $\beta_1$ as $\Pi$=
0.35, 0.67, 0.88, 0.96, and 0.99
 for $\beta_1$=
0.1, 0.3, 1.0, 3.0, and 10, respectively.
This indicates that small ${\cal M}$ and $\beta_1$ 
 (slow shock speed and strong magnetic field) induce the ratio $\Pi$ to decrease,
 or in other words, the effect of the magnetic field becomes evident.   

Using the Sedov solution,
 the expansion of the shock front is given as
\begin{equation}
R_s(t)=1.15\left(\frac{E_0}{\rho_0}\right)^{1/5}t^{2/5},
\end{equation}
for the radius and
\begin{equation}
V_s(t)=0.46\left(\frac{E_0}{\rho_0}\right)^{1/5}t^{-3/5},
\end{equation}
for the expansion speed. 
These give the following typical values for $R_s$ and $V_s$ as
\begin{eqnarray}
R_s&=&38.5{\rm pc}\left(\frac{E_0}{5\times 10^{50}{\rm erg}}\right)^{1/5}
      \left(\frac{n_0}{0.2{\rm cm^{-3}}}\right)^{-1/5}
      \left(\frac{t}{10^5{\rm yr}}\right)^{2/5},\\
V_s&=&151{\rm km\,s^{-1}}\left(\frac{E_0}{5\times 10^{50}{\rm erg}}\right)^{1/5}
      \left(\frac{n_0}{0.2{\rm cm^{-3}}}\right)^{-1/5}
      \left(\frac{t}{10^5{\rm yr}}\right)^{-3/5}.\label{eqn:Vs}
\end{eqnarray}
Using the Mach number of the shock expansion speed ${\cal M}=V_s/c_s$ 
and the plasma beta of the pre-shock
 ISM, the post-shock gas pressure is obtained in relation to the pre-shock pressure.

\clearpage
%%%%%%%%% TABLE %%%%%%%%%

\begin{table}[h]
\begin{center}
\caption{Model parameters.\label{tab:1}}
\begin{tabular}{lcccc}
\tableline
Model & $n_0$ & $T_0$ & $B_0$ & $\Gamma_0$ \\
      & $({\rm cm^{-3}})$ & (K) & $(\mu {\rm G})$ & $(\rm erg\,s^{-1})$\\
\tableline\tableline
A\ldots\ldots    & 0.2 & $10^4$ & 5 & $5\times 10^{-26}$\\
B\ldots\ldots    & 0.2 & $10^4$ & 3 & $5\times 10^{-26}$\\
C\ldots\ldots    & 0.2 & $10^4$ & 1 & $5\times 10^{-26}$\\
D\ldots\ldots    & 0.2 & $10^4$ & 0 & $5\times 10^{-26}$\\
AW\ldots\ldots    & 0.2 & $10^4$ & 5 & $5\times 10^{-27}$\\
BW\ldots\ldots    & 0.2 & $10^4$ & 3 & $5\times 10^{-27}$\\
CW\ldots\ldots    & 0.2 & $10^4$ & 1 & $5\times 10^{-27}$\\
DW\ldots\ldots    & 0.2 & $10^4$ & 0 & $5\times 10^{-27}$\\
E\ldots\ldots    & 1 & $10^4$ & 5 & $2.5\times 10^{-25}$\\
F\ldots\ldots    & 1 & $10^4$ & 3 & $2.5\times 10^{-25}$\\
G\ldots\ldots    & 1 & $10^4$ & 1 & $2.5\times 10^{-25}$\\
H\ldots\ldots    & 1 & $10^4$ & 0 & $2.5\times 10^{-25}$\\
EW\ldots\ldots    & 1 & $10^4$ & 5 & $5\times 10^{-27}$\\
FW\ldots\ldots    & 1 & $10^4$ & 3 & $5\times 10^{-27}$\\
GW\ldots\ldots    & 1 & $10^4$ & 1 & $5\times 10^{-27}$\\
HW\ldots\ldots    & 1 & $10^4$ & 0 & $5\times 10^{-27}$\\
\tableline
\end{tabular}
\end{center}
\end{table}
\clearpage
%
% FIG.1
%

%
% FIG 2
%
%
% FIG 3
%

%
% FIG.4
%

%
% FIG.5
%

%
% FIG.6
%

%
% FIG.7 
%

%
% FIG.8
%

%
% FIG.9
%

%
% FIG.10
%

\end{document}